\documentclass[12pt]{article}
\usepackage[dvips]{graphicx}

\newcommand{\beq}{\begin{equation}}
\newcommand{\eeq}{\end{equation}}
\newcommand{\bqa}{\begin{eqnarray}}
\newcommand{\eqa}{\end{eqnarray}}

\newcommand{\bra}[1]{\langle{#1}|}
\newcommand{\ket}[1]{|{#1}\rangle}

\title{Entangled states and collective nonclassical effects
in two-atom systems}
\author{Z. Ficek$^*$ and R. Tana\'s$^{\dagger}$}
\date{$^*$Department of Physics, School of Physical Sciences,
The University of Queensland, Brisbane, QLD 4072, Australia\\
\vspace{0.5cm} 
$^{\dagger}$Nonlinear Optics Division, Institute of Physics,
Adam Mickiewicz University, Umultowska 85, 61-614 Pozna\'n, Poland}

\begin{document}

\maketitle

\begin{abstract}
We propose a review of recent developments on entanglement and
non-classical effects in collective two-atom systems and present a
uniform physical picture of the many predicted
phenomena. The collective effects have brought into sharp focus some of
the most basic features of quantum theory, such as nonclassical states
of light and entangled states of multiatom systems. The entangled states
are linear superpositions of the internal states of the system which
cannot be separated into product states of the individual atoms.
This property is recognized as entirely quantum-mechanical effect and
have played a crucial role in many discussions of the nature of quantum
measurements and, in particular, in the developments of quantum
communications. Much of the fundamental interest in entangled states is
connected with its practical application ranging from quantum computation,
information processing, cryptography, and interferometry to atomic
spectroscopy.
\end{abstract}

\newpage

\section{Introduction}\label{ftsec1}

A central topic in the current studies of collective effects in multi-atom
systems are the theoretical investigations and experimental implementation
of entangled states to quantum computation and quantum information
processing~\cite{pv}. The term entanglement, one of the most intriguing
properties of multiparticle systems, was introduced by
Schr\"{o}dinger~\cite{schr} in his discussions of the foundations of
quantum mechanics. It describes a multiparticle system
which has the astonishing property that the results of a measurement on one
particle cannot be specified independently of the results of measurements on
the other particles. In recent years, entanglement has become of interest not
only for the basic understanding of quantum mechanics, but also because it
lies at the heart of many new applications ranging from quantum
information~\cite{beke,pok}, cryptography~\cite{ek} and quantum
computation~\cite{bar,gro} to atomic and molecular
spectroscopy~\cite{boll1,boll2}. These practical implementations all stem
from the
realization that we may control and manipulate quantum systems at the
level of single atoms and photons to store and transfer information in
a controlled way and with high fidelity.

All the implementations of entangled atoms must contend with the conflict
inherent to open systems. Entangling operations on atoms must provide
strong coherent coupling between the atoms, while shielding the atoms
from the environment in order to make the effect of decoherence and
dissipation negligible. The difficulty of isolating the atoms from the
environment is the main obstacle inhibiting practical applications
of entangled states. The environment consists of a continuum of
electromagnetic
field modes surrounding the atoms. This gives rise to decoherence that
leads to the loss of information stored in the system. However, it
has been recognised that the collective properties of multi-atom systems
can alter spontaneous emission compared with the single atom case.
As it was first pointed out by Dicke~\cite{dic}, the interaction
between the
atomic dipoles could cause the multiatom system to decay with two
significantly different, one enhanced and the other reduced,
spontaneous emission rates. The presence of the reduced spontaneous
emission rate induces a reduction of the linewidth of the spectrum of
spontaneous emission~\cite{ftk87,buz}. This reduced (subradiant)
spontaneous emission implies that the multi-atom system can decohere
slower compared with the decoherence of individual atoms.

Several physical realisations of entangled atoms have been proposed
involving trapping and cooling of a small number of ions or neutral
atoms~\cite{eich,deb,ber,tos}. This is the case with the lifetime of
the superradiant and subradiant states that have been demonstrated
experimentally with two barium ions confined in a spherical Paul
trap~\cite{eich,deb}. The reason for using cold trapped atoms or ions
is twofold. On the one hand, it has been realised that the trapped atoms
are essentially motionless and lie at a known and controllable distance
from one another, permitting qualitatively new studies of interatomic
interactions not accessible in a gas cell or an atomic beam~\cite{ag70}.
The advantage of the
trapped atoms is that it allows to separate collective effects,
arising from the correlations between the atoms, from the single-atom
effects. On the other hand it was discovered that cold trapped atoms
can be prepared in maximally entangled states that are isolated from
its environment~\cite{bbtk,afs,bcjd,bdj,tbk}.

An example of maximally entangled states in a two-atom system are the
superradiant and subradiant states, which correspond to the symmetric
and antisymmetric combinations of the atomic dipole moments,
respectively. These states are created by the interaction between the
atoms and are characterized by different spontaneous decay rates that
the symmetric state decays with an enhanced, whereas the antisymmetric
state decays with a reduced spontaneous emission rate. The reduced
spontaneous emission rate of the antisymmetric state implies that the
state is weakly coupled to the environment. For the case of the atoms
confined into the region much smaller than the optical wavelength
(Dicke model), the antisymmetric state is completely decoupled from
the environment, and therefore can be regarded as a decoherence-free
state.

Another particularly interesting entangled states of a two-atom
system are two-photon entangled states that are superpositions of
only those states of the two-atom system in which both or neither of
the atoms is excited. These states have been known for a long time as
pairwise atomic states or multi-atom squeezed
states~\cite{mas,pas1,pas2,pas3,pas4,pas5}.
The two-photon entangled states cannot be generated
by a coherent laser field coupled to the atomic dipole moments. The
states can be created by a two-photon excitation process with
nonclassical correlations that can transfer the population from the
two-atom ground state to the upper state without populating the
intermediate one-photon states. An obvious candidate for the creation
of the two-photon entangled states is a broadband squeezed vacuum
field which is characterised by strong nonclassical two-photon
correlations~\cite{park,fd1,fd2}.

One of the fundamental interests in collective atomic effects is to
demonstrate creation of entanglement on systems containing only two atoms.
A significant body of work on preparation of a two-atom system in an
entangled state has accumulated, and two-atom entangled states have
already been demonstrated experimentally using ultra cold trapped
ions in free space~\cite{deb,tur} and cavity quantum electrodynamics
(QED) schemes~\cite{hag,rbh}. In the free space situation, the collective
effects arise from the interaction between the atoms through the vacuum
field that the electromagnetic field produced by one of the atoms
influences the dipole moment of the another atom. This leads to an
additional damping and a shift of the atomic levels that both depend
on the interatomic separation. In the cavity QED
scheme, the atoms interact through the cavity mode and in a good cavity
limit, photons emitted by one of the atoms are almost immediately
absorbed by the another atom. In this case, the system behaves like
the Dicke model. Moreover, the strong coupling of the atoms to
the cavity mode prevents the atoms to emit photons to the vacuum
modes different from the cavity mode that reduces decoherence.

Recently, the preparation of correlated superposition
states in multi-atom system has been performed using a quantum nondemolition
(QND) measurement technique~\cite{kmb}. Osnaghi {\it et al.}~\cite{osna}
have demonstrated coherent control of two Rydberg
atoms in a non-resonant cavity environment. By adjusting the atom-cavity
detuning, the final entangled state could be controlled, opening the door
to complex entanglement manipulations~\cite{ymbrh}. Several proposals have
also been made for entangling atoms trapped in distant
cavities~\cite{mb1,mb2,mb3,mb4,mb5,prsg}, or in a Bose-Einstein
condensate~\cite{fg1,fg2}. In a very important experiment,
Schlosser {\it et al.}~\cite{schl} succeeded in confining single atoms in
microscopic traps, thus enhancing the possibility of further progress in
entanglement and quantum engineering.

This review is concerned primarily with two-atom systems, since
it is generally believed that entanglement of only two microscopic
quantum systems (two qubits) is essential to implement quantum
protocols such as quantum computation.
Some description of the theoretical tools required for prediction of
entanglement in atomic systems is appropriate. Thus, we propose to begin
the review with an overview of the mathematical apparatus necessary
for describing the interaction of atoms with the electromagnetic field.
We will present the master equation technique and, in addition, we also
describe a more general formalism based on the
quantum jump approach. We review theoretical and experimental schemes
proposed for the preparation of two two-level atoms in an entangled state.
We will also relate the atomic entanglement to nonclassical effects
such as photon antibunching, squeezing and sub-Poissonian photon
statistics. In particular, we consider different schemes of generation
of entangled and nonclassical states of two identical as well as
nonidentical atoms. The cases of maximally and non-maximally entangled
states will be considered and methods of detecting of particular entangled
and nonclassical state of two-atom systems are discussed. Next, we will
examine methods of preparation of a two-atom system in two-photon
entangled states. Finally, we will discuss methods of mapping of the
entanglement of light on atoms involving collective atomic
interactions and squeezing of the atomic dipole fluctuations.

\section{Time evolution of a collective atomic system}\label{ftsec2}

The standard formalism for the calculations of the time evolution and
correlation properties of a collective system of atoms is the master
equation method. In this approach, the dynamics are studied in terms of
the reduced density operator $\hat{\rho}_{A}$ of the atomic system
interacting with the quantized
electromagnetic (EM) field regarded as a reservoir~\cite{leh,lui,ag74}.
There are many possible realizations of reservoirs. The typical reservoir
to which atomic systems are coupled is the quantized three-dimensional
multimode field. The reservoir can be modelled as a vacuum field whose
the modes are in ordinary vacuum states, or in thermal states, or even
in squeezed vacuum states. The major advantage of the master equation is
that it allows us to consider the evolution of the atoms plus field
system entirely in terms of average values of atomic operators. We can
derive equations of motion for expectation values of an arbitrary
combination of the atomic operators, and solve these equations for
time-dependent averages or the steady-state. Another method is the
quantum jump approach. This is based on the theory of quantum
trajectories~\cite{car93}, which is equivalent to the Monte Carlo
wave-function approach~\cite{dcm92,pk98}, and allows to predict all
possible trajectories of a single quantum system which stochastically
emits photons. Both methods, the master equation and quantum jumps
approaches lead to the same final results of the dynamics of an atomic
system, and are widely used in quantum optics.

\subsection{Master equation approach}\label{ftsec21}

We first give an outline of the derivation of the master equation of
a system of $N$ non-identical nonoverlapping atoms coupled to the
quantized three-dimensional EM field. This derivation is a generalisation
of the master equation technique, introduced by Lehmberg~\cite{leh}, to
the case of non-identical atoms interacting with a squeezed vacuum field.
Useful references on the derivation of
the master equation of an atomic system coupled to an ordinary vacuum
are the books of Louisell~\cite{lui} and Agarwal~\cite{ag74}.
The atoms are modelled as two-level systems, with excited state
$\ket {e_{i}}$, ground state $\ket {g_{i}}$, transition frequency
$\omega_{i}$, and transition dipole moments $\vec{\mu}_{i}$. We assume
that the atoms are located at different points $\vec{r}_{1},\dots
\vec{r}_{N}$, have different transition frequencies $\omega_{1}\neq
\omega_{2}\neq \dots \neq \omega_{N}$, and different transition dipole
moments $\vec{\mu}_{1}\neq \vec{\mu}_{2}\neq \dots \neq \vec{\mu}_{N}$.

In the electric dipole approximation, the total Hamiltonian of the
combined system, the atoms plus the EM field, is given by
\begin{eqnarray}
        \hat{H} &=& \sum_{i=1}^{N}\hbar \omega_{i}S_{i}^{z} +
        \sum_{\vec{k}s} \hbar \omega_{k}\left(\hat {a}_{\vec{k}s}^{\dagger}
        \hat {a}_{\vec{k}s} +\frac{1}{2}\right) \nonumber \\
       &-& i\hbar\sum_{\vec{k}s}\sum_{i=1}^{N}\left[ \vec{\mu}_{i}\cdot
\vec{g}_{\vec{k}s}\left(\vec{r}_{i}\right) \left(
S_{i}^{+}+S_{i}^{-}\right)\hat{a}_{\vec{k}s}
-{\rm H.c.}\right]  \ ,\label{t1}
\end{eqnarray}
where $S_{i}^{+} = \ket {e_{i}}\bra {g_{i}}$ and $S_{i}^{-} =
\ket {g_{i}}\bra {e_{i}}$ are the dipole raising and lowering
operators, $S_{i}^{z} =\left(\ket {e_{i}}
\bra {e_{i}} -\ket {g_{i}}\bra {g_{i}}\right)/2$ is the energy operator
of the $i$th atom, $\hat {a}_{\vec{k}s}$ and
$\hat {a}_{\vec{k}s}^{\dagger}$ are the annihilation and creation
operators of the field mode~$\vec{ks}$, which has wave vector $\vec{k}$,
frequency $\omega_{k}$ and the index of polarization~$s$. The coupling
constant
\begin{equation}
\vec{g}_{\vec{k}s}\left( \vec{r}_{i}\right) =\left( \frac{\omega_{k}}{2
\epsilon_{0}\hbar V}\right) ^{\frac{1}{2}}
\bar{e}_{\vec{k}s}e^{i\vec{k}\cdot \vec{r}_{i}} \ , \label{t2}
\end{equation}
is the mode function of the three-dimensional vacuum field,
evaluated at the position $\vec{r}_{i}$ of the $i$th atom,
$V$ is the normalization volume, and $\bar{e}_{\vec{k}s}$ is the unit
polarization vector of the field.

The atomic dipole operators, appearing in Eq.~(\ref{t1}), satisfy
the well-known commutation and anticommutation relations
\begin{equation}
\left[S_{i}^{+}, S_{j}^{-}\right] =2S_{i}^{z}\delta_{ij} \ ,\quad
\left[S_{i}^{z}, S_{j}^{\pm}\right] =\pm S_{i}^{\pm}\delta_{ij} \
,\quad \left[S_{i}^{+}, S_{j}^{-}\right]_{+} =\delta_{ij}
     \ ,\label{t3}
\end{equation}
with $\left(S_{i}^{\pm}\right)^{2}\equiv 0$.

While this is straightforward, it is often the case that it is simpler
to work in the interaction picture in which the
Hamiltonian~(\ref{t1}) evolves in time according to the interaction
with the vacuum field. Therefore, we write the total Hamiltonian~(\ref{t1})
as
\begin{eqnarray}
        \hat{H} &=& \hat{H}_{0} +\hat{H}_{I} \ , \label{t4}
\end{eqnarray}
where
\begin{eqnarray}
        \hat{H}_{0} &=& \sum_{i=1}^{N}\hbar \omega_{i}S_{i}^{z}
        +\sum_{\vec{k}s} \hbar \omega_{k}\left(\hat {a}_{\vec{k}s}^{\dagger}
        \hat {a}_{\vec{k}s} +\frac{1}{2}\right) \ ,\label{t5}
\end{eqnarray}
is the Hamiltonian of the non-interacting atoms and the EM field, and
\begin{eqnarray}
        \hat{H}_{I} &=&
       -i\hbar\sum_{\vec{k}s}\sum_{i=1}^{N}\left[ \vec{\mu}_{i}\cdot
\vec{g}_{\vec{k}s}\left(\vec{r}_{i}\right)
\left(S_{i}^{+}+S_{i}^{-}\right)\hat{a}_{\vec{k}s}
-{\rm H.c.}\right] \ ,\label{t6}
\end{eqnarray}
is the interaction Hamiltonian between the atoms and the EM field.

The Hamiltonian $\hat{H}_{0}$ transforms the total Hamiltonian~(\ref{t1}) into
\begin{eqnarray}
        \hat{H}\left(t\right) &=& e^{i\hat{H}_{0}t/\hbar}
\left(\hat{H}-\hat{H}_{0}\right)e^{-i\hat{H}_{0}t/\hbar} =
\hat{V}\left(t\right) \ ,\label{t7}
\end{eqnarray}
where
\begin{eqnarray}
       \hat{V}\left(t\right) &=& -i\hbar\sum_{\vec{k}s}\sum_{i=1}^{N}
        \left\{ \vec{\mu}_{i}\cdot
\vec{g}_{\vec{k}s}\left(\vec{r}_{i}\right)
S_{i}^{+}\hat{a}_{\vec{k}s}e^{-i\left(\omega_{k}-\omega_{i}\right)t}\right.
\nonumber \\
&&\left. +\vec{\mu}_{i}\cdot
\vec{g}_{\vec{k}s}\left(\vec{r}_{i}\right)S_{i}^{-}\hat{a}_{\vec{k}s}
e^{-i\left(\omega_{k}+\omega_{i}\right)t}
-{\rm H.c.}\right\} \ .\label{t8}
\end{eqnarray}

We will consider the time evolution of the collection of atoms interacting
with the vacuum field in terms of the density operator $\hat{\rho}_{AF}$
characterizing the statistical state of the combined system of the
atoms and the vacuum field. The time evolution of the density operator
of the combined system obeys the equation
\begin{equation}
\frac{\partial}{\partial t}\hat{\rho}_{AF} =
\frac{1}{i\hbar}\left[\hat{H},\hat{\rho}_{AF}\right] \
.\label{t9}
\end{equation}

Transforming Eq.~(\ref{t9}) into the interaction picture with
\begin{equation}
\tilde{\hat{\rho}}_{AF}\left(t\right) = e^{i\hat{H}_{0}t/\hbar}
\hat{\rho}_{AF}e^{-i\hat{H}_{0}t/\hbar} \ ,\label{t10}
\end{equation}
we find that the transformed density operator satisfies the equation
\begin{equation}
\frac{\partial}{\partial t}\tilde{\hat{\rho}}_{AF}\left(t\right) =
\frac{1}{i\hbar}\left[ \hat{V}\left(t\right),
\tilde{\hat{\rho}}_{AF}\left(t\right)\right] \ , \label{t11}
\end{equation}
where the interaction Hamiltonian $\hat{V}\left(t\right)$ is given
in Eq.~(\ref{t8}).

Equation~(\ref{t11}) is a simple differential equation which can be
solved by the iteration method. For the initial time $t=0$, the
integration of Eq.~(\ref{t11}) leads to the following first-order
solution in $\hat{V}\left(t\right)$:
\begin{eqnarray}
        \tilde{\hat{\rho}}_{AF}\left(t\right) =
        \tilde{\hat{\rho}}_{AF}\left(0\right)
        + \frac{1}{i\hbar}\int_{0}^{t}dt^{\prime}\left[
        \hat{V}\left(t^{\prime}\right),
\tilde{\hat{\rho}}_{AF}\left(t^{\prime}\right)\right] \ . \label{t12}
\end{eqnarray}
Substituting Eq.~(\ref{t12}) into the right side of Eq.~(\ref{t11})
and taking the trace over the vacuum field variables, we find that to
the second order in $\hat{V}\left(t\right)$ the reduced density
operator of the atomic system $\hat{\rho}_{A}\left(t\right)={\rm
Tr}_{F}\tilde{\hat{\rho}}_{AF}\left(t\right)$ satisfies the
integro-differential equation
\begin{eqnarray}
\frac{\partial}{\partial t}\hat{\rho}_{A}\left(t\right) &=&
\frac{1}{i\hbar}{\rm Tr}_{F}\left[\hat{V}\left(t\right),
\tilde{\hat{\rho}}_{AF}\left(0\right)\right] \nonumber \\
&-& \frac{1}{\hbar^{2}}\int_{0}^{t} dt^{\prime} {\rm Tr}_{F}\left\{\left[
\hat{V}\left(t\right),\left[\hat{V}\left(t^{\prime} \right),
\tilde{\hat{\rho}}_{AF}\left(t^{\prime} \right)\right]\right]\right\}
\ .\label{t13}
\end{eqnarray}
We choose an initial state with no correlations between the atomic
system and the vacuum field, which allows us to factorize the initial
density operator of the combined system as
\begin{eqnarray}
        \tilde{\hat{\rho}}_{AF}\left(0\right)
        =\hat{\rho}_{A}\left(0\right)\hat{\rho}_{F}\left(0\right) \
        ,\label{t14}
\end{eqnarray}
where $\hat{\rho}_{F}$ is the density operator of the vacuum field.

We now employ the Born approximation~\cite{lui}, in which the interaction
between the atomic system and the field is supposed to be weak, and there
is no the back reaction effect of the atoms on the field. In this
approximation the state of the vacuum field does not change in time,
and we can write the density operator
$\tilde{\hat{\rho}}_{AF}\left(t^{\prime}\right)$, appearing in
Eq.~(\ref{t13}), as
\begin{eqnarray}
        \tilde{\hat{\rho}}_{AF}\left(t^{\prime}\right)
        =\hat{\rho}_{A}\left(t^{\prime}\right)\hat{\rho}_{F}\left(0\right)
        \ .\label{t15}
\end{eqnarray}
Under this approximation, and after changing the time variable to
$t^{\prime}=t -\tau$, Eq.~(\ref{t13}) simplifies to
\begin{eqnarray}
\frac{\partial}{\partial t}\hat{\rho}\left(t\right) &=&
\frac{1}{i\hbar}{\rm Tr}_{F}\left[\hat{V}\left(t\right),
\hat{\rho}\left(0\right)\hat{\rho}_{F}\left(0\right)\right] \nonumber \\
&-& \frac{1}{\hbar^{2}}\int_{0}^{t} d\tau {\rm Tr}_{F}\left\{\left[
\hat{V}\left(t\right),\left[\hat{V}\left(t -\tau \right),
\hat{\rho}\left(t-\tau \right)\hat{\rho}_{F}\left(0\right)
\right]\right]\right\}
\ ,\label{t16}
\end{eqnarray}
where we use a shorter notation $\hat{\rho}= \hat{\rho}_{A}$.

Substituting the explicit form of
$\hat{V}\left(t\right)$ into Eq.~(\ref{t16}), we find that the
evolution of the density operator depends on the first and second
order correlation functions of the vacuum field operators. We assume
that a part of the vacuum modes is in a squeezed vacuum state for
which the correlation functions are given by~\cite{park,fd1,fd2}
\begin{eqnarray}
{\rm Tr}_{F}\left[\hat{\rho}_{F}\left(0\right){\hat a}_{\vec{k}s}
\right] &=& {\rm Tr}_{F}\left[\hat{\rho}_{F}\left(0\right)
{\hat a}^{\dagger}_{\vec{k}s}\right] = 0 \ , \nonumber \\
{\rm Tr}_{F}\left[\hat{\rho}_{F}\left(0\right){\hat a}_{\vec{k}s}
{\hat a}^{\dagger}_{\vec{k}^{\prime}s^{\prime}}\right] &=&
\left[\left|D\left(\omega_{k}\right)\right|^{2}
N\left(\omega_{k}\right)+1\right]
\delta^{3}\left(\vec{k}-\vec{k^{\prime}}\right)\delta_{ss^{\prime}}
\ ,\nonumber \\
{\rm Tr}_{F}\left[\hat{\rho}_{F}\left(0\right){\hat a}^{\dagger}_{\vec{k}s}
{\hat a}_{\vec{k}^{\prime}s^{\prime}}\right] &=&
\left|D\left(\omega_{k}\right)\right|^{2} N\left(\omega_{k}\right)
\delta^{3}\left(\vec{k}-\vec{k^{\prime}}\right)\delta_{ss^{\prime}}
\ ,\nonumber \\
{\rm Tr}_{F}\left[\hat{\rho}_{F}\left(0\right){\hat a}_{\vec{k}s}
{\hat a}_{\vec{k}^{\prime}s^{\prime}}\right] &=&
D^{2}\left(\omega_{k}\right) M\left(\omega_{k}\right)
\delta^{3}\left(2\vec{k}_{s} -\vec{k}-\vec{k}^{\prime}\right)
\delta_{ss^{\prime}} \ ,\nonumber \\
{\rm Tr}_{F}\left[\hat{\rho}_{F}\left(0\right){\hat a}^{\dagger}_{\vec{k}s}
{\hat a}^{\dagger}_{\vec{k}^{\prime}s^{\prime}}\right] &=&
D^{2}\left(\omega_{k}\right) M^{\ast}\left(\omega_{k}\right)
\delta^{3}\left(2\vec{k}_{s} -\vec{k}-\vec{k}^{\prime}\right)
\delta_{ss^{\prime}} \ ,\label{t17}
\end{eqnarray}
where the parameters $N\left(\omega_{k}\right)$ and
$M\left(\omega_{k}\right)$ characterize squeezing in the vacuum field,
such that $N\left(\omega_{k}\right)$ is the number of photons in the
mode $\vec{k}$, $M\left(\omega_{k}\right)=
|M\left(\omega_{k}\right)|{\rm exp}(i\phi_{s})$ is the magnitude of
two-photon correlations between the vacuum modes, and
$\phi_{s}$ is the phase of the squeezed field. The two-photon
correlations are symmetric about the squeezing carrier frequency
$2\omega_{s}$, i.e.
$M\left(\omega_{k}\right) = M\left(2\omega_{s}-\omega_{k}\right)$, and
are related by the inequality
\begin{equation}
\left|M\left(\omega_{k}\right)\right|^{2} \leq
N\left(\omega_{k}\right)\left(N\left(2\omega_{s}-\omega_{k}\right)
+1\right) ,\label{t18}
\end{equation}
where the term $+1$ on the right-hand side arises from the quantum
nature of the squeezed field~\cite{fd1,fd2}. Such a field is often
called a quantum squeezed field. For a classical analogue
of squeezed field the two-photon correlations are given by the
inequality $\left|M\left(\omega_{k}\right)\right| \leq
N\left(\omega_{k}\right)$. Thus, two-photon correlations with
$0<\left|M\left(\omega_{k}\right)\right| \leq
N\left(\omega_{k}\right)$ may be generated by a classical field,
whereas correlations with $N\left(\omega_{k}\right)<
\left|M\left(\omega_{k}\right)\right| \leq
\sqrt{N\left(\omega_{k}\right)\left(N\left(2\omega_{s}
-\omega_{k}\right)+1\right)}$ can only be generated by a quantum
field which has no classical analog.

The parameter $D\left(\omega_{k}\right)$, appearing in Eq.~(\ref{t17}),
determines the matching of the squeezed modes to the three-dimensional
vacuum modes surrounding the atoms, and contains both the amplitude and
phase coupling. The explicit form of $D\left(\omega_{k}\right)$ depends
on the method of propagation and focusing the squeezed
field~\cite{pas5,pg89}. For perfect matching,
$\left| D\left(\omega_{k}\right)\right|^{2} =1$, whereas $\left|
D\left( \omega_{k}\right)\right|^{2} <1$ for an imperfect matching.
The perfect matching is an idealization as it is practically impossible
to achieve perfect matching in present experiments~\cite{kimb1,kimb2}.
In order to avoid the experimental difficulties, cavity situations
have been suggested. In this case, the parameter
$D\left(\omega_{k}\right)$ is identified as the cavity transfer function,
the absolute value square of which is the Airy function of the
cavity~\cite{bw,yar}. The function $|D\left(\omega_{k}\right)|^{2}$
exhibits a sharp peak centred at the cavity axis and all
the cavity modes are contained in a small solid angle around this
central mode. By squeezing of these modes we can achieve perfect
matching between the squeezed field and the atoms. In a realistic
experimental situation the input squeezed modes have a Gaussian profile
for which the parameter $D\left(\omega_{k}\right)$ is given by
\cite{yar,fde,fd91}
\begin{equation}
D\left(\omega_{k}\right) =\exp\left[- W_{0}\sin^{2}\theta_{k}-ikz_{f}\cos
\theta _{k}\right] \ ,\label{t19}
\end{equation}
where $\theta_{k}$ is an angle over which the squeezed mode $\vec{k}$
is propagated, and $W_{0} $ is the beam spot size at the focal point
$z_{f}$. Thus, even in the cavity situation, perfect matching could be
difficult to achieve in present experiments.

Before returning to the derivation of the master equation, we should
remark that in realistic experimental situations, the squeezed modes
cover only a small portion of the modes surrounding the atoms. The
squeezing modes lie inside a cone of angle $\theta_{k}<\pi $, and the
modes outside the cone are in their ordinary vacuum state. In fact, the
modes are in a finite temperature black-body state, which means that
inside the cone the modes are in mixed squeezed vacuum and black-body
states. However, this is not a serious practical problem as experiments are
usually performed at low temperatures where the black-body radiation is
negligible. In principle, we can include the black-body radiation effect
(thermal noise) to the problem replacing
$\left|D\left(\omega_{k}\right)\right|^{2} N\left( \omega_{k}\right)$
in Eq.~(\ref{t17})
by $\left|D\left(\omega_{k}\right)\right|^{2} N\left( \omega_{k}\right)
+\bar{N},$ where $\bar{N}$ is proportional to the photon number in
the black-body radiation.

We now return to the derivation of the master equation for the density
operator of the atomic system coupled to a squeezed vacuum field.
First, we change the sum over $\vec{k}s$ into an integral
\begin{eqnarray}
\sum_{\vec{k}s}\longrightarrow \frac{V}{\left(2\pi
c\right)^{3}}\sum_{s=1}^{2}\int_{0}^{\infty}d\omega_{k}\omega_{k}^{2}
\int d\Omega_{k} \ .\label{t20}
\end{eqnarray}
Next, with the correlation functions~(\ref{t17}) and after the
rotating-wave approximation (RWA)~\cite{ae}, in which we ignore all
terms oscillating at higher frequencies,
$2\omega_{i}, \omega_{i}+\omega_{j}$, the general master
equation~(\ref{t16}) can be written as
\begin{eqnarray}
\frac{\partial}{\partial t} \hat{\rho}\left( t\right)  =
&\sum_{i,j=1}^{N}& \left\{
\left[S_{j}^{-}\hat{X}_{ij}\left( t,\tau \right),S_{i}^{+}\right]
+\left[S_{j}^{-},\hat{X}_{ji}^{\dagger}\left( t,\tau
\right)S_{i}^{+}\right]\right. \nonumber \\
&&+ \left. \left[S_{j}^{+}\hat{Y}_{ij}\left( t,\tau
\right),S_{i}^{-}\right]
+\left[S_{j}^{+},\hat{Y}_{ji}^{\dagger}\left( t,\tau
\right)S_{i}^{-}\right]\right. \nonumber \\
&&+ \left. \left[S_{i}^{+}\hat{K}_{ij}\left( t,\tau
\right),S_{j}^{+}\right]
+\left[S_{i}^{+},\hat{K}_{ij}\left( t,\tau
\right)S_{j}^{+}\right]\right. \nonumber \\
&&+ \left. \left[S_{i}^{-}\hat{K}^{\dagger}_{ij}\left( t,\tau
\right),S_{j}^{-}\right]
+\left[S_{i}^{-},\hat{K}^{\dagger}_{ij}\left( t,\tau
\right)S_{j}^{-}\right]\right\} \ ,\label{t21}
\end{eqnarray}
where the two-time operators are
\begin{eqnarray}
\hat{X}_{ij}\left( t,\tau \right)  &=& \frac{V}{(2\pi c)^{3}}\int
d\omega_{k}\omega_{k}^{2} e^{-i\left( \omega _{i}-\omega_{j}\right) t}
\int d\Omega _{k}\sum_{s=1}^{2}\chi_{ij}^{(-)}\left( t,\tau \right)
     \ ,  \nonumber \\
\hat{Y}_{ij}\left( t,\tau \right)  &=&\frac{V}{(2\pi c)^{3}}
\int d\omega_{k}\omega_{k}^{2}e^{i\left( \omega _{i}-\omega_{j}\right) t}
\int d\Omega _{k}\sum_{s=1}^{2}\chi_{ij}^{(+)}\left( t,\tau \right)
     \ , \label{t22} \\
\hat{K}_{ij}\left( t,\tau \right)  &=&\frac{V}{(2\pi c)^{3}}\int
d\omega_{k}\omega_{k}\left( 2\omega_{s}-\omega_{k}\right)
e^{-i\left(2\omega_{s}- \omega _{i}-\omega_{j}\right) t} \nonumber \\
&\times&
\int_{\Omega_{s}} d\Omega _{k}\sum_{s=1}^{2}\chi_{ij}^{(M)}
\left( t,\tau \right) \ ,\nonumber
\end{eqnarray}
with
\begin{eqnarray}
\chi_{ij}^{(\pm)}\left( t,\tau \right)  &=&
\left[\left|D\left(\omega_{k}\right)\right|^{2}
N\left(\omega_{k}\right)+1\right]
\left[\vec{\mu}_{i}\cdot \vec{g}_{\vec{k}s}\left(\vec{r}_{i}\right) \right]
\left[\vec{\mu}_{j}^{\ast}\cdot \vec{g}_{\vec{k}s}^{\ast}
\left(\vec{r}_{j}\right) \right] \nonumber \\
&\times& \int_{0}^{t}d\tau \hat{\rho}\left(t-\tau
\right)e^{-i\left(\omega_{k}\pm \omega_{j}\right)\tau} \nonumber \\
&+&\left|D\left(\omega_{k}\right)\right|^{2} N\left(\omega_{k}\right)
\left[\vec{\mu}_{i}^{\ast}\cdot
\vec{g}_{\vec{k}s}^{\ast}\left(\vec{r}_{i}\right) \right]
\left[\vec{\mu}_{j}\cdot \vec{g}_{\vec{k}s}
\left(\vec{r}_{j}\right) \right] \nonumber \\
&\times& \int_{0}^{t}d\tau \hat{\rho}\left(t-\tau
\right)e^{i\left(\omega_{k}\mp \omega_{j}\right)\tau} \ ,\nonumber \\
\chi_{ij}^{\left( M\right) }\left(t\right) &=& M\left( \omega_{k}\right)
D^{2}\left(\omega_{k}\right)
\left[\vec{\mu}_{i}\cdot \vec{g}_{\vec{k}s}\left(
\vec{r}_{i}\right) \right] \left[\vec{\mu}_{j}\cdot
\vec{g}_{\vec{k}s}\left(\vec{r}_{j}\right) \right] \nonumber \\
&\times& \int_{0}^{t}d\tau \hat{\rho}\left(t-\tau
\right)e^{i\left(2\omega_{s}-\omega_{k}- \omega_{j}\right)\tau} \
,\label{t23}
\end{eqnarray}
and $\Omega_{s}$ is the solid angle over which the squeezed vacuum
field is propagated.

The master equation (\ref{t21}) with parameters (\ref{t22}) and
(\ref{t23}) is
quite general in terms of the matching of the squeezed modes to the vacuum
modes and the bandwidth of the squeezed field relative to the atomic
linewidths. The master equation is in the form of an integro-differential
equation, and can be simplified by employing the Markov
approximation~\cite{lui}.
In this approximation the integral over the time delay~$\tau $ contains
functions which decay to zero over a short correlation time~$\tau_{c}.$
This correlation time is of the order of the inverse bandwidth of the
squeezed field, and the short correlation time approximation is formally
equivalent to assume that squeezing bandwidths are much larger
than the atomic linewidths. Over this short time-scale the density operator
would
hardly have changed from $\hat{\rho} \left( t\right) ,$ thus we can replace
$\hat{\rho} \left( t-\tau \right)$ by $\hat{\rho }\left( t\right) $ in
Eq.~(\ref{t23}) and extend the integral to infinity. Under these
conditions, we can perform the integration over $\tau$ and
obtain~\cite{ae}
\begin{equation}
\lim_{t \to\infty}\int_{0}^{t}d\tau \hat{\rho} \left( t-\tau \right)
e^{i x \tau } \approx \hat{\rho}
\left( t\right) \left[ \pi \delta \left( x\right)
+i\frac{{\cal P}}{x}\right] \ ,\label{t24}
\end{equation}
where ${\cal P}$ indicates the principal value of the integral. Moreover,
for squeezing bandwidth much larger than the atomic linewidths, we
can approximate the squeezing parameters and the mode function
evaluated at $\omega_{k}$ by their maximal values evaluated at
$\omega_{s}$, i.e., we can take
$N\left(\omega_{k}\right) =N\left(\omega_{s}\right)$,
$M\left( \omega_{k}\right)= M\left( \omega_{s}\right)$, and
$D\left(\omega_{k}\right)= D\left(\omega_{s}\right)$.

Finally, to carry out the polarization sums and integrals over
$d\Omega _{k}$ in Eq.~(\ref{t22}), we assume that the dipole moments
of the atoms are parallel and use the spherical representation for the
propagation vector $\vec{k}$. The integral over $d\Omega_{k}$ contains
integrals over the spherical angular coordinates $\theta$ and $\phi$.
The angle $\theta$ is formed by $\vec{r}_{ij}$ and $\vec{k}$ directions,
so we can write
\begin{eqnarray}
        \vec{k} = \left|\vec{k}\right|\left[\sin\theta \cos \phi
        \ ,\sin\theta \sin \phi \ ,\cos\theta \right] \ .\label{t25}
\end{eqnarray}
In this representation, the unit polarization vectors $\bar{e}_{\vec{k}1}$
and $\bar{e}_{\vec{k}2}$ may be chosen as~\cite{lui}
\begin{eqnarray}
\bar{e}_{\vec{k}1} &=& \left[ -\cos\theta \cos\phi \ ,-\cos\theta
\sin\phi \ ,\sin\theta \right] \ ,  \nonumber \\
\bar{e}_{\vec{k}2} &=& \left[ \sin\phi \ ,-\cos\phi \ , 0 \right] \ ,
\label{t26}
\end{eqnarray}
and the orientation of the atomic dipole moments can be taken in
the $x$ direction
\begin{eqnarray}
        \vec{\mu}_{i} &=& \left|\vec{\mu}_{i}\right|\left[1 \ ,0 \ ,0\right] \
        ,\nonumber \\
        \vec{\mu}_{j} &=&
        \left|\vec{\mu}_{j}\right|\left[1 \ ,0 \ ,0\right] \ .\label{t27}
\end{eqnarray}
With this choice of the polarization vectors and the orientation of
the dipole moments, we obtain
\begin{eqnarray}
\hat{X}_{ij}\left( t,\tau \right)  &=& \left\{
\left[ 1+\tilde{N}\left(\omega_{s}\right)\right]\left(\frac{1}{2}\Gamma_{ij}
-i\Omega_{ij}^{(-)}\right) +i\tilde{N}\left(\omega_{s}\right)
\Omega_{ij}^{(+)}\right\}
\hat{\rho} \left(t\right) e^{-i\left( \omega_{i}-\omega _{j}\right) t}
\ ,\nonumber \\
\hat{Y}_{ij}\left( t,\tau \right)  &=& \left\{
\tilde{N}\left(\omega_{s}\right)\left(\frac{1}{2}\Gamma_{ij}
+i\Omega_{ij}^{(-)}\right) -i\left[ 1+\tilde{N}\left(\omega_{s}
\right)\right]\Omega_{ij}^{(+)}\right\}
\hat{\rho} \left(t\right) e^{i\left( \omega_{i}-\omega _{j}\right) t}
\ ,\nonumber \\
\hat{K}_{ij}\left( t,\tau \right)  &=& \tilde{M}\left(\omega_{s}\right)
\left(\frac{1}{2}\Gamma_{ij} +i\Omega_{ij}^{(M)}\right)
\hat{\rho} \left( t\right) e^{-i\left( 2\omega
_{s}-\omega _{i}-\omega _{j}\right) t} \ ,  \label{t28}
\end{eqnarray}
where
\begin{eqnarray}
        \tilde{N}\left(\omega_{s}\right) &=& N\left(\omega_{s}\right)
        \left|D\left(\omega_{s}\right)\right|^{2} v\left(\theta
        _{s}\right) \ ,\nonumber \\
        \tilde{M}\left(\omega_{s}\right) &=& M\left(\omega_{s}\right)
        \left|D\left(\omega_{s}\right)\right|^{2} v\left(\theta
        _{s}\right) \ ,\label{t29}
\end{eqnarray}
with
\begin{eqnarray}
v\left( \theta _{s}\right) =\frac{1}{2}\left[ 1-\frac{1}{4}\left( 3+\cos
^{2}\theta _{s}\right) \cos \theta _{s}\right] \ ,  \label{t30}
\end{eqnarray}
and $\theta _{s}$ is the angle over which the squeezed vacuum is
propagated.

The parameters $\Gamma_{ij}$, which appear in Eq.~(\ref{t28}),
are spontaneous emission rates, such that
\begin{eqnarray}
        \Gamma_{i}\equiv \Gamma_{ii} =
        \frac{\omega_{i}^{3}\mu_{i}^{2}}{3\pi \varepsilon_{o}\hbar
        c^{3}} \label{t31}
\end{eqnarray}
is the spontaneous emission rate of the $i$th atom, equal to the
Einstein $A$ coefficient for spontaneous emission, and
\begin{eqnarray}
\Gamma_{ij}=\Gamma_{ji}=
\sqrt{\Gamma_{i}\Gamma_{j}}F\left(k_{0}r_{ij}\right) \qquad (i\neq j)
\ ,\label{t32}
\end{eqnarray}
where
\begin{eqnarray}
F\left(k_{0}r_{ij}\right) &=&\frac{3}{2}
\left\{ \left[1 -\left( \bar{\mu}\cdot \bar{r}
_{ij}\right)^{2} \right] \frac{\sin \left( k_{0}r_{ij}\right)
}{k_{0}r_{ij}}\right.  \nonumber \\
&&\left. +\left[ 1 -3\left( \bar{\mu}\cdot
\bar{r}_{ij}\right)^{2} \right] \left[ \frac{\cos \left(
k_{0}r_{ij}\right) }{\left( k_{0}r_{ij}\right) ^{2}}-\frac{\sin \left(
k_{0}r_{ij}\right) }{\left( k_{0}r_{ij}\right) ^{3}}\right] \right\} \ ,
\label{t33}
\end{eqnarray}
are collective spontaneous emission rates arising from the coupling
between the atoms through the vacuum
field~\cite{ftk87,leh,ag74,mk74,mk75}. In the
expression~(\ref{t33}), $\bar{\mu}=\bar{\mu}_{i}=\bar{\mu}_{j}$ and
$\bar{r}_{ij}$ are unit vectors along the atomic transition dipole
moments and the vector $\vec{r}_{ij}=\vec{r}_{j}-\vec{r}_{i}$,
respectively. Moreover, $k_{0}=\omega_{0}/c$, where
$\omega_{0}=(\omega_{i}+\omega_{j})/2$, and we have assumed that
$(\omega_{i}-\omega_{j})\ll \omega_{0}$.

The remaining parameters $\Omega_{ij}^{(\pm)}$ and $\Omega_{ij}^{(M)}$,
that appear in Eq.~(\ref{t28}), will contribute to the shifts of the
atomic levels, and are given by
\begin{eqnarray}
       \Omega_{ij}^{(\pm)} &=& P\frac{\sqrt{\Gamma _{i}\Gamma _{j}}} {2\pi
       \omega_{0}^{3}}\int_{0}^{\infty}
       \frac{\omega_{k}^{3}F\left(\omega_{k}r_{ij}/c\right)}
       {\omega_{k}\pm \omega_{j}}d\omega_{k} \ ,\label{t34}
\end{eqnarray}
and
\begin{eqnarray}
       \Omega_{ij}^{(M)} &=& P\frac{\sqrt{\Gamma _{i}\Gamma _{j}}} {2\pi
       \omega_{0}^{3}}\int_{0}^{\infty}
       \frac{\omega_{k}^{2}\left(2\omega_{s}-\omega_{k}\right)
       F\left(\omega_{k}r_{0}/c\right)}
       {2\omega_{s}-\omega_{k}- \omega_{j}}d\omega_{k} \ ,\label{t35}
\end{eqnarray}
where $F\left(\omega_{k}r_{0}/c\right)$ is given in Eq.~(\ref{t33}) with
$k_{0}$ replaced by $\omega_{k}/c$, and $r_{ij}$ replaced by
$r_{0}=r_{i}+r_{j}$.

With the parameters (\ref{t28}), the master equation of the system of
non-identical atoms in a broadband squeezed vacuum, written in the
Schr\"{o}dinger picture, reads
\begin{eqnarray}
\frac{\partial \hat{\rho} }{\partial t} &=&
-\frac{1}{2}\sum_{i,j=1}^{N}\Gamma _{ij}
\left[ 1+\tilde{N}\left(\omega_{s}\right) \right] \left( \hat{\rho}
S_{i}^{+}S_{j}^{-}+S_{i}^{+}S_{j}^{-}\hat{\rho}
-2S_{j}^{-}\hat{\rho} S_{i}^{+}\right) \nonumber \\
&&-\frac{1}{2}\sum_{i,j=1}^{N}\Gamma _{ij}\tilde{N}\left(\omega_{s}\right)
\left( \hat{\rho} S_{i}^{-}S_{j}^{+}+S_{i}^{-}S_{j}^{+}\hat{\rho}
-2S_{j}^{+}\hat{\rho} S_{i}^{-}\right) \nonumber \\
&&+\frac{1}{2}\sum_{i,j=1}^{N}\left(\Gamma _{ij}
+i\Omega_{ij}^{(M)}\right)\tilde{M}\left( \omega_{s}\right)
\left( \hat{\rho} S_{i}^{+}S_{j}^{+}+S_{i}^{+}S_{j}^{+}\hat{\rho}
-2S_{j}^{+}\hat{\rho} S_{i}^{+}\right) \nonumber \\
&&+\frac{1}{2}\sum_{i,j=1}^{N}\left(\Gamma _{ij}-i\Omega_{ij}^{(M)}\right)
\tilde{M}^{\ast }\left(\omega_{s}\right)
\left( \hat{\rho} S_{i}^{-}S_{j}^{-}+S_{i}^{-}S_{j}^{-}\hat{\rho}
-2S_{j}^{-}\hat{\rho} S_{i}^{-}\right) \nonumber \\
&&-i\sum_{i=1}^{N}\left(\omega_{i}+\delta_{i}\right)\left[
S^{z}_{i},\hat{\rho} \right]
-i\sum_{i\neq j}^{N}\Omega_{ij}\left[ S^{+}_{i}S^{-}_{j},
\hat{\rho} \right] \ ,  \label{t36}
\end{eqnarray}
where
\begin{eqnarray}
        \delta_{i} &=&  \left[ 2\tilde{N}\left(\omega_{s}\right)
        +1\right]\left( \Omega_{ii}^{(+)} -\Omega_{ii}^{(-)}\right)
        \label{t37}
\end{eqnarray}
represent a part of the intensity dependent Lamb shift of
the atomic levels, while
\begin{eqnarray}
\Omega_{ij} =  -\left( \Omega_{ij}^{(+)} +\Omega_{ij}^{(-)}\right) \
,\qquad (i\neq j)
\label{t38}
\end{eqnarray}
represents the vacuum induced coherent (dipole-dipole) interaction
between the atoms. It is well known that to obtain a complete
calculation of the Lamb shift, it is necessary to extend the calculations
to a second-order multilevel Hamiltonian including electron mass
renormalisation~\cite{sz97}.

The parameters $\delta _{i}$ are usually absorbed into the atomic
frequencies $\omega _{i}$, by redefining the frequencies
$\tilde{\omega}_{i}=\omega _{i}+\delta_{i}$ and are not often
explicitly included in the master equations. The other parameters,
$\Omega_{ij}^{(M)}$ and $\Omega_{ij}$, do not appear as a shift of the
atomic levels. One can show by the calculation of the integral appearing in
Eq.~(\ref{t35}) that the parameter $\Omega_{ij}^{(M)}$ is
negligibly small when the carrier frequency of the squeezed field is tuned
close to the atomic frequencies~\cite{fd91,mil,oc,pk89}. On the other hand,
the parameter $\Omega_{ij}$ is independent of the squeezing parameters
$\tilde{N}\left(\omega_{s}\right)$
and $\tilde{M}\left(\omega_{s}\right)$, and arises from the interaction
between the atoms through the vacuum field. It can be seen that
$\Omega_{ij}$ plays a role of a coherent (dipole-dipole) coupling between
the atoms. Thus, the collective interactions between the atoms
give rise not only to the modified dissipative spontaneous
emission but also lead to a coherent coupling between the atoms.

Using the contours integration method, we find from Eq.~(\ref{t38})
the explicit form of $\Omega_{ij}$ as~\cite{ftk87,leh,ag74,st64,hh64}
\begin{eqnarray}
\Omega_{ij} &=&\frac{3}{4}\sqrt{\Gamma _{i}\Gamma _{j}}\left\{ -\left[
1-\left( \bar{\mu}\cdot \bar{r}%
_{ij}\right)^{2} \right] \frac{\cos \left( k_{0}r_{ij}\right)
}{k_{0}r_{ij}}\right.  \nonumber \\
&&\left. +\left[ 1 -3\left( \bar{\mu}\cdot
\bar{{r}}_{ij}\right)^{2} \right] \left[ \frac{\sin \left(
k_{0}r_{ij}\right) }{\left( k_{0}r_{ij}\right) ^{2}}+\frac{\cos \left(
k_{0}r_{ij}\right) }{\left( k_{0}r_{ij}\right) ^{3}}\right] \right\} \ .
\label{t39}
\end{eqnarray}

\begin{figure}[t]
\begin{center}
\includegraphics[width=13cm]{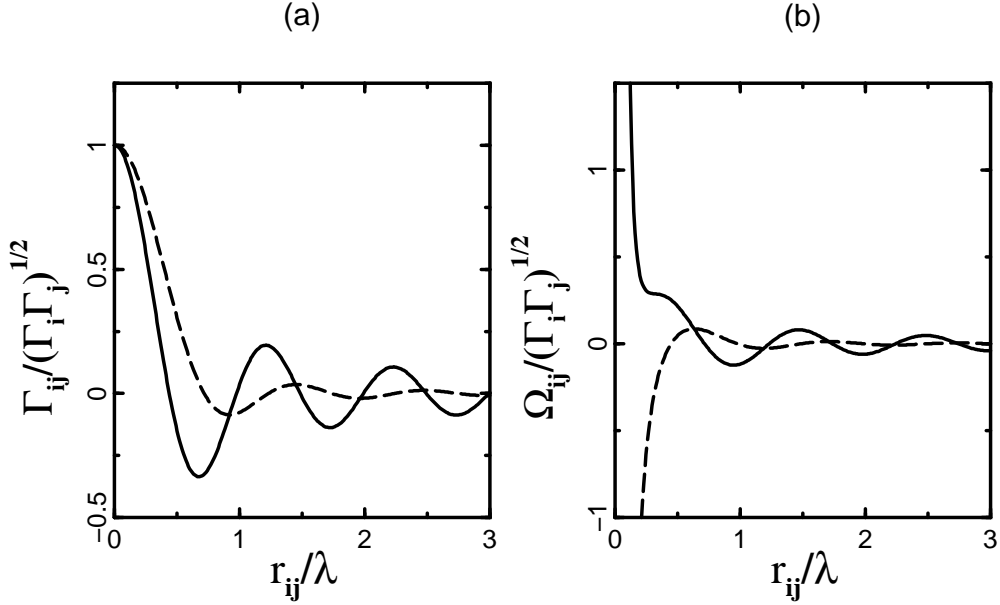}
\end{center}
\caption{(a) Collective damping
$\Gamma_{ij}/\sqrt{\Gamma_{i}\Gamma_{j}}$ and (b) the
dipole-dipole interaction
$\Omega_{ij}/\sqrt{\Gamma_{i}\Gamma_{j}}$ as a function of
$r_{ij}/\lambda$ for $\bar{\mu} \perp
\bar{r}_{ij}$ (solid line) and $\bar{\mu} \parallel
\bar{r}_{ij}$ (dashed line).}
\label{ftfig1}
\end{figure}

The collective parameters $\Gamma_{ij}$ and $\Omega_{ij}$, which both depend
on the interatomic separation, determine the collective properties of the
multiatom system. In Fig.~\ref{ftfig1}, we
plot $\Gamma_{ij}/\sqrt{\Gamma_{i}\Gamma_{j}}$ and
$\Omega_{ij}/\sqrt{\Gamma_{i}\Gamma_{j}}$ as a function of
$r_{ij}/\lambda$, where $\lambda$ is the resonant wavelength.
For large separations $(r_{ij}\gg \lambda)$ the parameters are
very small $(\Gamma _{ij}=\Omega_{ij}\approx 0)$, and become
important for $r_{ij}<\lambda/2$. For atomic separations much smaller than
the resonant wavelength (the small sample model), the parameters attain their
maximal values
\begin{equation}
\Gamma _{ij}=\sqrt{\Gamma _{i}\Gamma _{j}} \ ,\label{t40}
\end{equation}
and
\begin{equation}
\Omega_{ij}\approx \frac{3\sqrt{\Gamma _{i}\Gamma _{j}}}{4\left(
k_{0}r_{ij}\right) ^{3}}\left[ 1 -3\left(\bar{\mu}
\cdot \bar{r}_{ij}\right)^{2}\right] \ .  \label{t41}
\end{equation}
In this small sample model $\Omega_{ij}$ corresponds to the quasistatic
dipole-dipole interaction potential.

Equation (\ref{t36}) is the final form of the master equation that gives
us an elegant description of the physics involved in the dynamics of
interacting atoms. The collective parameters $\Gamma_{ij}$ and
$\Omega_{ij}$, which arise from the mutual interaction between the atoms,
significantly modify the master equation of a two-atom system. The parameter
$\Gamma_{ij}$ introduces a coupling between the atoms through the vacuum
field that the spontaneous emission from one of the atoms influences the
spontaneous emission from the other. The dipole-dipole interaction
term $\Omega_{ij}$ introduces a coherent coupling between the atoms. Owing
to the dipole-dipole interaction, the population is coherently
transferred back and forth from one atom to the other. Here, the
dipole-dipole interaction parameter $\Omega_{ij}$ plays a role
similar to that of the Rabi frequency in the atom-field interaction.

For the next few sections, we restrict ourselves to the interaction of
the atoms with the ordinary vacuum, $\tilde{M}(\omega_{s})
=\tilde{N}(\omega_{s}) =0$, and driven by an external coherent laser
field. In this case, the master equation~(\ref{t36}) can be written as
\begin{eqnarray}
\frac{\partial \hat{\rho} }{\partial t} &=&
-\frac{i}{\hbar}\left[\hat{H}_{s},\hat{\rho}\right]
-\frac{1}{2}\sum_{i,j=1}^{N}\Gamma _{ij}\left( \hat{\rho}
S_{i}^{+}S_{j}^{-}+S_{i}^{+}S_{j}^{-}\hat{\rho}
-2S_{j}^{-}\hat{\rho} S_{i}^{+}\right) \ ,  \label{t42}
\end{eqnarray}
where
\begin{eqnarray}
        \hat{H}_{s} &=& \hbar \sum_{i=1}^{N}\left(\omega_{i}+\delta_{i}\right)
        S^{z}_{i}
+\hbar \sum_{i\neq j}^{N}\Omega_{ij}S^{+}_{i}S^{-}_{j}
+\hat{H}_{L} \ ,\label{t43}
\end{eqnarray}
and
\begin{eqnarray}
       \hat{H}_{L}  &=&
        -\frac{1}{2}\hbar \sum_{i=1}^{N}\left[\Omega
       \left(\vec{r}_{i}\right)S_{i}^{+}
       e^{i\left(\omega_{L}t+\phi_{L}\right)} + \rm{H.c.}\right] \
       ,\label{t44}
\end{eqnarray}
is the interaction Hamiltonian of the atoms with a classical
coherent laser field of the Rabi frequency $\Omega \left(\vec{r}_{i}\right)$,
the angular frequency $\omega_{L}$ and phase $\phi_{L}$.

Note that the Rabi frequencies of the driving field are evaluated at
the positions of the atoms and are defined as~\cite{ae}
\begin{eqnarray}
\Omega \left(\vec{r}_{i}\right)\equiv \Omega_{i} =\vec{\mu}_{i}\cdot
\vec{E}_{L}e^{i\vec{k}_{L}
\cdot \vec{r}_{i}}/\hbar \ ,  \label{t45}
\end{eqnarray}
where $\vec{E}_{L}$ is the amplitude and $\vec{k}_{L}$ is the wave
vector of the driving field, respectively. The Rabi
frequencies depend on the positions of the atoms and can be
different for the atoms located at different points. For example,
if the dipole moments of the atoms are parallel, the Rabi
frequencies $\Omega_{i}$ and $\Omega_{j}$ of two arbitrary atoms
separated by a distance $r_{ij}$ are related by
\begin{eqnarray}
\Omega_{j} = \Omega_{i}
\frac{\left|\vec{\mu}_{j}\right|}
{\left|\vec{\mu}_{i} \right|}e^{i\vec{k}_{L}\cdot \vec{r}_{ij}}
\ ,\label{t46}
\end{eqnarray}
where $\vec{r}_{ij}$ is the vector in the
direction of the interatomic axis and $|\vec{r}_{ij}| =r_{ij}$ is the
distance between the atoms. Thus, for two identical atoms
$(|\vec{\mu}_{i}|= |\vec{\mu}_{j}|)$, the Rabi frequencies
differ by the phase factor exp$(i\vec{k}_{L}\cdot \vec{r}_{ij})$
arising from different position coordinates of the atoms. However,
the phase factor depends on the orientation
of the interatomic axis in respect to the direction of propagation of the
driving field, and therefore exp$(i\vec{k}_{L}\cdot \vec{r}_{ij})$ can be
equal to one, even for large interatomic separations $r_{ij}$. This happens
when the direction of propagation of the driving field is
perpendicular to the interatomic axis, $\vec{k}_{L}\cdot \vec{r}_{ij}=0$.
For directions different from perpendicular,
$\vec{k}_{L}\cdot \vec{r}_{ij}\neq 0$, and then the atoms are in
nonequivalent positions in the driving field, with
different Rabi frequencies~$(\Omega_{i}\neq \Omega_{j})$. For a very
special geometrical configuration of the atoms that are confined to a
volume with linear dimensions that are much smaller compared to the
laser wavelength, the phase factor exp$(i\vec{k}_{L}\cdot
\vec{r}_{ij}) \approx 1$, and then the Rabi frequencies are
independent of the atomic positions. This specific configuration of
the atoms is known as the small sample model or the Dicke model,
and do not correspond in general to the experimentally realised atomic
systems such as atomic beams or trapped atoms.

The formalism presented here for the derivation of the master
equation can be easily extended to the case of $N$ multi-level
atoms~\cite{kbr87,bgz00,bg02,ryf} and atoms interacting with colour (frequency
dependent) reservoirs~\cite{lm88,kky,fwd97,mtf99} or photonic band-gap
materials~\cite{lambro,dkw}. Freedhoff~\cite{hel87} has extended the master
equation formalism to electric quadrupole transitions in atoms.
In the following sections, we will apply the
master equations~(\ref{t36}) and (\ref{t42}) to a wide variety of cases
ranging from two identical as well as nonidentical atoms interacting with
the ordinary vacuum to atoms driven by a laser field and finally to atoms
interacting with a squeezed vacuum field.

\subsection{Quantum jump approach}\label{ftsec22}

The master equation is a very powerful tool for calculations of the
dynamics of Markovian systems which assume that the bandwidth of the
vacuum field is broadband. The Markovian master equation leads to
linear differential equations for the density matrix elements that
can be solved numerically or analytically by the direct integration.

An alternative to the master equation technique is quantum jump
approach. This technique is based on quantum
trajectories~\cite{car93} that are equivalent to the Monte Carlo
wave-function approach~\cite{dcm92,pk98}, and has been developed
largely in connection with problems involving prediction of all
possible evolution trajectories of a given system. This approach can
be used to predict all evolution trajectories of a single quantum
system which stochastically emits photons. Our review of this approach
will concentrate on the example considered by Beige and
Hegerfeldt~\cite{bh98} of two identical two-level atoms interacting
with the three-dimensional EM field whose the modes are in the
ordinary vacuum states.

In the quantum jump approach it is assumed that the probability
density for a photon emission is known for all times $t$, and
therefore the state of the atoms changes abruptly. After one photon
emission the system jumps into another state, which can be determined
with the help of the so called reset operator. The continuous time
evolution of the system between two successive photon emissions is
determined by the conditional Hamiltonian~$\hat{H}_{c}$. Suppose that
at time $t_{0}$ the state of the combined system of the atoms and EM
field is given by
\begin{eqnarray}
        \left|\Psi \right\rangle \left\langle \Psi \right|=
        \left|0 \right\rangle \hat{\rho}
        \left\langle 0\right| \ ,\label{t47}
\end{eqnarray}
where $\hat{\rho}$ is the density operator of the atoms and
$\left|0 \right\rangle$ is the vacuum state of the field. After a
time $\Delta t$ a photon is detected and then the state of the system
changes to
\begin{eqnarray}
        {\cal{P}}\hat{U}_{I}\left(t_{0}+\Delta t,t_{0}\right)
        \left|0 \right\rangle \hat{\rho} \left\langle 0\right|
        \hat{U}_{I}^{\dagger}\left(t_{0}+\Delta t,t_{0}\right)
        {\cal{P}} \ ,\label{t48}
\end{eqnarray}
where ${\cal{P}} =1-\left|0 \right\rangle \left\langle 0\right|$ is
the projection onto the one photon space, and
\begin{eqnarray}
        \hat{U}_{I}\left(t,t_{0}\right)=
        e^{-\frac{i}{\hbar}\hat{V}(t)(t-t_{0})} \label{t49}
\end{eqnarray}
is the evolution operator with the Hamiltonian $\hat{V}(t)$ given
in Eq.~(\ref{t8}).

The non-normalised state of the atomic system, denoted as
$R(\hat{\rho})\Delta t$, is obtained by taking trace of
Eq.~(\ref{t48}) over the field states
\begin{eqnarray}
        R(\hat{\rho})\Delta t = {\rm Tr}_{F}\left(
        {\cal{P}}\hat{U}_{I}\left(t_{0}+\Delta t,t_{0}\right)
        \left|0 \right\rangle \hat{\rho} \left\langle 0\right|
        \hat{U}_{I}^{\dagger}\left(t_{0}+\Delta t,t_{0}\right)
        {\cal{P}}\right) \ ,\label{t50}
\end{eqnarray}
where $R(\hat{\rho})$ is called the non-normalised reset state and
the corresponding operator $\hat{R}$ is called the reset
operator.

Using the perturbation theory and Eq.~(\ref{t8}), we find the
explicit form of $\hat{R}(\hat{\rho})$ for the two-atom system as
\begin{eqnarray}
        \hat{R}(\hat{\rho}) &=&
        \frac{1}{2}\left(C_{12}^{\ast}+C_{21}\right)S_{1}^{-}\hat{\rho}
        S_{2}^{+} +\frac{1}{2}\left(C_{12}+C_{21}^{\ast}\right)S_{2}^{-}
        \hat{\rho} S_{1}^{+} \nonumber \\
        && +\Gamma\left(S_{1}^{-}\hat{\rho} S_{1}^{+}
        +S_{2}^{-}\hat{\rho} S_{2}^{+}\right) \ ,\label{t51}
\end{eqnarray}
where
\begin{eqnarray}
        C_{ij} &=& -\frac{3}{2}i\Gamma e^{ik_{0}r_{ij}}\left\{
        \left[ 1 -\left( \bar{\mu}\cdot \bar{{r}}_{ij}\right)^{2}
        \right] \frac{1}{k_{0}r_{ij}}\right. \nonumber \\
        &&\left. +\left[ 1 -3\left( \bar{\mu}\cdot
\bar{{r}}_{ij}\right)^{2} \right]\left(\frac{i}{(k_{0}r_{ij})^{2}}
-\frac{1}{(k_{0}r_{ij})^{3}}\right)\right\} \ .\label{t52}
\end{eqnarray}
Note that ${\rm Re}C_{ij} =\Gamma_{ij}$ and ${\rm Im}C_{ij}
=2\Omega_{ij}$, where $\Gamma_{ij}$ and $\Omega_{ij}$ are the
collective atomic parameters, given in Eqs.~(\ref{t32}) and (\ref{t39}),
respectively.

The time evolution of the system under the condition that no photon is
emitted is described by the conditional Hamiltonian $\hat{H}_{c}$,
which is found from the relation
\begin{eqnarray}
        1 -\frac{i}{\hbar}\hat{H}_{c}\Delta t &=& \left\langle 0\right|
        \hat{U}_{I}\left(t_{0}+\Delta t,t_{0}\right)\left|0
        \right\rangle \ ,\label{t53}
\end{eqnarray}
where $\Delta t$ is a short evolution time such that $\Delta t <
1/\Gamma$. Using second order perturbation theory, we find from
Eq.~(\ref{t53}) that the conditional Hamiltonian for the two-atom
system is of the form
\begin{eqnarray}
        \hat{H}_{c} &=& \frac{\hbar}{2i}\left[\Gamma
        \left(S_{1}^{+}S_{1}^{-}+S_{2}^{+}S_{2}^{-}\right)
        +C_{12}S_{1}^{+}S_{2}^{-} +C_{21}S_{2}^{+}S_{1}^{-}\right] \
        .\label{t54}
\end{eqnarray}
Hence, between photon emissions the time evolution of the system is
given by an operator
\begin{eqnarray}
        \hat{U}_{c}\left(t_{0}+\Delta t,t_{0}\right)=
        e^{-\frac{i}{\hbar}\hat{H}_{c}(t-t_{0})} \ ,\label{t55}
\end{eqnarray}
which is nonunitary since $\hat{H}_{c}$ is non-Hermitian, and the
state vector of the system is
\begin{eqnarray}
        \left|\Psi_{\Delta t}\right\rangle &=&
        \hat{U}_{c}\left(t_{0}+\Delta t,t_{0}\right)
        \left|\Psi_{0}\right\rangle \ .\label{t56}
\end{eqnarray}
Then, the probability to detect no photon until time $t$ is given by
\begin{eqnarray}
        P\left(t;\left|\Psi_{0}\right\rangle \right) &=& \left|
        \hat{U}_{c}\left(t,t_{0}\right)
        \left|\Psi_{0}\right\rangle \right|^{2} \ .\label{t57}
\end{eqnarray}
The probability density $w_{1}\left(t;\left|\Psi_{0}\right\rangle
\right)$ of detecting a photon at time $t$ is defined as
\begin{eqnarray}
        w_{1}\left(t;\left|\Psi_{0}\right\rangle \right)= -\frac{d}{dt}
        P\left(t;\left|\Psi_{0}\right\rangle \right) ,\label{t58}
\end{eqnarray}
and is often called the waiting time distribution.

The results (\ref{t57}) and (\ref{t58}) show that in the quantum jump
method one calculates the times of the photon
detection stochastically. Starting at $t=t_{0}$ with a pure state, the
state develops according to $\hat{U}_{c}$ until the first emission at
some time $t_{1}$, determined from the waiting time $w_{1}$. Then the
state is reset, according to Eq.~(\ref{t51}), to a new density matrix
and the system evolves again according to $\hat{U}_{c}$ until the
second emission appearing at some time $t_{2}$, and the procedure
repeats until the final time $t_{n}$. In this way, we obtain a set of
trajectories of the atomic evolution. The ensemble of such
trajectories yields to equations of motion which are solved using the
standard analytical or numerical methods. As a practical matter,
individual trajectories are generally not observed. The ensemble average
over all possible trajectories leads to equations of motion which are
equivalent to the equations of motion derived from the master
equation of the system. Thus, the quantum jump approach is consistent
with the master equation method. However, the advantage of the
quantum jump approach over the master equation method is that it
allows to predict all possible trajectories of a single system. Using
this approach, it has been demonstrated that environment induced
measurements can assist in the realization of universal gates for
quantum computing~\cite{bbtk}. Cabrillo {\it et al.}~\cite{cabr} have
applied the method to demonstrate entangling between distant atoms by
interference. Sch\"{o}n and Beige~\cite{sb} have demonstrated the
advantage of the method in the analysis of a two-atom double-slit
experiment.

\section{Entangled atomic states}\label{ftsec3}

The modification of spontaneous emission by the collective damping
and in particular the presence of the dipole-dipole interaction
between the atoms suggest that the bare atomic states are no longer
the eigenstates of the atomic system. We will illustrate this on a
system of two identical as well as nonidentical atoms, and present
a general formalism for diagonalization of the Hamiltonian of the
atoms in respect to the dipole-dipole interaction.

In the absence of the dipole-dipole interaction and the driving laser
field, the space of the two-atom system is spanned by four
product states
\begin{eqnarray}
\ket {g_{1}}\ket {g_{2}} \ ,\quad \ket {e_{1}}\ket {g_{2}} \ ,\quad
\ket {g_{1}}\ket
{e_{2}} \ ,\quad \ket {e_{1}}\ket {e_{2}} \ ,\label{t59}
\end{eqnarray}
with corresponding energies
\begin{eqnarray}
        E_{gg}=-\hbar \omega_{0} \ ,\quad E_{eg}=-\hbar \Delta
        \ ,\quad E_{ge}=\hbar \Delta \ ,\quad E_{ee}=\hbar \omega_{0}
        \ ,\label{t60}
\end{eqnarray}
where $\omega_{0}=\frac{1}{2}\left(\omega_{1}+\omega_{2}\right)$ and
$\Delta =\frac{1}{2}\left(\omega_{2}-\omega_{1}\right)$.

The product states $\ket {e_{1}}\ket {g_{2}}$ and $\ket {g_{1}}\ket
{e_{2}}$ form a pair of nearly degenerated states. When we include the
dipole-dipole interaction between the atoms, the product states
combine into two linear superpositions (entangled states), with
their energies shifted from $\pm \hbar \Delta$ by the
dipole-dipole interaction energy. To see this, we begin with
the Hamiltonian of two atoms including the dipole-dipole interaction
\begin{eqnarray}
        \hat{H}_{aa} = \sum_{i=1}^{2}\hbar \omega_{i} S_{i}^{z}
        + \hbar \sum_{i\neq j}\Omega_{ij}S_{i}^{+}S_{j}^{-} \ .\label{t61}
\end{eqnarray}
In the basis of the product states~(\ref{t59}), the
Hamiltonian~(\ref{t61}) can be written in a matrix form as
\begin{eqnarray}
\hat{H}_{aa} &=& \hbar \left(
\begin{array}{cccc}
-\omega_{0} & 0 & 0 & 0 \\
0 & -\Delta & \Omega_{12} & 0 \\
0 & \Omega_{12} & \Delta & 0 \\
0 & 0 & 0 & \omega_{0}
\end{array}
\right) \ . \label{t62}
\end{eqnarray}

Evidently, in the presence of the dipole-dipole interaction
the matrix~(\ref{t62}) is not diagonal, which indicates that the product
states~(\ref{t59}) are not the eigenstates of the two-atom
system. We will diagonalize the matrix~(\ref{t62}) separately for the
case of identical $(\Delta =0)$ and nonidentical $(\Delta \neq 0)$
atoms to find eigenstates of the systems and their energies.

\subsection{Entangled states of two identical atoms}\label{ftsec31}

Consider first a system of two identical atoms $(\Delta =0)$.
In order to find energies and corresponding eigenstates of the system,
we have to diagonalize the matrix~(\ref{t62}). The resulting
energies and corresponding eigenstates of the system are~\cite{dic,leh}
\begin{eqnarray}
&& E_{g} =-\hbar \omega_{0} \ ,\qquad \ket g = \ket {g_{1}}\ket {g_{2}}
\ ,\nonumber \\
&& E_{s} = \hbar \Omega_{12} \ ,\qquad
\ket s = \frac{1}{\sqrt{2}}\left( \ket {e_{1}}\ket {g_{2}}
+\ket {g_{1}}\ket {e_{2}}\right) \ ,\nonumber \\
&& E_{a} = -\hbar \Omega_{12} \ ,\qquad
\ket a = \frac{1}{\sqrt{2}}\left( \ket {e_{1}}\ket {g_{2}}
-\ket {g_{1}}\ket {e_{2}}\right) \ ,\nonumber \\
&& E_{e} = \hbar \omega_{0} \ ,\qquad
\ket e = \ket {e_{1}}\ket {e_{2}} \ .\label{t63}
\end{eqnarray}

The eigenstates~(\ref{t63}), first introduced by Dicke~\cite{dic}, are
known as the collective states of two interacting
atoms. The ground state $\ket g$ and the upper state $\ket e$ are not
affected by the dipole-dipole interaction, whereas the states $\ket s$
and $\ket a$ are shifted from their unperturbed energies by the amount
$\pm \Omega_{12}$, the dipole-dipole energy.
The most important property of the collective states $\ket s$ and
$\ket a$ is that they are an example of maximally entangled states
of the two-atom system. The states are linear superpositions of the
product states which cannot be separated into product states of
the individual atoms.
\begin{figure}[t]
\begin{center}
\includegraphics[width=10cm]{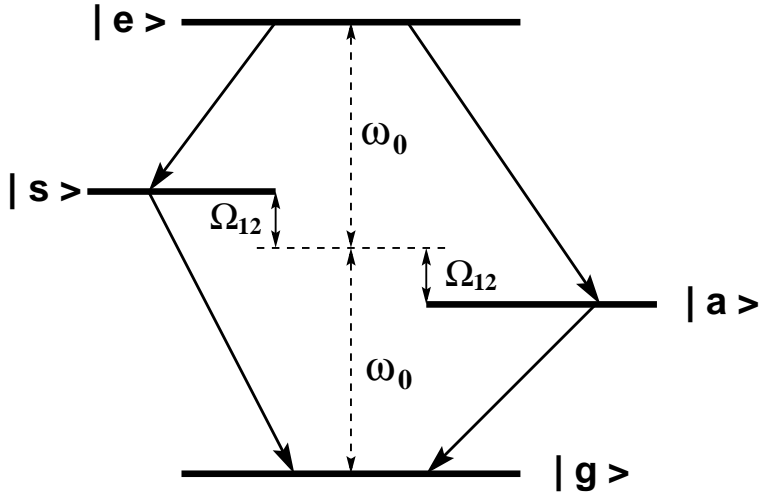}
\end{center}
\caption{Collective states of two identical atoms. The energies of
the symmetric and antisymmetric states are shifted by the dipole-dipole
interaction $\Omega_{12}$. The arrows indicate possible one-photon
transitions.}
\label{ftfig2}
\end{figure}

We show the collective states of two identical atoms in Fig.~\ref{ftfig2}.
It is seen that in the collective states representation, the two-atom system
behaves as a single four-level system, with the ground state $\ket g$, the
upper state $\ket e$, and two intermediate states: the symmetric
state $\ket s$ and the antisymmetric state $\ket a$. The energies of
the intermediate states depend on the dipole-dipole interaction and
these states suffer a large shift when the interatomic separation is
small. There are two transition channels $\ket e \rightarrow \ket s
\rightarrow \ket g$ and $\ket e \rightarrow \ket a \rightarrow \ket
g$, each with two cascade nondegenerate transitions. For two
identical atoms, these two channels are uncorrelated, but the
transitions in these channels are damped with significantly different
rates. To illustrate these features, we transform the master
equation~(\ref{t42})
into the basis of the collective states~(\ref{t63}). We define
collective operators $A_{ij}=\ket i\bra j$, where $i,j=e,a,s,g$, that
represent the energies $(i=j)$ of the collective
states and coherences $(i\neq j)$. Using Eq.~(\ref{t63}), we find that
the collective operators are related to the atomic operators
$S_{i}^{\pm}$ through the following identities
\begin{eqnarray}
S_{1}^{+} &=& \frac{1}{\sqrt{2}}\left(A_{es}-A_{ea}+A_{sg}
+A_{ag}\right) \ ,\nonumber \\
S_{2}^{+} &=& \frac{1}{\sqrt{2}}\left(A_{es}+A_{ea}+A_{sg}
-A_{ag}\right) \ .\label{t64}
\end{eqnarray}

Substituting the transformation identities into Eq.~(\ref{t42}),
we find that in the basis of the collective states the master equation
of the system can be written as
\begin{eqnarray}
        \frac{\partial}{\partial t}\hat{\rho} &=&
        -\frac{i}{\hbar}\left[\hat{H}_{as}, \hat{\rho}\right]
        +\left(\frac{\partial}{\partial t}\hat{\rho}\right)_{s}
        +\left(\frac{\partial}{\partial t}\hat{\rho}\right)_{a} \
        ,\label{t65}
\end{eqnarray}
where
\begin{eqnarray}
\hat{H}_{as} &=&
\hbar\left[\omega_{0}\left(A_{ee}-A_{gg}\right)
+\Omega_{12}\left(A_{ss}-A_{aa}\right)\right] \nonumber \\
&-& \frac{\hbar}{2\sqrt{2}}
\left(\Omega_{1} +\Omega_{2}\right)\left[\left(A_{es}+A_{sg}\right)
e^{i\left(\omega_{L}t+\phi_{L}\right)} +{\rm H.c.} \right] \nonumber \\
&-& \frac{\hbar}{2\sqrt{2}} \left(\Omega_{2}
-\Omega_{1}\right)\left[\left(A_{ea}-A_{ag}\right)
e^{i\left(\omega_{L}t+\phi_{L}\right)} +{\rm
H.c.} \right] \ ,\label{t66}
\end{eqnarray}
is the Hamiltonian of the interacting atoms and the driving laser
field,
\begin{eqnarray}
\left(\frac{\partial}{\partial t}\hat{\rho}\right)_{s} &=&
- \frac{1}{2}\left(\Gamma +\Gamma_{12}\right)
\left\{\left(A_{ee}+A_{ss}\right)\hat{\rho} +\hat{\rho} \left(A_{ee}
+A_{ss}\right)\right. \nonumber \\
&&- \left. 2\left(A_{se}+A_{gs}\right)\hat{\rho} \left(A_{es}
+A_{sg}\right)\right\}  \ ,\label{t67}
\end{eqnarray}
describes dissipation through the cascade $\ket e \rightarrow
\ket s \rightarrow \ket g$ channel involving the symmetric state
$\ket s$, and
\begin{eqnarray}
\left(\frac{\partial}{\partial t}\hat{\rho}\right)_{a} &=&
- \frac{1}{2}\left(\Gamma -\Gamma_{12}\right)
\left\{\left(A_{ee}+A_{aa}\right)\hat{\rho} +\hat{\rho} \left(A_{ee}
+A_{aa}\right)\right. \nonumber \\
&&-\left. 2\left(A_{ae}-A_{ga}\right)\hat{\rho} \left(A_{ea}
-A_{ag}\right)\right\}  \ ,\label{t68}
\end{eqnarray}
describes dissipation through the cascade
$\ket e \rightarrow \ket a \rightarrow \ket g$ channel involving the
antisymmetric state $\ket a$.

We will call the two cascade channels $\ket e \rightarrow \ket s
\rightarrow \ket g$ and $\ket e \rightarrow \ket a \rightarrow
\ket g$ as symmetric and antisymmetric transitions, respectively.
The first term in
$\hat{H}_{as}$ is the energy of the collective states, while the
second and third terms are the interactions of the laser field with
the symmetric and antisymmetric transitions, respectively.
One can see from Eqs.~(\ref{t65})-(\ref{t68}) that the symmetric and
antisymmetric transitions are uncorrelated and decay with different
rates; the symmetric transitions decay with an enhanced (superradiant)
rate $(\Gamma +\Gamma_{12})$, whereas the antisymmetric transitions
decay with a reduced (subradiant) rate $(\Gamma -\Gamma_{12})$. For
$\Gamma =\Gamma_{12}$, which appears when the interatomic separation
is much smaller than the resonant wavelength, the antisymmetric
transitions decouple from the driving field and does not decay. In this
case, the antisymmetric state is completely decoupled from the remaining
states and the system decays only through the symmetric channel.
Hence, for $\Gamma_{12}=\Gamma$ the system reduces to a three-level
cascade system, referred to as the small-sample model or two-atom Dicke
model~\cite{dic,leh,ag74}. The model assumes that the atoms are close
enough that we can ignore any effects resulting from different spatial
positions of the atoms. In other words, the phase factors $\exp
(i\vec{k}\cdot \vec{r}_{i})$ are assumed to have the same value for
all the atoms, and are set equal to one. This assumption may prove
difficult in experimental realization as the present atom trapping and
cooling techniques can trap two atoms at distances of the order of a
resonant wavelength~\cite{eich,deb,ber,tos}. At these distances the
collective damping parameter $\Gamma_{12}$ differs significantly from
$\Gamma$ (see Fig.~\ref{ftfig1}), and we cannot ignore the transitions
to and from the antisymmetric state. We can, however,
employ the Dicke model to spatially extended atomic systems. This
could be achieved assuming that the observation time of the atomic
dynamics is shorter than $\Gamma^{-1}$. The antisymmetric state $\ket
a$ decays on a time scale $\sim (\Gamma -\Gamma_{12})^{-1}$, which for
$\Gamma_{12}\approx \Gamma$ is much longer than $\Gamma^{-1}$. On the
other hand, the symmetric state decays on a time scale $\sim
(\Gamma +\Gamma_{12})^{-1}$, which is shorter than $\Gamma^{-1}$.
Clearly, if we consider short observation times, the antisymmetric
state does not participate in the dynamics and the system can be
considered as evolving only between the Dicke states.

Although the symmetric and antisymmetric transitions of the collective
system are uncorrelated, the dynamics of the four-level system
may be significantly different from the three-level Dicke model.
As an example, consider the total intensity of the fluorescence field
emitted from a two-atom system driven by a resonant coherent laser
field $(\omega_{L}=\omega_{0})$.
We make two simplifying assumptions in order to obtain a simple
analytical solution: Firstly, we limit our calculations to the
steady-state intensity. Secondly, we take $\vec{k}_{L}\cdot
\vec{r}_{12}=0$ that corresponds to the direction of propagation of
the driving field perpendicular to the interatomic axis. We emphasize
that these assumptions do not limit qualitatively the physics of the
system, as experiments are usually performed in the steady-state,
and with $\vec{k}_{L}\cdot \vec{r}_{12}=0$ the interatomic separation
$r_{12}$ may still be any size relative to the resonant wavelength.

We consider the radiation intensity $I(\vec{R},t)$ detected at a
point $\vec{R}$ at the moment of time $t$. If the detection point
$\vec{R}$ is in the far-field zone of the radiation emitted by the atomic
system, then the intensity can be expressed in terms of the first-order
correlation functions of the atomic dipole operators as~\cite{leh,ag74}
\begin{eqnarray}
I\left(\vec{R},t\right) = u(\vec{R})\sum_{i,j=1}^{2}\langle
S_{i}^{+}\left(t-R/c\right)
S_{j}^{-}\left(t-R/c\right)\rangle e^{ik\bar{R}\cdot \vec{r}_{ij}}
\ ,\label{t69}
\end{eqnarray}
where
\begin{equation}
        u(\vec{R}) = \left(\omega_{0}^{4}\mu^{2}/2R^{2}c^{4}\pi
        \varepsilon_{0}\right)\sin^{2}\varphi \label{t70}
\end{equation}
is a constant which depends on the
geometry of the system, $\varphi$ the angle between the
observation direction $\vec{R}=R\bar{R}$ and the atomic dipole
moment $\vec{\mu}$.

On integrating over all directions, Eq.~(\ref{t69}) yields the total
radiation intensity given in photons per second as
\begin{eqnarray}
I\left(t\right) = \sum_{i,j=1}^{2}\Gamma_{ij}\langle
S_{i}^{+}\left(t-R/c\right)
S_{j}^{-}\left(t-R/c\right)\rangle \ .\label{t71}
\end{eqnarray}
The atomic correlation functions, appearing in Eq.~(\ref{t71}),
are found from the master equation~(\ref{t42}). There are,
however, two different steady-state solutions of the master
equation~(\ref{t42}) depending on whether the collective damping
rates $\Gamma_{12}=\Gamma$ or $\Gamma_{12}\neq
\Gamma$~\cite{ftk81,ftk83,hsf82}.

For $\Gamma_{12}\neq \Gamma$ and $\vec{k}_{L}\cdot \vec{r}_{12}=0$,
the steady-state solutions for the atomic correlation functions are
\begin{eqnarray}
        \left\langle S_{1}^{+}S_{1}^{-}\right\rangle =
        \left\langle S_{2}^{+}S_{2}^{-}\right\rangle &=&
        \frac{2\Omega^{4} +\Gamma^{2}\Omega^{2}}{4D}
        \ ,\nonumber \\
         \left\langle S_{1}^{+}S_{2}^{-}\right\rangle =
        \left\langle S_{2}^{+}S_{1}^{-}\right\rangle &=&
        \frac{\Gamma^{2}\Omega^{2}}{4D} \ ,\label{t72}
\end{eqnarray}
where
\begin{eqnarray}
        D = \Omega^{4}+\left(\Omega^{2}
        +\Omega_{12}^{2}\right)\Gamma^{2} +\frac{1}{4}\Gamma^{2}\left(\Gamma
        +\Gamma_{12}\right)^{2}  \ .\label{t73}
\end{eqnarray}
If we take $\Gamma_{12} =\Gamma$ and $\Omega_{12}=0$, that corresponds
to the two-atom Dicke model, the steady-state solutions for the atomic
correlation functions are of the following form
\begin{eqnarray}
        \left\langle S_{1}^{+}S_{1}^{-}\right\rangle =
        \left\langle S_{2}^{+}S_{2}^{-}\right\rangle &=&
        \frac{3\Omega^{4} +2\Omega^{2}\Gamma^{2}}{2D^{\prime}}
        \ ,\nonumber \\
         \left\langle S_{1}^{+}S_{2}^{-}\right\rangle =
        \left\langle S_{2}^{+}S_{1}^{-}\right\rangle &=&
        \frac{\Omega^{4}+2\Omega^{2}\Gamma^{2}}{2D^{\prime}} \ ,\label{t74}
\end{eqnarray}
where
\begin{eqnarray}
        D^{\prime} = 3\Omega^{4}+4\Gamma^{2}\Omega^{2}+4\Gamma^{4}
        \ .\label{t75}
\end{eqnarray}

In the limit of a strong driving field, $\Omega \gg \Gamma$, the
steady-state total radiation intensity from the two-atom Dicke model
is equal to $4\Gamma/3$. However, for the spatially separated atoms
$I_{ss}=\lim_{t\rightarrow \infty}I\left(t\right) =\Gamma$, which is
twice of the intensity from a single atom~\cite{mw}. There is no
additional enhancement of the intensity.

Note that in the limit of $r_{12}\rightarrow 0$, the steady-state
solution~(\ref{t72}) does not reduce to that of the Dicke model,
given in Eq.~(\ref{t74}). This fact is connected with conservation of
the total spin $S^{2}$, that $S^{2}$ is a constant of motion for the
Dicke model and $S^{2}$ not being a constant of motion for a spatially
extended system of atoms~\cite{ftk81,ftk83}. We can explain it by
expressing the square of the total spin of the two-atom system in
terms of the density matrix elements of the collective system as
\begin{eqnarray}
        S^{2}\left(t\right) = 2 -2\rho_{aa}\left(t\right) \ .\label{t76}
\end{eqnarray}
It is clear from Eq.~(\ref{t76}) that $S^{2}$ is conserved only in
the Dicke model, in which the antisymmetric state is ignored. For a
spatially extended system the antisymmetric state participates fully
in the dynamics and $S^{2}$ is not conserved. The Dicke model reaches
steady state between the triplet states $\ket e$, $\ket s$, and $\ket
g$, while the spatially extended two-atom system reaches steady state
between the triplet and the antisymmetric states.

Amin and Cordes~\cite{ac78} calculated the total radiation intensity
from an $N$-atom Dicke model and showed the intensity is $N(N+2)/3$
times that for a single atom, which they called "scaling factor". The
above calculations show that the scaling factor is characteristic of
the small sample model and does not exist in spatially extended atomic
systems. Thus, in physical systems the antisymmetric state plays
important role and as we have shown its presence affects the
steady-state fluorescence intensity. The antisymmetric state can also
affect other phenomena, for example, photon
antibunching~\cite{ftk81a}, and purity of two-photon entangled
states, that is discussed in Sec.~\ref{ftsec10}.

\subsection{Collective states of two nonidentical atoms}\label{ftsec32}

For two identical atoms, the dipole-dipole interaction leads to the
maximally entangled symmetric and antisymmetric states that decay
independently with different damping rates. Furthermore, in the case
of the small sample model of two atoms the antisymmetric state decouples
from the external coherent field and the environment, and consequently
does not decay. The decoupling of the antisymmetric state from the
coherent field prevents the state from the external coherent interactions.
This is not, however, an useful property from the point of view of quantum
computation where it is required to prepare entangled states which are
decoupled from the external environment and simultaneously should be
accessible by coherent processes. This requirement can be achieved if
the atoms are not identical, and we will discuss here some consequences
of the fact that the atoms could have different transition frequencies
or different spontaneous emission rates. To make our discussion more
transparent, we will concentrate on two specific cases:
(1) $\Delta \neq 0$ and $\Gamma_{1}=\Gamma_{2}$, and (2)
$\Delta =0$ and $\Gamma_{1}\neq \Gamma_{2}$.

\subsubsection{ The case $\Delta \neq 0$ and
$\Gamma_{1}=\Gamma_{2}$}\label{ftsec321}

When the atoms are nonidentical with different
transition frequencies, the states~(\ref{t63}) are no longer the
eigenstates of the Hamiltonian~(\ref{t60}). The diagonalization of the
matrix~(\ref{t62}) with $\Delta \neq 0$ leads to the following
energies and corresponding eigenstates~\cite{ftk86}
\begin{eqnarray}
&& E_{g} =-\hbar \omega_{0} \ ,\qquad
\ket g = \ket {g_{1}}\ket {g_{2}} \ ,\nonumber \\
&& E_{s^{\prime}} = \hbar w \ ,\qquad
\ket {s^{\prime}} = \beta \ket {e_{1}}\ket {g_{2}}
+\alpha \ket {g_{1}}\ket {e_{2}} \ ,\nonumber \\
&& E_{a^{\prime}} = -\hbar w \ ,\qquad
\ket {a^{\prime}} = \alpha \ket {e_{1}}\ket {g_{2}}
-\beta \ket {g_{1}}\ket {e_{2}} \ ,\nonumber \\
&& E_{e} = \hbar \omega_{0} \ ,\qquad
\ket e = \ket {e_{1}}\ket {e_{2}} \ ,\label{t77}
\end{eqnarray}
where
\begin{equation}
\alpha = \frac{d}{\sqrt{d^{2}+\Omega_{12}^{2}}} \ ,\quad
\beta = \frac{\Omega_{12}}{\sqrt{d^{2}+\Omega_{12}^{2}}} \ ,\quad
w= \sqrt{\Omega_{12}^{2}+\Delta^{2}} \ ,\label{t78}
\end{equation}
and $d = \Delta
+\sqrt{\Omega_{12}^{2}+\Delta^{2}}$.

The energy level structure of the collective system of two nonidentical
atoms is similar to that of the
identical atoms, with the ground state $\ket g$, the upper state
$\ket e$, and two intermediate states $\ket {s^{\prime}}$ and
$\ket {a^{\prime}}$. The effect of the frequency
difference $\Delta$ on the collective atomic states
is to increase the splitting between the intermediate levels, which
now is equal to $w=\sqrt{\Omega_{12}^{2}+\Delta^{2}}$.
However, the most dramatic effect of the detuning $\Delta$ is on the
degree of entanglement of the intermediate
states $\ket {s^{\prime}}$ and $\ket {a^{\prime}}$ that in the case
of nonidentical atoms the states are no longer maximally entangled states.
For $\Delta =0$ the states $\ket {s^{\prime}}$
and $\ket {a^{\prime}}$ reduce to the maximally entangled states $\ket
s$ and $\ket a$, whereas for $\Delta \gg \Omega_{12}$ the entangled
states reduce to the
product states  $\ket {e_{1}}\ket {g_{2}}$ and $-\ket {g_{1}}\ket
{e_{2}}$, respectively.

Using the same procedure as for the case of identical atoms, we
rewrite the master equation~(\ref{t42}) in terms of the collective
operators $A_{ij}=\ket i\bra j$, where now the collective states $\ket
i$ are given in Eq.~(\ref{t77}).  First, we find that in the case of
nonidentical atoms the atomic dipole operators can be written in
terms of the linear combinations of the collective operators as
\begin{eqnarray}
S_{1}^{+} &=& \alpha A_{es^{\prime}}-\beta A_{ea^{\prime}}+\beta
A_{s^{\prime}g}+\alpha A_{a^{\prime}g} \ ,\nonumber \\
S_{2}^{+} &=& \beta A_{es^{\prime}}+\alpha A_{ea^{\prime}}+\alpha
A_{s^{\prime}g}-\beta A_{a^{\prime}g} \ .\label{t79}
\end{eqnarray}
Hence, in terms of the collective operators $A_{ij}$, the master equation
takes the form
\begin{equation}
\frac{\partial}{\partial t}\hat{\rho}
=-\frac{i}{\hbar}\left[\hat{H}_{s^{\prime}},\hat{\rho}
\right] +{\cal{L}}\hat{\rho}  \ ,\label{t80}
\end{equation}
where
\begin{eqnarray}
\hat{H}_{s^{\prime}} &=& \hbar\left[\omega_{0}\left(A_{ee}
-A_{gg}\right)
+w\left(A_{s^{\prime}s^{\prime}}- A_{a^{\prime}a^{\prime}}\right)\right]
\nonumber \\
&-& \frac{\hbar}{2} \left\{\left[\left(\alpha \Omega_{1} +\beta
\Omega_{2}\right)A_{es^{\prime}}+
\left(\beta \Omega_{1} +\alpha
\Omega_{2}\right)A_{s^{\prime}g}\right]
e^{i\left(\omega_{L}t+\phi_{L}\right)}\right. \nonumber \\
&+& \left. \left[\left(\alpha \Omega_{2} -\beta
\Omega_{1}\right)A_{ea^{\prime}}
-\left(\beta \Omega_{2} -\alpha
\Omega_{1}\right)A_{a^{\prime}g}\right]
e^{i\left(\omega_{L}t+\phi_{L}\right)}+{\rm H.c.}\right\}
\label{t81}
\end{eqnarray}
is the Hamiltonian of the system in the collective states basis, and
the Liouville operator
${\cal{L}}\hat{\rho}$ describes the dissipative part of the
evolution. The dissipative part is composed of three terms
\begin{eqnarray}
        {\cal{L}}\hat{\rho} &=&
        \left(\frac{\partial}{\partial t}\hat{\rho}\right)_{s}
        +\left(\frac{\partial}{\partial t}\hat{\rho}\right)_{a}
        +\left(\frac{\partial}{\partial t}\hat{\rho}\right)_{I} \
        ,\label{t82}
\end{eqnarray}
where
\begin{eqnarray}
\left(\frac{\partial}{\partial t}\hat{\rho}\right)_{s} &=&
-\Gamma_{s^{\prime}}\left\{\left(A_{ee}
+A_{s^{\prime}s^{\prime}}\right)\hat{\rho} +\hat{\rho}\left(
A_{ee}+A_{s^{\prime}s^{\prime}}\right)\right. \nonumber \\
&&\left. -2\left(A_{s^{\prime}e}\hat{\rho} A_{es^{\prime}}
+A_{gs^{\prime}}\hat{\rho} A_{s^{\prime}g}\right)\right\} \nonumber \\
&& -\left(\alpha \beta
\Gamma +\Gamma_{12}\right)
\left(A_{s^{\prime}e}\rho A_{s^{\prime}g} + A_{gs^{\prime}}\rho
A_{es^{\prime}}\right) \ ,\label{t83}\\
\left(\frac{\partial}{\partial t}\hat{\rho}\right)_{a} &=&
-\Gamma_{a^{\prime}}\left\{\left(A_{ee}
+A_{a^{\prime}a^{\prime}}\right)\hat{\rho} +\hat{\rho}\left(
A_{ee}+A_{a^{\prime}a^{\prime}}\right)\right. \nonumber \\
&&\left. -2\left(A_{a^{\prime}e}\hat{\rho} A_{ea^{\prime}}
+A_{ga^{\prime}}\hat{\rho} A_{a^{\prime}g}\right)\right\} \nonumber \\
&& -\left(\alpha \beta
\Gamma -\Gamma_{12}\right)
\left(A_{a^{\prime}e}\hat{\rho} A_{a^{\prime}g}
+ A_{ga^{\prime}}\hat{\rho}A_{ea^{\prime}}\right) \ ,\label{t84}
\end{eqnarray}
and
\begin{eqnarray}
\left(\frac{\partial}{\partial t}\hat{\rho}\right)_{I} &=&
-\Gamma_{a^{\prime}s^{\prime}}\left\{\left(
A_{a^{\prime}s^{\prime}}+A_{s^{\prime}a^{\prime}}\right)\hat{\rho} +
\hat{\rho} \left(A_{a^{\prime}s^{\prime}}+A_{s^{\prime}a^{\prime}}
\right)\right. \nonumber \\
&&-\left. 2\left(A_{ga^{\prime}}\hat{\rho} A_{s^{\prime}g}
+A_{gs^{\prime}}\hat{\rho} A_{a^{\prime}g}+A_{s^{\prime}e}\hat{\rho}
A_{ea^{\prime}} + A_{a^{\prime}e}\hat{\rho}
A_{es^{\prime}}\right) \right\} \nonumber \\
&&+ \left(\alpha^{2}-\beta^{2}\right)\Gamma \left\{
A_{a^{\prime}e}\hat{\rho} A_{s^{\prime}g} + A_{gs^{\prime}}\hat{\rho}
A_{ea^{\prime}}\right. \nonumber \\
&&\left. +A_{s^{\prime}e}\hat{\rho} A_{a^{\prime}g} + A_{ga^{\prime}}
\hat{\rho} A_{es^{\prime}}\right\} \ ,\label{t85}
\end{eqnarray}
with the damping coefficients
\begin{eqnarray}
\Gamma_{s^{\prime}} &=& \frac{1}{2}\left(\Gamma
+2\alpha \beta \Gamma_{12}\right) \ ,\qquad
\Gamma_{a^{\prime}} = \frac{1}{2}\left(\Gamma
-2\alpha \beta \Gamma_{12}\right) \ ,\nonumber \\
\Gamma_{a^{\prime}s^{\prime}} &=&
\frac{1}{2}\left(\alpha^{2}-\beta^{2}\right)\Gamma_{12} \
.\label{t86}
\end{eqnarray}

The dissipative part of the master equation is very extensive and
unlike the case of identical atoms, contains the interference term
between the symmetric and antisymmetric transitions.
The terms~(\ref{t83}) and (\ref{t84}) describes spontaneous
transitions in the symmetric and antisymmetric channels, respectively.
The coefficients $\Gamma_{s^{\prime}}$, and $\Gamma_{a^{\prime}}$
are the spontaneous emission rates of the transitions. The
interference term~(\ref{t85}) results from spontaneously induced
coherences between the symmetric and antisymmetric transitions.
This term appears only in systems of atoms with different transition
frequencies $(\Delta \neq 0)$, and reflects the fact that, as the
system decays from the state $\ket {s^{\prime}}$, it drives the
antisymmetric state, and vice versa. Thus, in contrast to the case of
identical atoms, the symmetric and antisymmetric transitions are no longer
independent and are correlated due to the presence of the detuning $\Delta$.
Moreover, for nonidentical atoms the damping rate of the antisymmetric state
cannot be reduced to zero. In the case of interatomic separations much
smaller than the optical wavelength (the small sample model), the damping
rate reduces to
\begin{equation}
\Gamma_{a^{\prime}} = \frac{1}{2}\Gamma \left(\alpha -\beta
\right)^{2} \ ,\label{t87}
\end{equation}
that is different from zero, unless $\Delta =0$.
\begin{figure}[t]
\begin{center}
\includegraphics[width=10cm]{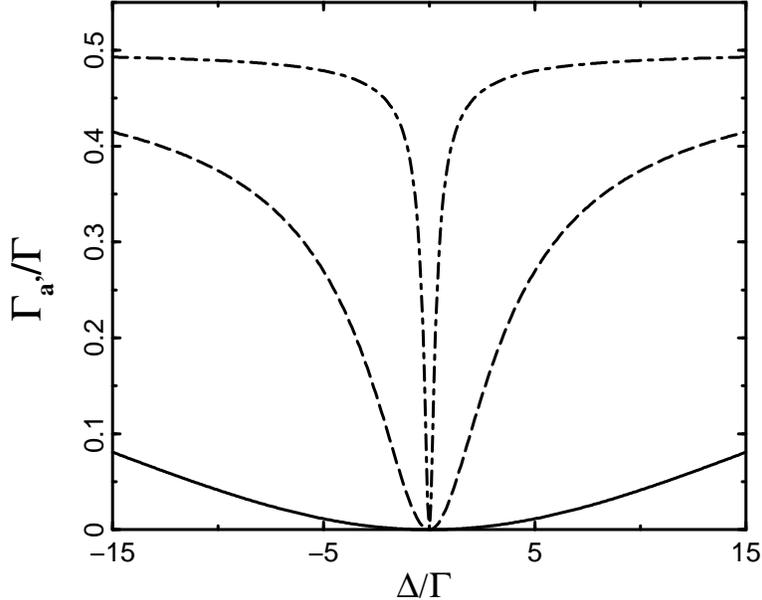}
\end{center}
\caption{The spontaneous emission damping rate $\Gamma_{a^{\prime}}$
as a function of $\Delta$ for $\vec{\mu}\perp \bar{r}_{12}$, and
different interatomic separations: $r_{12}/\lambda =0.05$ (solid line),
$r_{12}/\lambda =0.1$ (dashed line), $r_{12}/\lambda =0.5$
(dashed-dotted line).}
\label{ftfig3}
\end{figure}

In Fig.~\ref{ftfig3}, we plot the damping rate $\Gamma_{a^{\prime}}$
as a function
of $\Delta$ for different interatomic separations. The damping rate
vanishes for $\Delta =0$ independent of the interatomic separation,
but for small interatomic separations there is a significant range of
$\Delta$ for which $\Gamma_{a^{\prime}}\ll \Gamma$.

\subsubsection{The case $\Delta =0$ and $\Gamma_{1}\neq
\Gamma_{2}$}\label{sec322}

The choice of the collective states~(\ref{t77}) as a basis leads to a
complicated dissipative part of the master equation. A different
choice of collective states is proposed here, which allows to
obtain a simple master equation of the system with only the uncorrelated
dissipative parts of the symmetric and antisymmetric
transitions~\cite{afs}. Moreover, we will show that it is possible to
create an entangled state in the system of two nonidentical atoms which
can be decoupled from the external environment and, at the same time,
the state exhibits a strong coherent coupling with the remaining states.

To illustrate this, we introduce superposition operator
$S_{s}^{\pm}$ and $S_{a}^{\pm}$ which are linear combinations of the
atomic dipole operators $S_{1}^{\pm}$ and $S_{2}^{\pm}$ as
\begin{eqnarray}
S_{s}^{+} &=& uS_{1}^{+}+vS_{2}^{+} \ ,\quad S_{s}^{-} =
u^{\ast}S_{1}^{-} + v^{\ast}S_{2}^{-} \ ,\nonumber \\
S_{a}^{+} &=& vS_{1}^{+} -u S_{2}^{+} \ ,\quad S_{a}^{-} =
v^{\ast}S_{1}^{-} - u^{\ast}S_{2}^{-} \ ,\label{t88}
\end{eqnarray}
where $u$ and $v$ are the transformation coefficients which are in
general complex numbers. The coefficients satisfy the condition
\begin{equation}
        \left|u\right|^{2} +\left|v\right|^{2} =1 \ .\label{t89}
\end{equation}

The operators $S_{s}^{\pm}$ and $S_{a}^{\pm}$ represent,
respectively, symmetric and antisymmetric superpositions of the
atomic dipole operators.
In terms of the superposition operators, the dissipative
part of the master equation~(\ref{t42}) can be written as
\begin{eqnarray}
{\cal {L}}\hat{\rho} &=& -\Gamma_{ss}\left(S^{+}_{s}S^{-}_{s}\hat{\rho}
+\hat{\rho}
S^{+}_{s}S^{-}_{s}-2S^{-}_{s}\hat{\rho} S^{+}_{s} \right)  \nonumber \\
&&- \Gamma_{aa}\left(S^{+}_{a}S^{-}_{a}\hat{\rho} +\hat{\rho}
S^{+}_{a}S^{-}_{a}-2S^{-}_{a}\hat{\rho} S^{+}_{a} \right)  \nonumber \\
&& -\Gamma_{sa}\left(S^{+}_{s}S^{-}_{a}\hat{\rho} +\hat{\rho}
S^{+}_{s}S^{-}_{a} -2S^{-}_{a}\hat{\rho} S^{+}_{s} \right)  \nonumber \\
&& -\Gamma_{as}\left(S^{+}_{a}S^{-}_{s}\hat{\rho} +\hat{\rho}
S^{+}_{a}S^{-}_{s} -2S^{-}_{s}\hat{\rho} S^{+}_{a} \right) \ ,
\label{t90}
\end{eqnarray}
where the coefficients $\Gamma_{mn}$ are
\begin{eqnarray}
\Gamma_{ss} &=& \left|u\right|^{2}\Gamma_{1}
+\left|v\right|^{2}\Gamma_{2} +\left(uv^{\ast}
+u^{\ast}v\right)\Gamma_{12} ,\nonumber \\
\Gamma_{aa} &=& \left|v\right|^{2}\Gamma_{1}
+\left|u\right|^{2}\Gamma_{2} -\left(uv^{\ast}
+u^{\ast}v\right)\Gamma_{12} ,\nonumber \\
\Gamma_{as} &=& uv^{\ast}\Gamma_{1} -u^{\ast}v\Gamma_{2}
-\left(\left|u\right|^{2} -\left|v\right|^{2}\right)\Gamma_{12}
,\nonumber \\
\Gamma_{sa} &=& u^{\ast}v\Gamma_{1} -uv^{\ast}\Gamma_{2}
-\left(\left|u\right|^{2} -\left|v\right|^{2}\right)\Gamma_{12}
.\label{t91}
\end{eqnarray}
The first two terms in Eq.~(\ref{t90}) are familiar spontaneous
emission terms of the symmetric and antisymmetric transitions, and
the parameters $\Gamma_{ss}$ and $\Gamma_{aa}$ are spontaneous emission
rates of the transitions,
respectively. The last two terms are due to coherence between the
superposition states and the parameters
$\Gamma_{as}$ and $\Gamma_{sa}$ describes cross-damping rates between
the superpositions.

If we make the identification
\begin{equation}
u = \sqrt{\frac{\Gamma_{1}}{\Gamma_{1}+\Gamma_{2}}} \ ,\quad
v = \sqrt{\frac{\Gamma_{2}}{\Gamma_{1}+\Gamma_{2}}} \ ,\label{t92}
\end{equation}
then the damping coefficients~(\ref{t91}) simplify to
\begin{eqnarray}
\Gamma_{ss} &=& \frac{1}{2}\left(\Gamma_{1}+\Gamma_{2}\right)
+\frac{\sqrt{\Gamma_{1}\Gamma_{2}}\left(\Gamma_{12}
-\sqrt{\Gamma_{1}\Gamma_{2}}\right)}{\Gamma_{1}+\Gamma_{2}} \ , \nonumber \\
\Gamma_{aa} &=& \frac{\left(\sqrt{\Gamma_{1}\Gamma_{2}}-\Gamma_{12}\right)
\sqrt{\Gamma_{1}\Gamma_{2}}}{\Gamma_{1}+\Gamma_{2}} \ , \nonumber \\
\Gamma_{sa} &=& \Gamma_{as} = \frac{1}{2} \frac{\left(\Gamma_{1}
-\Gamma_{2}\right)\left(\sqrt{\Gamma_{1}\Gamma_{2}} -\Gamma_{12}\right)}
{\Gamma_{1}+\Gamma_{2}} \ .\label{t93}
\end{eqnarray}
When the damping rates of the atoms are equal $(\Gamma_{1}=\Gamma_{2})$,
the cross-damping terms $\Gamma_{as}$ and $\Gamma_{sa}$ vanish. Furthermore,
if $\Gamma_{12}=\sqrt{\Gamma_{1}\Gamma_{2}}$ then the spontaneous emission
rates $\Gamma_{aa}$, $\Gamma_{as}$ and $\Gamma_{sa}$ vanish regardless of
the ratio between the $\Gamma_{1}$ and $\Gamma_{2}$. In this case,
which corresponds to interatomic separations much
smaller than the optical wavelength, the antisymmetric superposition
does not decay and also decouples from the symmetric superposition.

An interesting question arises as to whether the nondecaying
antisymmetric superposition can still be coupled to the symmetric
superposition through the coherent interactions $\Omega_{12}$ and
$\Omega$ contained in the Hamiltonian $\hat{H}_{s}$. These interactions
can coherently transfer population between the superpositions.
To check it, we first transform the
Hamiltonian~(\ref{t43}) into the interaction picture and next rewrite
the transformed Hamiltonian in terms of the $S_{s}^{\pm}$ and $S_{a}^{\pm}$
operators as
\begin{eqnarray}
\hat{H}_{s} &=&-\hbar \Delta_{L}\left[\left(
S_{s}^{+}S_{s}^{-}+S_{a}^{+}S_{a}^{-}\right)
+\left(v^{\ast}u-vu^{\ast}\right)\left(
S_{s}^{+}S_{a}^{-}-S_{a}^{+}S_{s}^{-}\right)\right] \nonumber \\
&&+\hbar \Omega_{12}\left\{\left(vu^{\ast}+v^{\ast}u\right)
\left( S_{s}^{+}S_{s}^{-}-S_{a}^{+}S_{a}^{-}
\right)\right. \nonumber \\
&&\left. +\left(|v|^{2}-|u|^{2}\right) \left(
S_{s}^{+}S_{a}^{-}+S_{a}^{+}S_{s}^{-}\right) \right\}  \nonumber \\
&&-\frac{1}{2}\hbar \left[ \left( u\Omega _{1}+v\Omega _{2}\right)
S_{s}^{+}
+\left( v\Omega _{1}-u\Omega _{2}\right)S_{a}^{+}
+{\rm H.c.}\right] \ ,\label{t94}
\end{eqnarray}
where $\Delta_{L}=\omega_{L}-\omega_{0}$.

In the above equation, the first term arises from the atomic Hamiltonian and
shows that in the absence of the interatomic interactions the symmetric and
antisymmetric states have the same energy.
The second term in Eq.~(\ref{t94}), proportional to the dipole-dipole
interaction between the atoms, has two
effects on the dynamics of the symmetric and antisymmetric
superpositions. The first is a shift of the energies and the second is the
coherent interaction between the superpositions. It is seen from
Eq.~(\ref{t94}) that the contribution of $\Omega_{12}$ to the coherent
interaction between the superpositions vanishes for $\Gamma_{1}=\Gamma_{2}$
and then the effect of $\Omega_{12}$ is only the shift of the energies from
their unperturbed values. Note that the dipole-dipole interaction
$\Omega_{12}$ shifts the energies in the opposite directions.
The third term in Eq.~(\ref{t94}) represents the interaction of the
superpositions with the driving laser field. We see that the symmetric
superposition couples to the laser field with an effective Rabi
frequency proportional to $u\Omega _{1}+v\Omega _{2}$, whereas the Rabi
frequency of the antisymmetric superposition is proportional to $v\Omega
_{1}-u\Omega _{2}$ and vanishes for $v\Omega _{1}=u\Omega _{2}$.

Alternatively, we may write the Hamiltonian (\ref{t94}) in a more
transparent form which shows explicitly the presence of the coherent
coupling between the symmetric and antisymmetric states
\begin{eqnarray}
\hat{H}_{s} &=& -\hbar \left[\left(\Delta_{L}-\Delta^{\prime}\right)
S_{s}^{+}S_{s}^{-} + \left(\Delta_{L}+\Delta^{\prime}
\right)S_{a}^{+}S_{a}^{-} +\Delta_{c}S_{s}^{+}S_{a}^{-}
+\Delta_{c}^{\ast}S_{a}^{+}S_{s}^{-}\right]  \nonumber \\
&&- \frac{1}{2}\hbar \left[\left(u\Omega_{1}+v\Omega_{2}\right)
S^{+}_{s}
+ \left(v\Omega_{1}-u\Omega_{2}\right) S^{+}_{a}
+{\rm H.c.}\right] \ ,  \label{t95}
\end{eqnarray}
where $\Delta^{\prime}$ and $\Delta_{c}$ are given by
\begin{eqnarray}
\Delta^{\prime} = \left(vu^{\ast}+v^{\ast}u\right)\Omega_{12} \ ,\quad
\Delta_{c} = \left(|u|^{2}-|v|^{2}\right)\Omega_{12}
+ \left(v^{\ast}u-vu^{\ast}\right)\Delta_{L} \ .\label{t96}
\end{eqnarray}

The parameters $\Delta^{\prime}$ and $\Delta_{c}$ allow us to gain
physical insight into how the dipole-dipole interaction $\Omega_{12}$
and the unequal damping rates $\Gamma_{1}\neq \Gamma_{2}$ can modify the
dynamics of the
two-atom system. The parameter $\Delta^{\prime}$ appears as a shift of the
energies of the superposition systems, while
$\Delta_{c}$ determines the magnitude of the coherent interaction between the
superpositions. For identical atoms the shift $%
\Delta^{\prime}$ reduces to $\Omega_{12}$ that is the dipole-dipole
interaction shift of the energy levels.
In contrast to the shift $\Delta^{\prime}$, which is different from zero for
identical as well as nonidentical atoms, the coherent coupling $\Delta_{c}$
can be different from zero only for nonidentical atoms.

Thus, the condition $\Gamma_{12}=\sqrt{\Gamma_{1}\Gamma_{2}}$ for
suppression of spontaneous emission from the antisymmetric state is valid
for identical as well as non-identical atoms, whereas the coherent interaction
between the superpositions appears only for nonidentical atoms with different
spontaneous damping rates.

It should be noted that this treatment is valid with only a minor
modification for a number of other schemes of two-atom systems. For
example, it can be applied to the case of two identical atoms that
experience different intensities and phases of the driving
field~\cite{fs,rfd,lm93}.

In what follows, we will illustrate how the interference term in the
master equation of two nonidentical atoms results in quantum beats
and transfers of the population to the antisymmetric state even if
the antisymmetric state does not decay. Of particular interest is the
temporal dependence of the total radiation intensity of the
fluorescence field emitted by two interacting atoms.

\section{Quantum beats}\label{sec4}

The objective of this section is to give an account of interference
effects resulting from the direct correlations between the symmetric
and antisymmetric states. We will first analyse the simplest model
of spontaneous emission from two nonidentical atoms and consider the
time dependence of the total radiation intensity. After this, we will
consider the time evolution of the fluorescence intensity emitted by
two identical atoms that are not in the equivalent positions in the
driving field.

\subsection{Quantum beats in spontaneous emission from two
nonidentical atoms}\label{sec41}

For two nonidentical atoms the master equation~(\ref{t42}), in the
absence of the driving field $(\Omega_{i}=0)$, leads to a closed set
of five equations of motion for the expectation values of the atomic
dipole operators~\cite{ftk86}. This set of equations can be written
in a matrix form as
\begin{eqnarray}
        \frac{d}{dt}\vec{X}\left(t\right) = A\vec{X}\left(t\right) \
        ,\label{t97}
\end{eqnarray}
where $\vec{X}\left(t\right)$ is a column vector with components
\begin{eqnarray}
X_{1} &=& \langle S_{1}^{+}(t)S_{1}^{-}(t)\rangle \ ,\quad
X_{2} = \langle S_{2}^{+}(t)S_{2}^{-}(t)\rangle \ ,\nonumber \\
X_{3} &=& \langle S_{1}^{+}(t)S_{2}^{-}(t)\rangle \ ,\quad
X_{4} = \langle S_{2}^{+}(t)S_{1}^{-}(t)\rangle \ ,\nonumber \\
X_{5} &=& \langle S_{1}^{+}(t)S_{2}^{+}(t)S_{1}^{-}(t)
S_{2}^{-}(t)\rangle \ ,\label{t98}
\end{eqnarray}
and $A$ is the $5\times 5$ matrix
\begin{eqnarray}
A &=& \left(
\begin{array}{ccccc}
-\Gamma_{1} & 0 & \kappa &
\kappa^{\ast} & 0\\
0 & -\Gamma_{2} & \kappa^{\ast} &
\kappa & 0 \\
\kappa & \kappa^{\ast} & -\frac{1}{2}(\Gamma_{T}-4i\Delta)
& 0 & 2\Gamma_{12}\\
\kappa^{\ast} &
\kappa & 0 &
-\frac{1}{2}(\Gamma_{T}+4i\Delta) & 2\Gamma_{12}\\
0 & 0 & 0 & 0 & -\Gamma_{T}
\end{array}
\right) \ , \label{t99}
\end{eqnarray}
with $\kappa =-\frac{1}{2}(\Gamma_{12}+i\Omega_{12})$ and
$\Gamma_{T}=\Gamma_{1}+\Gamma_{2}$.

It is seen from Eq.~(\ref{t99}) that the equation of motion for the
second-order correlation function $\langle S_{1}^{+}(t)S_{2}^{+}(t)
S_{1}^{-}(t)S_{2}^{-}(t)\rangle$ is decoupled from the remaining four
equations. This allows for an exact solution of the set of
equations~(\ref{t96}). The exact solution is given in
Ref.~\cite{ftk86}. Here, we will focus on two special cases of
$\Delta \neq 0, \Gamma_{1}=\Gamma_{2}$ and $\Delta =0, \Gamma_{1}\neq
\Gamma_{2}$, and calculate the time evolution of the total
fluorescence intensity, defined in Eq.~(\ref{t71}). We will assume
that initially $(t=0)$ atom "1" was in its excited state $\ket {e_{1}}$
and atom "2" was in its ground state $\ket {g_{2}}$.

\subsubsection{The case $\Delta \neq 0$, $\Gamma_{1}=\Gamma_{2}=\Gamma$
and $\Omega_{12}\gg \Delta$}\label{sec411}

In this case the atoms have the same spontaneous damping rates but
different transition frequencies that, for simplicity, are taken much
smaller than the dipole-dipole interaction potential. In this limit,
the approximate solution of Eq.~(\ref{t97}) leads to the following
total radiation intensity
\begin{eqnarray}
        I\left(t\right) = e^{-\Gamma
        t}\left[\frac{\Delta}{2\Omega_{12}}\Gamma_{12}\cos 2wt
        +\Gamma \cosh \Gamma_{12}t -\Gamma_{12}\sinh \Gamma_{12}t
        \right] \ ,\label{t100}
\end{eqnarray}
where $w=\sqrt{\Omega_{12}^{2}+\Delta^{2}}$.

\begin{figure}[t]
\begin{center}
\includegraphics[width=10cm]{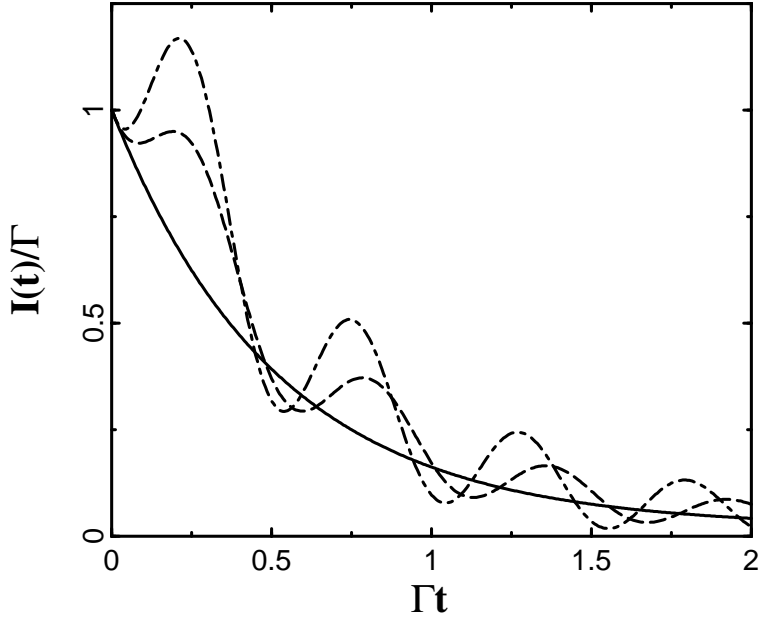}
\end{center}
\caption{Time evolution of the total radiation intensity for
$r_{12}=\lambda/12, \Gamma_{1}=\Gamma_{2},
\bar{\mu}\perp \bar{r}$, and different $\Delta$: $\Delta =0$ (solid
line), $\Delta = -2\Gamma$ (dashed line), $\Delta =-3\Gamma$
(dashed-dotted line).}
\label{ftfig4}
\end{figure}

The total radiation intensity exhibits sinusoidal modulation (beats)
superimposed on exponential decay with the damping rates $\Gamma \pm
\Gamma_{12}$. The amplitude of the oscillations is proportional to
$\Delta$ and vanishes for identical atoms. The damping rate $\Gamma
+\Gamma_{12}$ describes the spontaneous decay from the state
$\ket {s^{\prime}}$ to the ground state $\ket g$, while $\Gamma -
\Gamma_{12}$ is the decay rate of the
$\ket {a^{\prime}} \rightarrow \ket g$ transition. The frequency
$2w$ of the oscillations is equal to the frequency difference between
the $\ket {s^{\prime}}$ and $\ket {a^{\prime}}$ states. The
oscillations reflect the spontaneously induced correlations between
the $\ket {s^{\prime}} \rightarrow \ket g$ and
$\ket {a^{\prime}} \rightarrow \ket g$ transitions. According to
Eq.~(\ref{t86}) the amplitude of the spontaneously induced
correlations is equal to $\Gamma_{a^{\prime}s^{\prime}}$, which in
the limit of $\Omega_{12}\gg \Delta$ reduces to
$\Gamma_{a^{\prime}s^{\prime}}= \Delta \Gamma_{12}/(2\Omega_{12})$.
Hence, the amplitude of the oscillations appearing in Eq.~(\ref{t100})
is exactly equal to the amplitude of the spontaneously induced
correlations. Fig.~\ref{ftfig4} shows the temporal dependence of the
total radiation intensity
for interatomic separation $r_{12}=\lambda/12, \Gamma_{1}=\Gamma_{2},
\bar{\mu}\perp \bar{r}$, and different $\Delta$. As predicted by
Eq.~(\ref{t100}), the intensity exhibits quantum beats whose the
amplitude increases with increasing $\Delta$. Moreover, at short
times, the intensity can become greater than its initial value
$I\left(0\right)$. This effect is known as a superradiant
behavior and is absent in the case of two identical atoms.
Thus, the spontaneously induced correlations between the
$\ket {s^{\prime}} \rightarrow \ket g$ and $\ket {a^{\prime}}
\rightarrow \ket g$ transitions can induce quantum beats and
superradiant effect in the intensity of the emitted field.

The superradiant effect is characteristic of a large number of
atoms~\cite{bl,gh,er}, and it is quite surprising to obtain this
effect in the system of two atoms. Coffey and Friedberg~\cite{cf78}
and Richter~\cite{rich81} have shown that the superradiant effect can
be observed in some special cases of the atomic configuration of a
three-atom system. Blank {\it et al.}~\cite{bla} have shown that this
effect, for atoms located in an equidistant linear chain, appears for
at least six atoms. Recently, DeAngelis {\it et al.}~\cite{dea} have
experimentally observed the superradiant effect in the radiation
from two identical dipoles located inside a planar symmetrical
microcavity.

Quantum beats predicted here for spontaneous emission from two
nonidentical atoms are fully equivalent to the quantum beats
predicted recently by Zhou and Swain~\cite{zs98} in a single
three-level $V$ system with correlated spontaneous transitions. For
the initial conditions used here that initially only one of the atoms
was excited, the initial population distributes equally between the
states $\ket {s^{\prime}}$ and $\ket {a^{\prime}}$. Since the
transitions are correlated through the dissipative term
$\Gamma_{a^{\prime}s^{\prime}}$, the system of two nonidentical atoms
behaves as a three-level $V$ system with spontaneously correlated
transitions.

\subsubsection{The case of $\Delta =0, \Gamma_{1}\neq \Gamma_{2}$
and $\Omega_{12}\gg \Gamma_{1},\Gamma_{2}$}\label{sec412}

\begin{figure}[t]
\begin{center}
\includegraphics[width=10cm]{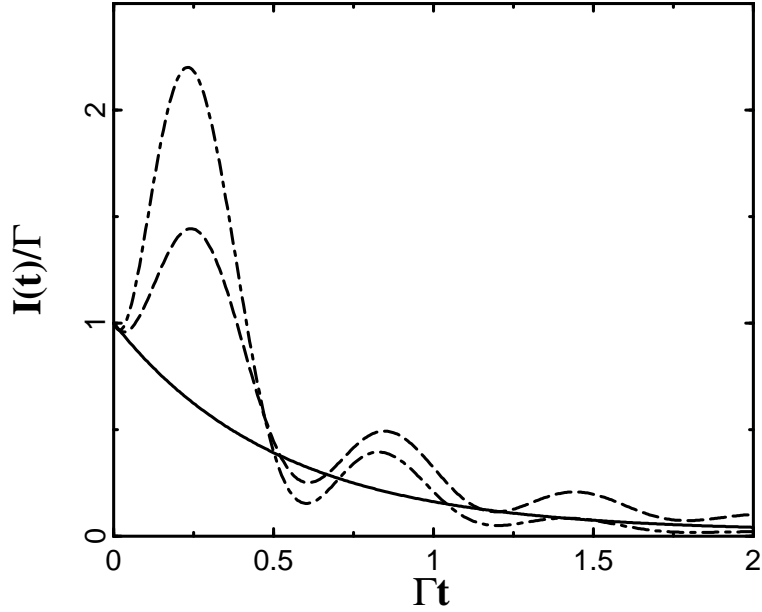}
\end{center}
\caption{Time evolution of the total radiation intensity for
$r_{12}=\lambda/12, \Delta =0,
\bar{\mu}\perp \bar{r}_{12}$, and different $\Gamma_{2}/\Gamma_{1}$:
$\Gamma_{2}/\Gamma_{1}=1$ (solid
line), $\Gamma_{2}/\Gamma_{1} = 2.5$ (dashed line),
$\Gamma_{2}/\Gamma_{1} =5$ (dashed-dotted line).}
\label{ftfig5}
\end{figure}

We now wish to show how quantum beats can be obtained in two
nonidentical atoms that have equal frequencies but different damping
rates. According to Eqs.~(\ref{t93}) and (\ref{t96}), the symmetric
and antisymmetric transitions are correlated not only through the
spontaneously induced coherences $\Gamma_{as}$, but also through the
coherent coupling $\Delta_{c}$. One can see from Eq.~(\ref{t93}) that
for small interatomic separations $\Gamma_{as}\approx 0$. However, the
coherent coupling parameter $\Delta_{c}$, which is proportional to
$\Omega_{12}$, is very large, and we will
show that the coherent coupling $\Delta_{c}$ can also lead to quantum
beats and the superradiant effect. In the case of $\Delta =0,
\Gamma_{1}\neq \Gamma_{2}$ and $\Omega_{12}\gg \Gamma_{1},\Gamma_{2}$,
the approximate solution of Eq.~(\ref{t97}) leads to the following
expression for the total radiation intensity
\begin{eqnarray}
        I\left(t\right) &=&
        e^{-\frac{1}{2}\left(\Gamma_{1}+\Gamma_{2}\right)t}
        \left\{\frac{1}{2}\left(\Gamma_{1}-\Gamma_{2}\right)\cos
        2\Omega_{12}t \right. \nonumber \\
       &&\left. +\frac{1}{2}\left(\Gamma_{1}+\Gamma_{2}\right) \cosh
       \Gamma_{12}t -\Gamma_{12}\sinh \Gamma_{12}t
        \right\} \ .\label{t101}
\end{eqnarray}
The intensity displays quantum-beat oscillations at frequency
$2\Omega_{12}$ corresponding to the frequency splitting between the
$\ket {s^{\prime}}$ and $\ket {a^{\prime}}$ states. The amplitude of
the oscillations is equal to $\left(\Gamma_{1}-\Gamma_{2}\right)/2$
that is proportional to the coherent coupling $\Delta_{c}$. For
$\Gamma_{1}=\Gamma_{2}$ the coherent coupling parameter $\Delta_{c}=0$
and no quantum beats occur. In this case the intensity exhibits pure
exponential decay. This is shown in Fig.~\ref{ftfig5}, where we plot the
time evolution of $I\left(t\right)$ for interatomic separation
$r_{12}=\lambda/12$, and different ratios
$\Gamma_{2}/\Gamma_{1}$. Similar to the case discussed in
Sec.~\ref{sec411}, the intensity exhibits quantum beats and the
superradiant effect. For $r_{12}=\lambda/12$ the collective damping
$\Gamma_{12}\approx \sqrt{\Gamma_{1}\Gamma_{2}}$, and then the
parameter $\Gamma_{as}\approx 0$, indicating that the
quantum beats and the superradiant effect result from the coherent
coupling between the $\ket {s^{\prime}}$ and $\ket {a^{\prime}}$
states.

\subsubsection{Two identical atoms in nonequivalent positions in a
driving field}\label{sec413}

Quantum beats and superradiant effect induced by interference between
different transitions in the system of two nonidentical atoms also occur
in other situations. For example, quantum beats can appear in a system of
two identical atoms that experience different amplitude or phase of a
coherent driving field~\cite{fs,rfd}.

Consider the Hamiltonian~(\ref{t44}) of the interaction between
coherent laser field and two identical atoms. In the interaction
picture, the Hamiltonian can be written as
\begin{eqnarray}
       \hat{H}_{L}  &=&
        -\frac{1}{2}\hbar \left[\left(\Omega_{1}S_{1}^{+}
        +\Omega_{2}S_{2}^{+}\right)
        + \rm{H.c.}\right] \ ,\label{t102}
\end{eqnarray}
where $\Omega_{i}$ is the Rabi frequency of the driving field at the
position of the $i$th atom.

For the atoms in a running-wave laser field with $\vec{k}_{L}
\cdot \vec{r}_{i}\neq 0$, the Rabi frequency is a
complex parameter, which may be written as
\begin{eqnarray}
\Omega_{i} =\Omega e^{i\vec{k}_{L}
\cdot \vec{r}_{i}} \ ,  \label{t103}
\end{eqnarray}
where $\Omega =|\vec{\mu}_{i}\cdot \vec{E}_{L}|/\hbar$ is the maximum
Rabi frequency and $\vec{k}_{L}$ is the wave vector of the driving
field. Thus, in the running-wave laser field the atoms experience
different phases of the driving field.

For the atoms in a standing-wave laser field and $\vec{k}_{L}
\cdot \vec{r}_{i}\neq 0$, the Rabi frequency is a
real parameter, which may be written as
\begin{eqnarray}
\Omega_{i} =\Omega \cos\left(\vec{k}_{L}
\cdot \vec{r}_{i}\right) \ .  \label{t104}
\end{eqnarray}
Hence, in the standing-wave laser field the atoms experience
different amplitudes of the driving field.

In the following, we choose the reference frame such that the atoms
are at the positions $\vec{r}_{1}=(r_{1},0,0)$ and
$\vec{r}_{2}=(r_{2},0,0)$ along the $x$-axis, with distance $r_{12}$
apart. In this case,
\begin{eqnarray}
    \Omega_{1} =\Omega e^{i\vec{k}_{L}
\cdot \vec{r}_{1}} \ ,\qquad
\Omega_{2} =\Omega e^{i\vec{k}_{L}
\cdot \vec{r}_{2}} \ ,  \label{t105}
\end{eqnarray}
for the atoms in the running-wave field, and
\begin{eqnarray}
\Omega_{1} =\Omega \cos\left(\vec{k}_{L}
\cdot \vec{r}_{1}\right) \ ,\qquad
\Omega_{2} =\Omega \cos\left(\vec{k}_{L}
\cdot \vec{r}_{2}\right) \ ,  \label{t106}
\end{eqnarray}
for the atoms in the standing-wave field.

With the above choice of the Rabi frequencies, the
Hamiltonian~(\ref{t102}) takes the form
\begin{eqnarray}
       \hat{H}_{L}  &=&
        -\frac{1}{2}\hbar \left(\Omega S_{s}^{+}
        + \rm{H.c.}\right) \ ,\label{t107}
\end{eqnarray}
where $S_{s}^{+}= S_{1}^{+}\exp(i\vec{k}_{L}\cdot \vec{r}_{1}) +
S_{2}^{+}\exp(i\vec{k}_{L}\cdot \vec{r}_{2})$ for the running-wave
field, and
$S_{s}^{+}= S_{1}^{+}\cos(\vec{k}_{L}\cdot \vec{r}_{1}) +
S_{2}^{+}\cos(\vec{k}_{L}\cdot \vec{r}_{2})$ for the standing-wave
field. The operator $S_{s}^{+}$
corresponds to the symmetric superposition operator defined in
Eq.(\ref{t88}). Following the procedure, we developed in
Sec.~\ref{sec322}, we find that the transformation coefficients
$u$ and $v$ are
\begin{eqnarray}
        u &=& \frac{e^{i\vec{k}_{L}\cdot \vec{r}_{1}}}{\sqrt{2}} \ ,\qquad
        v =\frac{e^{i\vec{k}_{L}\cdot \vec{r}_{2}}}{\sqrt{2}} \
        ,\label{t108}
\end{eqnarray}
for the running-wave field, and
\begin{eqnarray}
        u &=& \frac{\cos (\vec{k}_{L}\cdot \vec{r}_{1})}
        {\sqrt{\cos^{2}(\vec{k}_{L}\cdot \vec{r}_{1})
        +\cos^{2}(\vec{k}_{L}\cdot \vec{r}_{2})}}
        \ ,\nonumber \\
        v &=& \frac{\cos(\vec{k}_{L}\cdot \vec{r}_{2})}
        {\sqrt{\cos^{2}(\vec{k}_{L}\cdot \vec{r}_{1})
        +\cos^{2}(\vec{k}_{L}\cdot \vec{r}_{2})}} \ ,\label{t109}
\end{eqnarray}
for the standing-wave field.

Using the transformation coefficients~(\ref{t108}) and~(\ref{t109}),
we find that the spontaneously induced coherences $\Gamma_{as}$ and
the coherent coupling $\Delta_{c}$ between the symmetric and
antisymmetric transitions are
\begin{eqnarray}
        \Gamma_{as} &=& -i\Gamma \sin(\vec{k}_{L}\cdot \vec{r}_{12})
        \ ,\quad
        \Delta_{c} = i\Delta_{L}\sin(\vec{k}_{L}\cdot \vec{r}_{12})
        \ ,\label{t110}
\end{eqnarray}
for the running-wave field, and
\begin{eqnarray}
        \Gamma_{as} &=& -\Gamma_{12}\frac{\sin^{2}(\vec{k}_{L}\cdot
        \vec{r}_{12})}{1+\cos^{2}(\vec{k}_{L}\cdot \vec{r}_{12})}
        \ ,\quad
        \Delta_{c} = \Omega_{12}\frac{\sin^{2}(\vec{k}_{L}\cdot
        \vec{r}_{12})}{1+\cos^{2}(\vec{k}_{L}\cdot \vec{r}_{12})}
        \ ,\label{t111}
\end{eqnarray}
for the standing-wave field, where, for simplicity, we have chosen
the reference frame such that  $r_{1}=0$ and $r_{2}=r_{12}$.

First, we note that no quantum beats can be obtained for the
direction of propagation of the laser field perpendicular to the
interatomic axis, because $\sin(\vec{k}_{L}
\cdot \vec{r}_{12})= 0$; however, quantum beats occur for directions
of propagation different from the perpendicular to $\vec{r}_{12}$.
One can see from Eqs.~(\ref{t110}) and (\ref{t111}) that in the case
of the running-wave field and $\Delta_{L}=0$, the symmetric and antisymmetric
transitions are correlated only through the spontaneously induced
coherences $\Gamma_{as}$. In the case of the standing-wave field, both
coupling parameters $\Gamma_{as}$ and $\Delta_{c}$ are different from
zero. However, for interatomic separations $r_{12}<\lambda$, the
parameter $\Gamma_{as}$ is much smaller than $\Delta_{c}$, indicating
that in this case the coherent coupling dominates over the
spontaneously induced coherences. These simple analysis of the
parameters $\Gamma_{as}$ and $\Delta_{c}$ show that one should obtain
quantum beats in the total radiation intensity of the fluorescence
field emitted from two identical atoms.
\begin{figure}[t]
\begin{center}
\includegraphics[width=10cm]{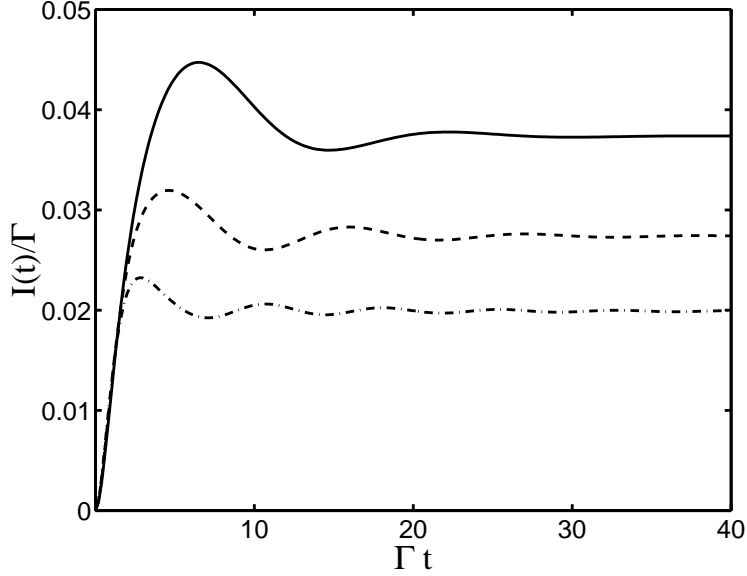}
\end{center}
\caption{Time evolution of the total radiation intensity for
the running-wave driving field with $\Omega =0.2\Gamma,
\vec{k}_{L}\parallel \vec{r}_{12}$
and different interatomic separations; $r_{12}=0.2\lambda$ (solid
line), $r_{12}=0.16\lambda$ (dashed line), $r_{12}=0.14\lambda$
(dashed-dotted line).}
\label{ftfig6}
\end{figure}
Figures~\ref{ftfig6} and~\ref{ftfig7} show the time evolution of the
total radiation
intensity, obtained by numerical solutions of the equations of motion
for the atomic correlation functions. The equations are found from the
master equation~(\ref{t42}), which in the case of the running- or
standing-wave driving field leads to a closed set of fifteen equations
of motion for the atomic correlation functions~\cite{fs,rfd}. In
Fig.~\ref{ftfig6}, we present the time-dependent total radiation
intensity for the running-wave driving field with $\Omega
=0.2\Gamma, \vec{k}_{L}\parallel \vec{r}_{12}$
and different interatomic separations. Fig.~\ref{ftfig7} shows the total
radiation intensity for the same parameters as in Fig.~\ref{ftfig6}, but
the standing-wave driving field. As predicted by Eqs.~(\ref{t110}) and
(\ref{t111}), the intensity exhibits quantum beats. The amplitude and
frequency of the oscillations is dependent on the interatomic
interactions and vanishes for large interatomic separations as well as
for separations very small compared with the resonant wavelength.
This is easily explained in the framework of collective states of a
two-atom system. For a weak driving field, the population oscillates
between the intermediate states $\ket s$, $\ket a$ and the ground
state $\ket g$. When interatomic separations are large, $\Omega_{12}$ is
approximately zero, and then the transitions $\ket s \rightarrow \ket
g$ and $\ket a \rightarrow \ket g$ have the same frequency. Therefore,
there are no quantum beats in the emitted field. On the other hand,
for very small interatomic separations, $\vec{k}_{L}\cdot
\vec{r}_{12}\approx 0$, and then the coupling parameters
$\Gamma_{as}$ and $\Delta_{c}$ vanish, resulting in the disappearance
of the quantum beats.
\begin{figure}[t]
\begin{center}
\includegraphics[width=10cm]{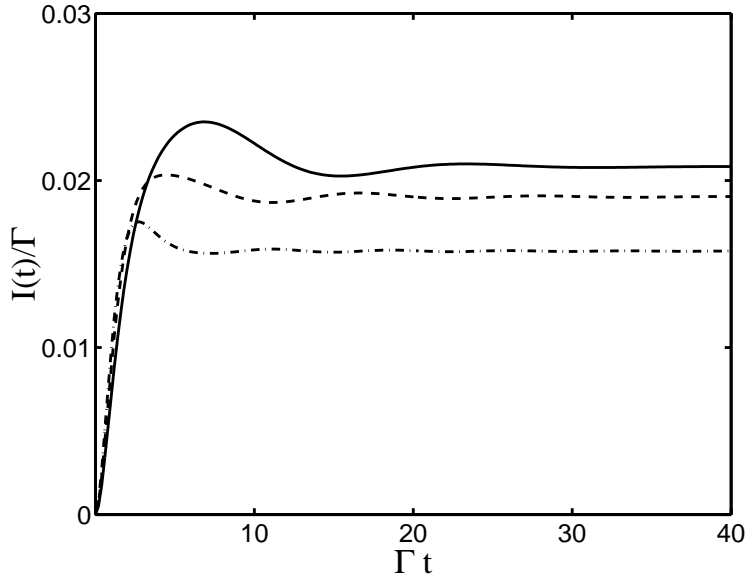}
\end{center}
\caption{Time evolution of the total radiation intensity for
the same parameters as in Fig.~\ref{ftfig6}, but the standing-wave
driving field.}
\label{ftfig7}
\end{figure}

\section{Nonclassical states of light}\label{ftsec5}

The interaction of light with atomic systems can lead to unique
phenomena such as photon antibunching and squeezing. These effects
are examples of a nonclassical light field, that is a field for which
quantum mechanics is essential for its description. Photon antibunching
is characteristic of a radiation in which the variance of the number of
photons is less than the mean number of photons, i.e. the photons
exhibit sub-Poissonian statistics. Squeezing is characteristic of a
field with phase-sensitive quantum fluctuations, which in one of the
two phase components are reduced below the vacuum (shot-noise) level.
Since photon antibunching and squeezing are distinguishing features of
light, it is clearly of interest to identify situations in which such
fields can be generated. Photon antibunching has been predicted
theoretically for the first time in resonance fluorescence of a
two-level atom~\cite{cw76,km76}. Since then, a number of papers have
appeared analyzing various schemes for generating photon antibunching
offered by nonlinear optics~\cite{wal79,lo80,per80,paul82}. Squeezing
has been extensively studied since the theoretical work by Walls and
Zoller~\cite{wz81} and Mandel~\cite{man82} on reduction of noise and
photon statistics in resonance fluorescence of a two-level atom.
Several experimental groups have been successful in producing
nonclassical light. Photon antibunching has been observed in resonance
fluorescence from a dilute atomic beam of sodium atoms driven by a
coherent laser field~\cite{kdm77,kdm78,cre}. More recently, beautiful
measurements of photon antibunching have been made
on trapped atoms~\cite{dw87}, and a cavity QED system~\cite{fost00}.
On the other hand, squeezed light was first observed by Slusher {\it et
al.}~\cite{slush} in four-wave mixing experiments. After that observation
squeezed light has been observed in many other nonlinear processes,
with a recent development being the availability of a tunable source
of squeezed light exhibiting a noise reduction of $\sim 70\%$ below
the shot-noise level. The experimental observation of photon antibunching
and squeezing have provided direct evidence of the quantum nature of light,
and these two phenomena were precursors of much of the present work on
nonclassical light fields. An extensive literature on various aspects of
photon antibunching and squeezing now exists and is reviewed in
several articles~\cite{jmo,josab,dodo}.

The objective of this section is to concentrate on collective two-atom
systems as a potential source for photon antibunching and squeezing.
We understand collective effects in a broad sense, that for
two or more atoms all effects that cannot be explained by the properties
of individual atoms are considered as collective. This definition of
collective effects thus includes, for example, both the resonance
fluorescence from a system of two atoms in free space and also collective
behaviour of two atoms strongly coupled to the same cavity mode in the
good cavity limit. Moreover, we emphasize the role of the
interatomic interactions in the generation of nonclassical light.
We also relate the nonclassical effects to the degree of entanglement
in the system.

\subsection{Photon antibunching}\label{ftsec51}

Photon antibunching is described through the normalized second-order
correlation function, defined as~\cite{mw}
\begin{eqnarray}
g^{(2)}\left(\vec{R}_{1},t_{1};\vec{R}_{2},t_{2}\right) =
\frac{G^{(2)}\left(\vec{R}_{1},t_{1};\vec{R}_{2},t_{2}\right)}
{G^{(1)}\left(\vec{R}_{1},t_{1}\right)G^{(1)}\left(\vec{R}_{2},t_{2}\right)}
\ ,\label{t112}
\end{eqnarray}
where
\begin{eqnarray}
G^{(2)}&&\left(\vec{R}_{1},t_{1}; \vec{R}_{2},t_{2}\right) \nonumber
\\
&&= \langle \vec{E}^{(-)}\left(\vec{R}_{1},t_{1}\right)\vec{E}^{(-)}
\left(\vec{R}_{2},t_{2}\right)\vec{E}^{(+)}\left(\vec{R}_{2}, t_{2}\right)
\vec{E}^{(+)}\left(\vec{R}_{1}, t_{1}\right)\rangle \ ,\label{t113}
\end{eqnarray}
is the two-time second-order correlation function of the EM field
detected at a point $\vec{R}_{1}$ at time $t_{1}$ and at a point
$\vec{R}_{2}$ at time $t_{2}$, and
\begin{eqnarray}
G^{(1)}\left(\vec{R}_{i},t_{i}\right) &=& \langle
\vec{E}^{(-)}\left(\vec{R}_{i},t_{i}\right)
\vec{E}^{(+)}\left(\vec{R}_{i},t_{i}\right)\rangle \ ,\label{t114}
\end{eqnarray}
is the first-order correlation function of the field (intensity) detected
at a point $R_{i}$ at time $t_{i} (i=1,2)$.

The correlation function $G^{(2)}(\vec{R}_{1},t_{1};
\vec{R}_{2},t_{2})$ is proportional to a joint probability of
finding one photon around the direction $\vec{R}_{1}$ at time $t_{1}$
and another photon around the direction $\vec{R}_{2}$ at the moment of
time $t_{2}$. For a coherent light, the probability of finding a
photon around $\vec{R}_{1}$ at time $t_{1}$ is independent of the
probability of finding another photon around $\vec{R}_{2}$ at time
$t_{2}$, and then $G^{(2)}(\vec{R}_{1},t_{1};
\vec{R}_{2},t_{2})$ simply factorizes into
$G^{(1)}(\vec{R}_{1},t_{1})
G^{(1)}(\vec{R}_{2},t_{2})$ giving
$g^{(2)}(\vec{R}_{1},t_{1};\vec{R}_{2},t_{2})=1$. For a
chaotic (thermal) field the second-order correlation function for
$t_{1}=t_{2}$ is greater than for $t_{2}-t_{1}=\tau >0$ giving
$g^{(2)}(\vec{R}_{1},t_{1};\vec{R}_{2},t_{1})
>g^{(2)}(\vec{R}_{1},t_{1};\vec{R}_{2},t_{1}+\tau )$. This
is a manifestation of the tendency of photons to be emitted in
correlated pairs, and is called photon bunching. Photon antibunching,
as the name implies, is the opposite of bunching, and describes a
situation in which fewer photons appear close together than further
apart. The condition for photon antibunching is
$g^{(2)}(\vec{R}_{1},t_{1};\vec{R}_{2},t_{1})
<g^{(2)}(\vec{R}_{1},t_{1};\vec{R}_{2},t_{1}+\tau )$ and
implies that the probability of detecting two photons at the same
time $t$ is smaller than the probability of detecting two photons at
different times $t$ and $t+\tau$. Moreover, the fact that there is a
small probability of detecting photon pairs with zero time separation
indicated that the one-time correlation function
$g^{(2)}(\vec{R}_{1},t;\vec{R}_{2},t)$ is smaller than
one. This effect is called photon anticorrelation. The normalized
one-time second-order correlation function carries also information
about photon statistics, which is given by the Mandel's $Q$ parameter
defined as~\cite{man82}
\begin{eqnarray}
        Q = qT\left[g^{(2)}\left(\vec{R}_{1},t; \vec{R}_{2},t\right)
        -1\right] \ ,\label{t115}
\end{eqnarray}
where $q$ is the quantum efficiency of the detector and $T$ is the
photon counting time.

We can relate the field correlation functions~(\ref{t113})
and~(\ref{t114}) to the correlation functions of the atomic operators,
which will allow us to apply directly the master equation~(\ref{t42})
to calculate photon antibunching in a collective atomic system.
The relation between the positive frequency part of the electric field
operator at a point $\vec{R}=R\bar{R}$, in the far-field zone,
and the atomic dipole operators $S_{i}^{-}$, is given by the well-known
expression~\cite{leh,ag74}
\begin{eqnarray}
\vec{E}^{(+)}\left(\vec{R},t\right) &=&
\vec{E}^{(+)}_{0}\left(\vec{R},t\right) \nonumber \\
&-& \sum_{i=1}^{2}\frac{\omega_{i}}{c^{2}}
\frac{\bar{R}\times \left(\bar{R}\times
\vec{\mu}_{i}\right)}{R}
S_{i}^{-}\left(t -\frac{R}{c}\right) \exp
\left(-ik\bar{R}\cdot \vec{r}_{i}\right) \ ,\label{t116}
\end{eqnarray}
where $\omega_{i}$ is the angular frequency of the $i$th atom located
at a point $\vec{r}_{i}$, and $\vec{E}^{(+)}_{0}\left(\vec{R},t\right)$
denotes the positive frequency part of the field in the absence of the
atoms.

If we assume that initially the field is in the vacuum state,
then the free-field part $\vec{E}^{(+)}_{0}(\vec{R},t)$ does
not contribute to the expectation values of the normally ordered
operators. Hence, substituting Eq.~(\ref{t116}) into Eqs.~(\ref{t113})
and~(\ref{t114}), we obtain
\begin{eqnarray}
G^{\left( 2\right)}\left( \vec{R},t;\vec{R},t+\tau \right)
&=& u(\vec{R}_{1})u(\vec{R}_{2})
\sum_{i,j,k,l=1}^{N}\left(\Gamma _{i}\Gamma_{j}\Gamma
_{k}\Gamma_{l}\right)^{\frac{1}{2}} \nonumber \\
&\times& \left\langle S^{+}_{i}\left( t\right) S^{+}_{k}\left(t+\tau \right)
S^{-}_{l}\left( t+\tau\right) S^{-}_{j}\left(t\right) \right\rangle
\nonumber \\
&\times& \exp \left[ik\left(\bar{R}_{1}\cdot \vec{r}_{ij}+\bar{R}_{2}
\cdot \vec{r}_{kl}\right)\right] \ ,\label{t117}\\
G^{\left( 1\right) }\left(\vec{R},t\right)
&=& u(\vec{R})
\sum_{i,j=1}^{N}\left(\Gamma_{i}\Gamma_{j}\right)^{\frac{1}{2}}\left\langle
S_{i}^{+}\left( t\right) S^{-}_{j}\left( t\right) \right\rangle
\nonumber \\
&\times& \exp \left(ik\bar{R}\cdot \vec{r}_{ij}\right)
\ ,\label{t118}
\end{eqnarray}
where $\tau =t_{2}-t_{1}$, $\Gamma_{i}$ is the damping rate of the
$i$th atom, and $u(\vec{R})$ is a constant given in
Eq.~(\ref{t70}). The second-order correlation function~(\ref{t117})
involves two-time atomic correlation function that can be calculated
from the master equation~(\ref{t36}) or~(\ref{t42}) and applying the
quantum regression theorem~\cite{lax}. From the quantum regression
theorem, it is well known that for $\tau >0$ the two-time correlation
function $\langle S^{+}_{i}\left( t\right) S^{+}_{k}\left(t+\tau
\right)S^{-}_{l}\left( t+\tau\right) S^{-}_{j}\left(t\right)
\rangle$ satisfies the same equation of motion as the one-time
correlation function $\langle S_{k}^{+}\left( t\right)
S^{-}_{l}\left( t\right) \rangle$.

We shall first of all consider the simplest collective system for
photon antibunching; two identical atoms in the Dicke model. Whilst
this model is not well satisfied with the present sources of two-atom
systems, it does enable analytic treatments that allow to understand
the role of the collective damping in the generation of nonclassical
light.

For the two-atom Dicke model the master equation~(\ref{t42}) reduces
to
\begin{eqnarray}
        \frac{\partial \hat{\rho}}{\partial t} &=& \frac{1}{2}i\Omega
        \left[S^{+}+S^{-}, \hat{\rho}\right] -\frac{1}{2}\Gamma
        \left(S^{+}S^{-}\hat{\rho} +\hat{\rho}S^{+}S^{-}
        -2S^{-}\hat{\rho}S^{+}\right) \ ,\label{t119}
\end{eqnarray}
where $S^{\pm}=S^{\pm}_{1}+S^{\pm}_{2}$ and $S^{z}=S^{z}_{1}+S^{z}_{2}$
are the collective atomic operators and $\Omega$ is the Rabi frequency
of the driving field, which in the Dicke model is the same for both
atoms. For simplicity, the laser frequency $\omega_{L}$ is taken to
be exactly equal to the atomic resonant frequency $\omega_{0}$.

\begin{figure}[t]
\begin{center}
\includegraphics[width=10cm]{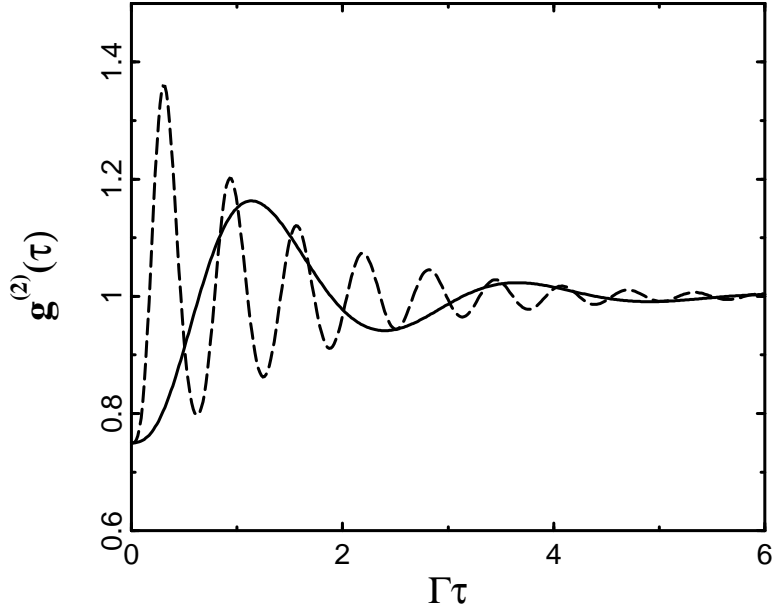}
\end{center}
\caption{The normalised second-order correlation function
$g^{(2)}\left(\tau \right)$ as a function of $\tau$ and different
$\Omega$; $\Omega =2.5\Gamma$ (solid line), $\Omega =10\Gamma$ (dashed
line).}
\label{ftfig8}
\end{figure}

The secular approximation technique has been suggested by
Agarwal {\it et al.}~\cite{aga79} and Kilin~\cite{kil80},
which greatly simplifies the master equation~(\ref{t119}). Hassan {\it
et al.}~\cite{has82} and Cordes~\cite{cor82,cor87} have generalised the
method to include non-zero detuning of the laser field and the
quasistatic dipole-dipole potential. The technique is a modification of
a collective dressed-atom approach developed by Freedhoff~\cite{fr79} and
is valid if the Rabi frequency of the driving field
is much greater than the damping rates of the atoms, $\Omega \gg
\Gamma$. To implement the technique, we transform the collective
operators into new (dressed) operators
\begin{eqnarray}
        S^{\pm} &=& \pm \frac{1}{2}i\left(R^{+} +R^{-}\right) +R^{z} \
        ,\nonumber \\
        S^{z} &=& -\frac{1}{2}i\left(R^{+} -R^{-}\right) \ .\label{t120}
\end{eqnarray}
The operators $R$ are a rotation of the operators $S$. For a strong
driving field, the operators $R^{\pm}$ vary rapidly with time,
approximately as $\exp (\pm i\Omega t)$, while $R^{z}$ varies slowly
in time. By expressing the operators $S^{\pm}$ and $S^{z}$ in terms of
the operators $R^{\pm}$ and $R^{z}$, and substituting into the master
equation~(\ref{t119}), we find that certain terms are slowly varying
in time while others oscillate rapidly. The secular approximation then
involves dropping the rapidly oscillating terms that results in an
approximate master equation of the form
\begin{eqnarray}
        \frac{\partial \hat{\rho}}{\partial t} = i\Omega
        \left[R^{z}, \hat{\rho}\right] &-& \frac{1}{2}\Gamma \left\{
        \left(R^{z}R^{z}\hat{\rho} +\hat{\rho}R^{z}R^{z}
        -2R^{z}\hat{\rho}R^{z}\right)\right. \nonumber \\
        &&\left. +\frac{1}{4}\left(R^{+}R^{-}\hat{\rho}
        +\hat{\rho}R^{+}R^{-} -2R^{-}\hat{\rho}R^{+}\right)\right.
        \nonumber \\
        &&\left. +\frac{1}{4}\left(R^{-}R^{+}\hat{\rho}
        +\hat{\rho}R^{-}R^{+} -2R^{+}\hat{\rho}R^{-}\right)\right\}
        \ .\label{t121}
\end{eqnarray}
The master equation~(\ref{t121}) enables to obtain equations of motion
for the expectation value of an arbitrary combination of the
transformed operators $R$. In particular, the master equation leads to
simple equations of motion for the expectation values required to calculate
the normalized second-order correlation function. The required equations of
motion are given by
\begin{eqnarray}
        \frac{d}{dt}\left\langle R^{z}\right\rangle &=& -\frac{1}{2}\Gamma
        \left\langle R^{z}\right\rangle \ ,\nonumber \\
        \frac{d}{dt}\langle R^{\pm}\rangle &=&
        -\left(\frac{3}{4}\Gamma \pm i\Omega \right)
        \langle R^{\pm}\rangle \ ,\nonumber \\
        \frac{d}{dt}\langle R^{+}R^{+}\rangle &=&
        -\left(\frac{5}{2}\Gamma +2i\Omega \right)
        \langle R^{+}R^{+}\rangle \ .\label{t122}
\end{eqnarray}
The solution of these decoupled differential equations is
straightforward. Performing the integration and applying the quantum
regression theorem~\cite{lax}, we obtain from
Eqs.~(\ref{t122}) and (\ref{t112}) the following solution for the
normalized second-order correlation function~\cite{ftk83,hds}
\begin{eqnarray}
        g^{(2)}\left(\tau \right) &\equiv&
        \lim_{t\rightarrow \infty} g^{(2)}\left(\vec{R}_{1},t;\vec{R}_{2},
        t+\tau \right) = 1
        +\frac{1}{32}\exp \left(-\frac{3}{2}\Gamma \tau \right) \nonumber
        \\
        && +\frac{3}{32}\exp \left(-\frac{5}{2}\Gamma \tau \right)\cos
        \left(2\Omega \tau \right)
        -\frac{3}{8}\exp \left(-\frac{3}{4}\Gamma \tau \right) \cos
        \left(\Omega \tau \right) \ .\label{t123}
\end{eqnarray}
The correlation function $g^{(2)}\left(\tau \right)$ is shown in
Fig.~\ref{ftfig8}
as a function of $\tau$ for different $\Omega$. For $\tau =0$, the
correlation function $g^{(2)}\left(0 \right)=0.75$, showing the photon
anticorrelation in the emitted fluorescence field. As $\tau$
increases, the correlation function increases $(g^{(2)}\left(\tau
\right) > g^{(2)}\left(0 \right))$, which reflects photon
antibunching in the emitted field. However, the photon anticorrelation
in the two-atom fluorescence field is reduced compared to that for a
single atom, for which $g^{(2)}\left(0 \right)=0$. This result
indicates that the collective damping reduces the photon
anticorrelations in the emitted fluorescence field.

As we have mentioned above, in the Dicke model the dipole-dipole
interaction between the atoms is
ignored. This approximation has no justification, since for small
interatomic separations the dipole-dipole parameter $\Omega_{12}$,
which varies as $(k_{0}r_{12})^{-3}$, is very large and goes to
infinity as $r_{12}$ goes to
zero (see Fig.~\ref{ftfig1}). Moreover, the Dicke model does not
correspond to the experimentally realistic systems in which atoms are
separated by distances comparable to the resonant wavelength.
Ficek {\it et al.}~\cite{ftk83} and Lawande {\it et al.}~\cite{law85}
have shown that the dipole-dipole
interaction does not considerably affect the anticorrelation effect
predicted in the Dicke model. Richter~\cite{rich1} has shown that the
value $g^{(2)}\left(0 \right)=0.75$ can in fact be reduced such that
even the complete photon anticorrelation $g^{(2)}\left(0 \right)=0$ can
be obtained, if the dipole-dipole is included and the laser frequency
is detuned from the atomic transition frequency. To show this, we
calculate the normalized second-order correlation function~(\ref{t112})
for the steady-state fluorescence field from two identical atoms $(N=2)$,
and $\tau =0$. In this case, the correlation function~(\ref{t112}) with
Eqs.~(\ref{t117}) and (\ref{t118}) can be written as
\begin{eqnarray}
        g^{(2)}\left(0 \right) =\frac{2U
        \left\{1+\cos\left[k\vec{r}_{12}\cdot
        \left(\bar{R}_{1}-\bar{R}_{2}\right)\right]\right\}}{\left[
        1+W\cos\left(k\vec{r}_{12}\cdot
        \bar{R}_{1}\right)\right] \left[
        1+W\cos\left(k\vec{r}_{12}\cdot
        \bar{R}_{2}\right)\right]} \ ,\label{t124}
\end{eqnarray}
where $U$ and $W$ are the steady-state atomic correlation functions
\begin{eqnarray}
        U = \frac{\langle
        S^{+}_{1}S^{+}_{2}S^{-}_{1}S^{-}_{2}\rangle}{\langle
        S^{+}_{1}S^{-}_{1} +S^{+}_{2}S^{-}_{2}\rangle^{2}} \ ,\qquad
        W = \frac{\langle
        S^{+}_{1}S^{-}_{2} +S^{+}_{2}S^{-}_{1}\rangle}{\langle
        S^{+}_{1}S^{-}_{1} +S^{+}_{2}S^{-}_{2}\rangle} \ .\label{t125}
\end{eqnarray}
The steady-state correlation functions are easily obtained from the
master equation~(\ref{t42}). We can simplify the solutions assuming
that the atoms are in equivalent positions in the driving field,
which can be achieved by propagating the laser field in the direction
perpendicular to the
interatomic axis. In this case we get analytical solutions, otherwise
for $\vec{k}_{L}\cdot \vec{r}_{12}\neq 0$ numerical methods are
more appropriate~\cite{fs,rfd,rich2}. With $\vec{k}_{L}\cdot
\vec{r}_{12}=0$ the master equation~(\ref{t42}) leads to a closed set
of nine equations of motion for the atomic correlation functions.
This set of equations can be solved exactly in the
steady-state~\cite{ftk84}, and the solutions for $U$ and $W$ are
\begin{eqnarray}
        U &=& \frac{\Omega^{4}
        +\left(\Gamma^{2}+4\Delta_{L}^{2}\right)\Omega^{2}
        +\left(\Gamma^{2}+4\Delta_{L}^{2}\right)\left[ \frac{1}{4}
        \left(\Gamma +\Gamma_{12}\right)^{2}
        +\left(\Delta_{L}-\Omega_{12}\right)^{2}\right]}
        {\left(\Gamma^{2} +4\Delta_{L}^{2} +2\Omega^{2}\right)^{2}}
        \ ,\nonumber \\
        W &=&  \frac{\left(\Gamma^{2}+4\Delta_{L}^{2}\right)}
        {\left(\Gamma^{2}+4\Delta_{L}^{2}+2\Omega^{2}\right)} \
        .\label{t126}
\end{eqnarray}
One can see from Eqs.~(\ref{t124}) and (\ref{t126}) that there are two
different processes which can lead to the total anticorrelation,
$g^{(2)}\left(0 \right)=0$. The
first one involves an observation of the fluorescence field with two
detectors located at different points. If the correlation function is
measured using two detectors, $\bar{R}_{1}\neq \bar{R}_{2}$, and then
we obtain $g^{(2)}\left(0 \right)=0$ whenever the positions of the
detectors are such that
\begin{eqnarray}
        \left\{1+\cos\left[k\vec{r}_{12}\cdot
        \left(\bar{R}_{1}-\bar{R}_{2}\right)\right]\right\} =0 \
        ,\label{t127}
\end{eqnarray}
which happens when
\begin{eqnarray}
k\vec{r}_{12}\cdot \left(\bar{R}_{1}-\bar{R}_{2}\right) =
\left(2n+1\right)\pi \ ,\quad n=0,\pm 1,\pm 2 \ldots \label{t128}
\end{eqnarray}
In other words, two photons can never be simultaneously detected at two
points separated by an odd number of $\lambda/2r_{12}$, despite the fact
that one photon can be detected anywhere. This complete anticorrelation
effect is due to spatial interference between different photons and
reflects the fact that one photon must have come from one source and
one from the other, but we cannot tell which came from which.

It should be emphasised that this effect is independent of the interatomic
interactions and the Rabi frequency of the
driving field. The vanishing of $g^{(2)}\left(0 \right)$ for two photons at
widely separated points $\vec{R}_{1}$ and $\vec{R}_{2}$ is an example of
quantum-mechanical nonlocality, that the outcome of a detection measurement
at $\vec{R}_{1}$ appears to be influenced by where we have chosen to locate
the $\vec{R}_{2}$ detector. At certain positions $\vec{R}_{2}$ we can never
detect a photon at $\vec{R}_{1}$ when there is a photon detected at
$\vec{R}_{2}$, whereas at other position $\vec{R}_{2}$ it is possible.
The photon correlation argument shows clearly that quantum theory does not
in general describe an objective physical reality independent of
observation.

The second process involves the shift of the collective atomic states
due to the dipole-dipole interaction that can lead
to $g^{(2)}\left(0 \right)=0$ even if the correlation function is
measured with a single detector $(\bar{R}_{1}=\bar{R}_{2})$ or two
detectors in configurations different from that given by
Eq.~(\ref{t128}). For a weak driving field $(\Omega \ll \Gamma)$ and
large detunings such that $\Delta_{L} =\Omega_{12}\gg \Gamma$, the
correlation function~(\ref{t112}) with $\bar{R}_{1}=\bar{R}_{2}$
simplifies to
\begin{eqnarray}
        g^{(2)}\left(0 \right) \approx \frac{\left(\Gamma
        +\Gamma_{12}\right)^{2}}{4\Delta_{L}^{2}} \ .\label{t129}
\end{eqnarray}
\begin{figure}[t]
\begin{center}
\includegraphics[width=10cm]{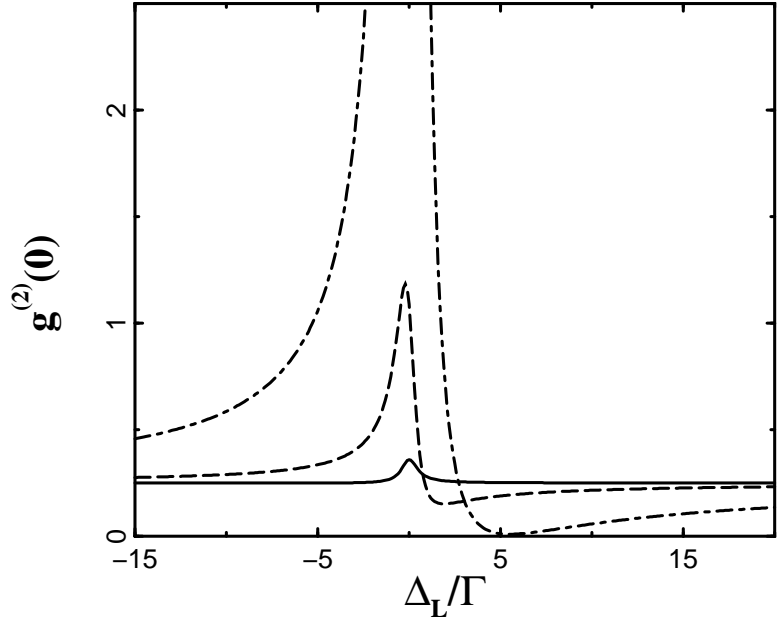}
\end{center}
\caption{The normalised second-order correlation function
$g^{(2)}\left(0 \right)$ as a function of $\Delta_{L}$ for
$\bar{R}_{1}=\bar{R}_{2}=\bar{R}, \vec{r}_{12}\perp \bar{R},
\bar{\mu}\perp \bar{r}_{12}$,
$\Omega =0.5\Gamma$ and different $r_{12}$; $r_{12} =10\lambda$
(solid line), $r_{12} =0.15\lambda$ (dashed line), $r_{12}=0.08\lambda$
(dashed-dotted line).}
\label{ftfig9}
\end{figure}
Thus, a pronounced photon anticorrelation, $g^{(2)}\left(0
\right)\approx 0$, can be obtained for large detunings such that
$\Delta_{L} =\Omega_{12}$, i.e., when the dipole-dipole interaction
shift of the collective states and the
detuning cancel out mutually. The correlation function
$g^{(2)}\left(0 \right)$ of the steady-state fluorescence field is
illustrated graphically in Fig.~\ref{ftfig9} as a function of $\Delta_{L}$
for the single detector configuration with $\bar{R}_{1}=\bar{R}_{2}=\bar{R}$,
and different $r_{12}$. The graphs show that $g^{(2)}\left(0 \right)$ strongly
depends on $\Delta_{L}$, and the total photon anticorrelation can be
obtained for $\Delta_{L} =\Omega_{12}$. Referring to Fig.~\ref{ftfig2},
the condition $\Delta_{L} =\Omega_{12}$ corresponds to the laser frequency
tuned to the resonance with the $\ket g \rightarrow \ket s$
transition. Since the other levels are far from the resonance, the
two-atom system behaves like a single two-level system with the
ground state $\ket g$ and the excited state $\ket s$.

\subsection{Squeezing}\label{ftsec52}

To understand squeezed light, recall that the electric
field amplitude $\vec{E}\left( \vec{r}\right) $ may
be expressed by positive- and negative-frequency parts
\begin{equation}
\vec{E}\left( \vec{r}\right) =\vec{E}^{\left( +\right) }\left( \vec{r}
\right) +\vec{E}^{\left( -\right) }\left( \vec{r}\right) \ ,
\label{t130}
\end{equation}
where
\begin{equation}
\vec{E}^{\left( +\right) }\left( \vec{r}\right) =\left( \vec{E}^{\left(
-\right) }\left( \vec{r}\right) \right) ^{\dagger }=-i\sum_{\vec{k}s}%
\left( \hbar \omega _{k}/2\varepsilon_{o}
V\right) ^{1/2}\bar{e}_{\vec{k}s}\hat{a}_{\vec{k}s}e^{i\vec{k} \cdot
\vec{r}} \ ,\label{t131}
\end{equation}
and $\omega_{k}=c\left| \vec{k}\right| $ is the angular
frequency of the mode $\vec{k}$.

We introduce two Hermitian combinations (quadrature components) of the field
components that are $\pi /2$ out of phase as
\begin{eqnarray}
\vec{E}_{\theta } &=&\vec{E}^{\left( +\right) }\left(
\vec{R}\right) e^{i\theta }+\vec{E}^{\left( -\right) }\left( \vec{R}\right)
e^{-i\theta } \ ,  \nonumber \\
\vec{E}_{\theta -\pi /2} &=&-i\left( \vec{E}^{\left(
+\right) }\left( \vec{R}\right) e^{i\theta }-\vec{E}^{\left( -\right)
}\left( \vec{R}\right) e^{-i\theta }\right) \ ,  \label{t132}
\end{eqnarray}
where
\begin{equation}
\theta =\omega t-\vec{k}\cdot \vec{R} \ ,  \label{t133}
\end{equation}
and $\omega $ is the angular frequency of the quadrature components.

The quadrature components do not commute, satisfying the commutation
relation
\begin{equation}
\left[ \vec{E}_{\theta } ,\vec{E}_{\theta -\pi/2} \right] =2iC \
,\label{t134}
\end{equation}
where $C$ is a positive number
\begin{equation}
C=\sum_{\vec{k}s}\left| \hbar \omega _{k}/2\varepsilon_{o}
V \right| \ .\label{t135}
\end{equation}
Hence the two quadrature components cannot be simultaneously precisely
measured, and from the Heisenberg uncertainty principle, we find that the
variances $\left\langle \Delta \vec{E}_{\theta }^{2} \right\rangle$ and
$\left\langle \Delta \vec{E}_{\theta -\pi /2}^{2}
\right\rangle $ satisfy the inequality
\begin{equation}
\left\langle \Delta \vec{E}_{\theta }^{2} \right\rangle
\left\langle \Delta \vec{E}_{\theta -\pi /2}^{2}
\right\rangle \geq C^{2} \ ,  \label{t136}
\end{equation}
where the equality holds for a minimum uncertainty state of the field.

The variances $\left\langle \Delta \vec{E}_{\theta }^{2}\right\rangle $
and $\left\langle \Delta \vec{E}_{\theta -\pi/2}^{2} \right\rangle $ depend on
the state of the field and can be larger or smaller than $C$. A
chaotic state of
the field leads to the variances in both components larger than $C$:
\begin{eqnarray}
\left\langle \Delta \vec{E}_{\theta }^{2}\right\rangle \geq C
\quad  &\mbox{and}& \quad  \left\langle \Delta \vec{E}_{\theta -\pi /2}^{2}
\right\rangle \geq C \ .\label{t137}
\end{eqnarray}
If the field is in a coherent or vacuum state
\begin{equation}
\left\langle \Delta \vec{E}_{\theta }^{2}\right\rangle
= \left\langle \Delta \vec{E}_{\theta -\pi /2}^{2}\right\rangle =C
\ ,\label{t138}
\end{equation}
which is an example of a minimum uncertainty state.

A squeezed state of the field is defined to be one in which the
variance in one of the two quadrature components is less than that
for the vacuum field
\begin{eqnarray}
\left\langle \Delta \vec{E}_{\theta }^{2}\right\rangle
<C \quad &\mbox{or}& \quad \left\langle \Delta \vec{E}_{\theta -\pi
/2}^{2}\right\rangle < C \ .  \label{t139}
\end{eqnarray}
The variances can be expressed as
\begin{eqnarray}
\left\langle \Delta \vec{E}_{\theta }^{2} \right\rangle
     &=& C + \left\langle :\Delta \vec{E}_{\theta}^{2} :\right\rangle \
     ,\nonumber \\
\left\langle \Delta \vec{E}_{\theta -\pi /2}^{2}
\right\rangle &=&C+\left\langle :\Delta \vec{E}_{\theta
-\pi /2}^{2}:\right\rangle \ ,\label{t140}
\end{eqnarray}
where the colon stands for normal ordering of the operators.

As the squeezed state has been defined by the requirement that either $%
\left\langle \Delta \vec{E}_{\theta }^{2}\right\rangle
$ or $\left\langle \Delta \vec{E}_{\theta -\pi /2}^{2}
\right\rangle $ be below the vacuum level $C$, it follows immediately from
Eq.~(\ref{t140}) that either
\begin{eqnarray}
\left\langle :\Delta \vec{E}_{\theta }^{2}
:\right\rangle < 0 \quad &\mbox{or}& \quad \left\langle :\Delta
\vec{E}_{\theta -\pi/2}^{2} :\right\rangle < 0  \label{t141}
\end{eqnarray}
for the field in a squeezed state.

We now determine the relation between variances in the field and the
atomic dipole operators. Using Eq.~(\ref{t116}), which relates the field
operators to the atomic dipole operators, we obtain
\begin{eqnarray}
        \left\langle :\Delta \vec{E}_{\alpha }^{2}:\right\rangle =
        u\left(\vec{R}\right)\left[\langle \left(\Delta
        S_{\alpha}\right)^{2}\rangle +\frac{1}{2}\langle
        S_{3}\rangle\right] \ ,\label{t142}
\end{eqnarray}
where $\alpha =\theta ,\theta -\pi/2$, $S_{\alpha}$ and $S_{3}$ are
real (phase) operators defined as
\begin{eqnarray}
        S_{\theta} &=& \frac{1}{2}\left(S_{\theta}^{+}
        +S_{\theta}^{-}\right) \ ,\qquad
        S_{\theta -\pi/2} = \frac{1}{2i}\left(S_{\theta}^{+}
        -S_{\theta}^{-}\right) \ ,\label{t143}
\end{eqnarray}
and
\begin{eqnarray}
        S_{3} = \frac{1}{2}\left[S_{\theta}^{+},
        S_{\theta}^{-}\right] \ ,\label{t144}
\end{eqnarray}
with
\begin{eqnarray}
        S_{\theta}^{\pm} =\sum_{i=1}^{N}S_{i}^{\pm}\exp \left[\pm
        i\left(k\bar{R}\cdot \vec{r}_{i} -\theta\right) \right] \
        .\label{t145}
\end{eqnarray}

\begin{figure}[t]
\begin{center}
\includegraphics[width=10cm]{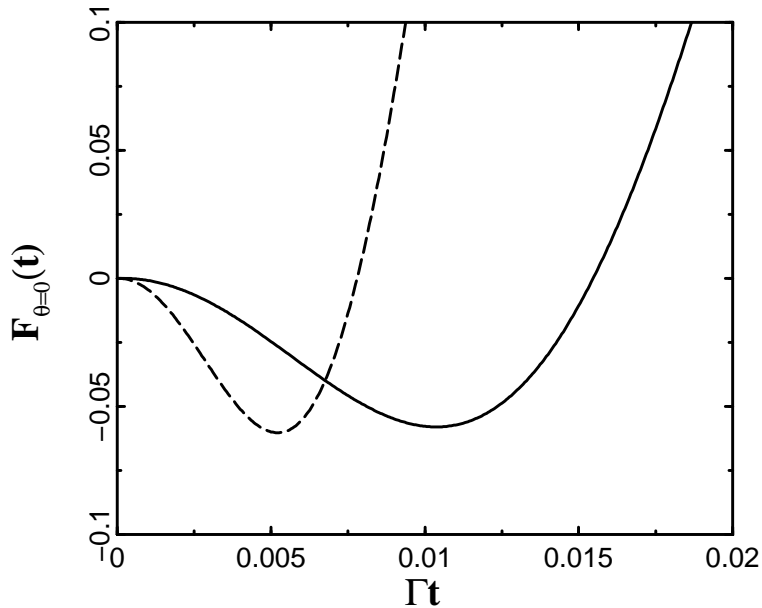}
\end{center}
\caption{The variance $F_{\theta =0}\left( t\right)$ as a function
of time for different $\Omega$; $\Omega =100\Gamma$
(solid line), $\Omega = 200\Gamma$ (dashed line).}
\label{ftfig10}
\end{figure}

We first consider quantum fluctuations in the fluorescence field emitted
by two identical atoms in the Dicke model. To simplify the
calculations we will treat only the case of zero detuning, $\Delta_{L}=0$.
Assuming that initially $(t=0)$ the atoms were in their ground states, we
find from Eqs.~(\ref{t122}) and (\ref{t142}) the following
expressions for the time-dependent variances~\cite{ftk87s}
\begin{eqnarray}
        F_{\theta =0}\left( t\right) &\equiv&
        \left\langle :\Delta \vec{E}_{\theta =0 }^{2}:\right\rangle /\left(
        2u(\vec{R})\right) = \frac{1}{3} -\frac{1}{8}\exp
        \left(-\frac{5}{2}\Gamma t\right)\cos\left(2\Omega t\right)
        \nonumber \\
        &+& \frac{1}{24}\exp \left(-\frac{3}{2}\Gamma t\right)
        -\frac{1}{2}\exp\left(-\frac{3}{2}\Gamma t\right)
        \sin^{2}\left(\Omega t\right) \nonumber \\
        &-& \frac{1}{4}\exp\left(-\frac{3}{4}\Gamma
        t\right)\cos\left(\Omega t\right) \ ,\label{t146}
\end{eqnarray}
and
\begin{eqnarray}
        F_{\theta =\pi/2}\left( t\right) &\equiv&
        \left\langle :\Delta \vec{E}_{\theta =\pi/2 }^{2}:\right\rangle
        /\left(2u(\vec{R})\right) = \frac{1}{3}
        -\frac{1}{12}\exp \left(-\frac{3}{2}\Gamma t\right) \nonumber \\
        &-& \frac{1}{4}\exp\left(-\frac{3}{4}\Gamma
        t\right)\cos\left(\Omega t\right) \ .\label{t147}
\end{eqnarray}
In writing Eqs.~(\ref{t146}) and (\ref{t147}), we have assumed that the
angular frequency of the quadrature components is equal to the laser
frequency, $\omega =\omega_{L}$, and we have normalised the variances
such that $F(t)$ determines fluctuations per atom.
It is easily to show that the variance $F_{\theta =\pi/2}\left(
t\right)$ is positive for all times $t$, and squeezing
$(F_{\theta}<0)$ can be observed in the variance $F_{\theta =0}\left(
t\right)$. The time dependence of the variance $F_{\theta =0}\left(
t\right)$ is shown in Fig.~\ref{ftfig10} for two different values of
the Rabi frequency. It is seen that squeezing appears in the transient
regime of resonance fluorescence and its maximum value (minimum of $F$)
moves towards shorter times as $\Omega$ increases. The optimum squeezing
reaches a value of $-1/16$ at a very short time. This value is equal
to the maximum possible squeezing in a single two-level
atom~\cite{wz81,ftk84a,bk88,dfk94}. Thus,
the collective damping does not affect squeezing in the two-atom
Dicke model. This is in contrast to the photon anticorrelation
effect which is greatly reduced by the collective damping.

\begin{figure}[t]
\begin{center}
\includegraphics[width=10cm]{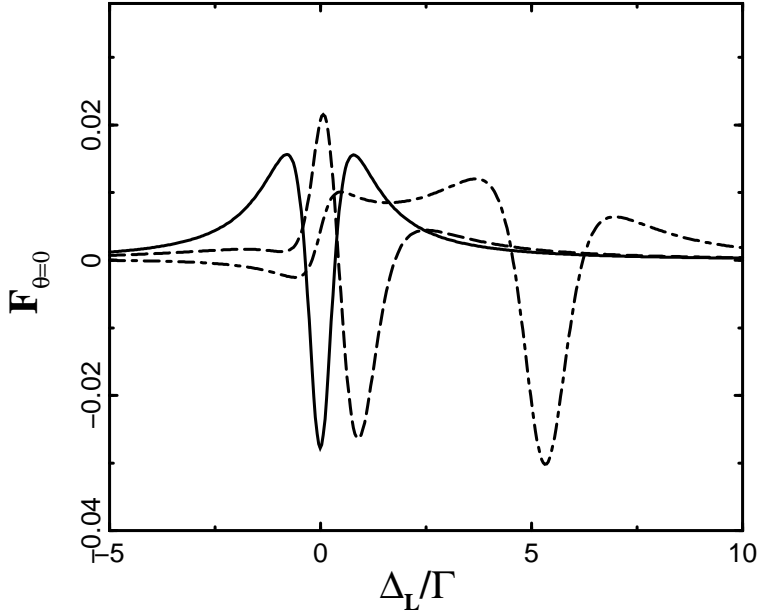}
\end{center}
\caption{The steady-state variance $F_{\theta =0}$ as a function
of $\Delta_{L}$ for
$\vec{r}_{12}\perp \bar{R}$, $\bar{\mu}\perp \bar{r}_{12}$,
$\Omega =0.5\Gamma$,
$\vec{k}_{L}\cdot \vec{r}_{12}= 0$ and different $r_{12}$;
$r_{12} =10\lambda$ (solid line), $r_{12} =0.15\lambda$ (dashed line),
$r_{12}=0.08\lambda$ (dashed-dotted line).}
\label{ftfig11}
\end{figure}

Figure~\ref{ftfig10} shows that in the two-atom Dicke model
there is no squeezing in the steady-state resonance fluorescence when
the atoms are excited by a strong laser field. Ficek {\it et
al.}~\cite{ftk84} and Richter~\cite{rich3} have shown that similarly
as in the case of photon anticorrelations, a large squeezing can
be obtained in the steady-state resonance fluorescence from a
strongly driven two-atom system, if the dipole-dipole interaction is included
and the laser frequency is detuned from the atomic transition
frequency. This is shown in Fig.~\ref{ftfig11}, where we plot $F_{\theta =0}$,
calculated from Eq.~(\ref{t142}) and the master equation~(\ref{t42}),
for the steady-state resonance fluorescence from two identical atoms,
with $\vec{r}_{12}\perp \bar{R}$, $\Omega =0.5\Gamma$,
$\vec{k}_{L}\cdot \vec{r}_{12}= 0$ and different~$r_{12}$. It is
evident from Fig.~\ref{ftfig11} that a large squeezing can be obtained
for a finite $\Delta_{L}$ and its maximum shifts towards larger
$\Delta_{L}$ as the interatomic separation decreases. Similar as in
the case of photon anticorrelations, the maximum squeezing appears
at $\Delta_{L}=\Omega_{12}$, and can again be attributed to the
shift of the collective energy states due to the dipole-dipole
interaction.

The variance $F_{\theta =0}$, shown in Fig.~\ref{ftfig11}, exhibits not
only the large squeezing at finite detuning $\Delta_{L}$, but also a small
squeezing near $\Delta_{L}=0$. In contrast to the
squeezing at finite $\Delta_{L}$, which has a clear physical
interpretation, the source of squeezing at $\Delta_{L}$ is not easy
to understand. To find the source of squeezing at $\Delta_{L}=0$,
we simplify the calculations assuming that the angular frequency of the
quadrature components $\omega =\omega_{L}$ and the fluorescence field
is observed in the direction perpendicular to the interatomic axis,
$\bar{R}\perp \vec{r}_{12}$. In this case, the variance
$\left\langle :\Delta \vec{E}_{\alpha }^{2}:\right\rangle$, written
in terms of the density matrix elements of the collective system, is
given by~\cite{ft94}
\begin{eqnarray}
        F_{\alpha} \equiv
        \left\langle :\Delta \vec{E}_{\alpha}^{2}:\right\rangle /\left(
        2u(\vec{R})\right) &=& \frac{1}{4}\left\{2\rho_{ee} +2\rho_{ss}
        +\rho_{eg}e^{2i\alpha}
        +\rho_{ge}e^{-2i\alpha}\right. \nonumber \\
        &-&\left. \left[\left(\rho_{es}+\rho_{sg}\right)e^{i\alpha}
        +\left(\rho_{se}+\rho_{gs}\right)e^{-i\alpha}\right]^{2}\right\}
        \ .\label{t148}
\end{eqnarray}
This equation shows that the variance depends on phase $\alpha$
not only through the one-photon coherences $\rho_{es}$ and
$\rho_{sg}$, but also through the two-photon coherences $\rho_{eg}$
and $\rho_{ge}$. This dependence suggests that there are two
different processes that can lead to squeezing in the two-atom
system. The one-photon coherences cause squeezing near one-photon
resonances $\ket e \rightarrow \ket s$ and $\ket s \rightarrow \ket
g$, whereas the two-photon coherences cause squeezing near the
two-photon resonance $\ket g \rightarrow \ket e$.

To show this, we calculate the steady-state populations and coherences
from the master equation~(\ref{t42}). We use the set of the collective
states~(\ref{t63}) as an appropriate representation for the density
operator
\begin{equation}
\hat{\rho} = \sum_{ij} \rho_{ij}\ket i\bra j ,\quad i,j=g,s,a,e
\ ,\label{t149}
\end{equation}
where $\rho_{ij}$ are the density matrix elements in the basis of the
collective states.

After transforming to the collective state basis, the master
equation~(\ref{t42}) leads to a closed system of fifteen equations
of motion for the density matrix elements~\cite{ftk83}. However, for
a specifically chosen geometry for the driving field, namely that the field
is propagated perpendicularly to the atomic axis
$(\vec{k}_{L}\cdot \vec{r}_{12}=0)$, the system of equations decouples into
nine equations for symmetric and six equations for antisymmetric combinations
of the density matrix elements~\cite{ftk83,hsf82,fs,rfd,rich1,rich2}. In this
case, we can solve the system analytically, and find that the steady-state
values of the populations and coherences are~\cite{ftk83,rich1}
\begin{eqnarray}
\rho_{ee} &=& \rho_{aa} =\frac{\tilde{\Omega}^{4}}{Z} \ ,\qquad
\rho_{ss} = \frac{\tilde{\Omega}^{2}\left(\Gamma^{2}
+4\Delta^{2}_{L}\right) +\tilde{\Omega}^{4}}{Z} \ ,\nonumber \\
\rho_{es} &=& i\tilde{\Omega}^{3}\left(\Gamma+2i\Delta_{L}\right)/Z \
        ,\nonumber \\
        \rho_{sg} &=& -i\tilde{\Omega}\left\{\Gamma \tilde{\Omega}
        \left(\tilde{\Omega}
        +2i\Delta_{L}\right)\right. \nonumber \\
        &&\left. +\left(\Gamma^{2}+4\Delta_{L}^{2}\right)
        \left[\frac{1}{2}\left(\Gamma +\Gamma_{12}\right)
        +i\left(\Delta_{L}-\Omega_{12}\right)\right]\right\}/Z \
        ,\nonumber \\
        \rho_{eg} &=& \tilde{\Omega}^{2}\left(\Gamma +2i\Delta_{L}\right)
        \left[\frac{1}{2}\left(\Gamma +\Gamma_{12}\right)
        +i\left(\Delta_{L}-\Omega_{12}\right)\right]/Z \ ,\label{t150}
\end{eqnarray}
where
\begin{equation}
Z= 4\tilde{\Omega}^{4}+\left(\Gamma^{2}+4\Delta^{2}_{L}\right)
\left\{2\tilde{\Omega}^{2}
+\left[\frac{1}{4}\left(\Gamma +\Gamma_{12}\right)^{2}
+\left(\Delta_{L} -\Omega_{12}\right)^{2}\right]\right\} \ ,\label{t151}
\end{equation}
and $\tilde{\Omega}=\Omega/\sqrt{2}$.

Near the one-photon resonance $\ket s \rightarrow \ket g$ the
detuning $\Delta_{L}=\Omega_{12}$, and assuming that $\Omega_{12}\gg
\Omega, \Gamma$, the coherences reduce to
\begin{eqnarray}
        \rho_{es} &=& \frac{\tilde{\Omega}}{8\Omega_{12}} \ ,\qquad
        \rho_{sg} = \frac{-i\left(\Gamma
        +\Gamma_{12}\right)}{4\tilde{\Omega}} \ ,\qquad
        \rho_{eg} = \frac{i\left(\Gamma +\Gamma_{12}\right)}
        {8\Omega_{12}} \ .\label{t152}
\end{eqnarray}
It is clear from Eq.~(\ref{t152}) that near the $\ket g \rightarrow
\ket s$ resonance the coherence $\rho_{sg}$ is large, whereas the
two-photon coherence is of order of $\Omega^{-1}_{12}$ and thus is
negligible for large $\Omega_{12}$.

Near two-photon resonance, $\Delta_{L}\approx 0$, and it follows
from Eq.~(\ref{t150}) that in the limit of $\Omega_{12}\gg
\Omega, \Gamma$ the coherences reduce to
\begin{eqnarray}
        \rho_{es} = \frac{i\tilde{\Omega}^{3}}{\Gamma \Omega_{12}^{2}}
        \ ,\qquad
        \rho_{sg} = -\frac{\tilde{\Omega}}{\Omega_{12}} \ ,\qquad
        \rho_{eg} = -\frac{i\tilde{\Omega}^{2}}{\Gamma \Omega_{12}}
        \ .\label{t153}
\end{eqnarray}
In this regime, the coherences $\rho_{sg}$ and $\rho_{eg}$ are of order
of magnitude $\Omega^{-1}_{12}$, but $\rho_{eg}$ dominates over the
one-photon coherence $\rho_{sg}$ when the driving field is
strong~\cite{va92}.
\begin{figure}[t]
\begin{center}
\includegraphics[width=10cm]{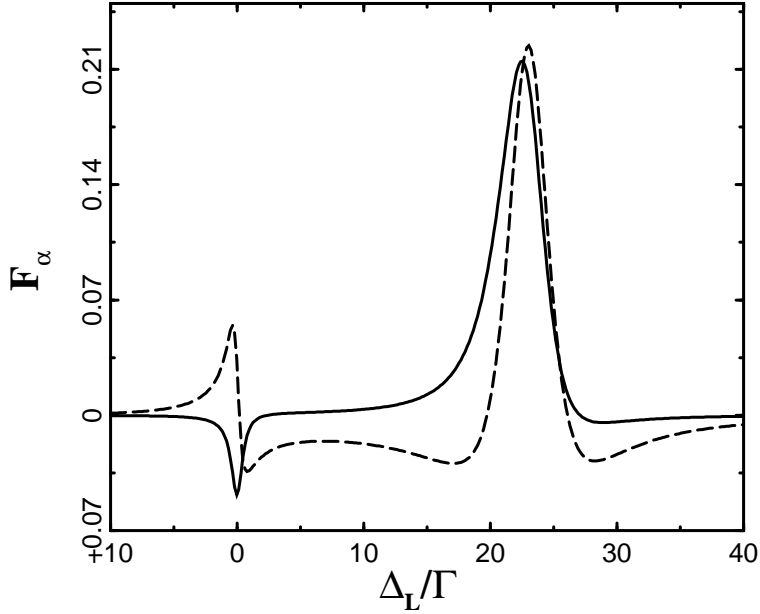}
\end{center}
\caption{The steady-state variance $F_{\alpha}$ as a
function of $\Delta_{L}$ for $r_{12}=0.05\lambda$, $\Omega
=3\Gamma$, $\bar{R}\perp \vec{r}_{12}$,
$\bar{\mu}\perp \bar{r}_{12}$ and different phases $\alpha$:
$\alpha =\pi/2$ (dashed line), $\alpha =3\pi/4$ (solid line).}
\label{ftfig12}
\end{figure}

The steady-state variance $F_{\alpha}$, calculated from Eqs.~(\ref{t148})
and~(\ref{t150}), is plotted in Fig.~\ref{ftfig12} as a function of
$\Delta_{L}$ for $r_{12}=0.05\lambda$, $\Omega =3\Gamma$,
$\bar{R}\perp \vec{r}_{12}$ and different phases $\alpha$.
The variance shows a strong dependence on~$\alpha$ near the one- and
two-photon resonances. Moreover, a large squeezing is found at these
resonances. It is also seen that near the two-photon resonance a
change by $\pi/4$ of the phase $\alpha$ changes a dispersion-like
structure of $F_{\alpha}$ into an absorption-like type. According to
Eqs.~(\ref{t148}) and (\ref{t150}), the variance $F_{\alpha}$ for
$\Omega_{12}\gg \Omega \gg \Gamma$, can be written as
\begin{eqnarray}
        F_{\alpha} = \frac{\Omega^{2}}{\Omega_{12}}\left[
        \frac{\Delta_{L}}{\left(\Gamma^{2} +4\Delta_{L}^{2}\right)} \cos
        2\alpha + \frac{\Gamma}{\left(\Gamma^{2} +4\Delta_{L}^{2}\right)}
        \sin 2\alpha \right] \ ,\label{t154}
\end{eqnarray}
where we retained only those terms which contribute near the
two-photon resonance. Equation~(\ref{t154}) predicts a dispersion-like
structure for $\alpha =0$ or $\pi/2$, and an absorption-like
structure for $\alpha =\pi/4$. Moreover, we see that the presence of
the dipole-dipole interaction is essential to obtain squeezing near
the two-photon resonance. The emergence of an additional dipole-dipole
interaction induced squeezing is a clear indication of a totally
different process which can appear in a two-atom system. The
dipole-dipole interaction shifts the collective states that induces
two-photon transitions responsible for the origin of the two-photon
coherence.

\section{Quantum interference of optical fields}\label{ftsec6}

In the classical theory of optical interference the EM field is
represented by complex vectorial amplitudes $\vec{E}(\vec{R},t)$
and $\vec{E}^{\ast}(\vec{R},t)$, and the first- and second-order
correlation functions are defined in a similar way as the correlation
functions~(\ref{t113}) and~(\ref{t114}) of the field operators
$\vec{E}^{(+)}(\vec{R},t)$ and $\vec{E}^{(-)}(\vec{R},t)$.
This could suggest that the only difference between the classical
and quantum correlation functions is the classical amplitudes
$\vec{E}^{\ast}(\vec{R},t)$ and $\vec{E}(\vec{R},t)$ are
replaced by the field operators $\vec{E}^{(-)}(\vec{R},t)$ and
$\vec{E}^{(+)}(\vec{R},t)$. This is true as long as the first-order
correlation functions (coherences) are considered, where the interference
effects do not distinguish between the quantum and classical theories of
the EM field~\cite{bsp}. However, there are significant differences
between the classical and quantum descriptions of the field in the
properties of the second-order correlation function~\cite{mw,rich4}.

\subsection{First-order interference}\label{ftsec61}

The simplest system in which the first-order interference can be
demonstrated is the Young's double slit experiment in which two
light beams of amplitudes $\vec{E}_{1}(\vec{r}_{1}, t_{1})$ and
$\vec{E}_{2}(\vec{r}_{2}, t_{2})$, produced at two slits located at
$\vec{r}_{1}$ and $\vec{r}_{2}$, respectively, incident on a detector
located at a point $\vec{R}$ far away from the slits. The resulting
average intensity of the two fields measured by the detector can be
written as~\cite{mw}
\begin{eqnarray}
\langle I(\vec{R},t)\rangle &=& \sigma \left\{\langle
I_{1}(\vec{R}_{1},t-t_{1})\rangle +\langle
I_{2}(\vec{R}_{2},t-t_{2})\rangle\right. \nonumber \\
&&\left. + 2{\rm Re}\left\langle \vec{E}_{1}^{\ast}(\vec{R}_{1},t-t_{1})
\vec{E}_{2}(\vec{R}_{2},t-t_{2})\right\rangle
\right\} \ ,\label{t155}
\end{eqnarray}
where $\sigma$ is a constant that depends on the geometry and the size
of the slits, and $I_{i}(\vec{R}_{i},t-t_{i})=
\left\langle \vec{E}_{i}^{\ast}(\vec{R}_{i},t-t_{i})
\vec{E}_{i}(\vec{R}_{i},t-t_{i})\right\rangle$.

If the observation point $\vec{R}$ lies in the far field zone of the
radiation emitted by the slits, the fields at the observation point can be
approximated by plane waves for which we can write
\begin{eqnarray}
\vec{E}_{i}(\vec{R}_{i}, t-t_{i}) &\approx&
\vec{E}_{i}(\vec{R},t)\exp \left[-i\left(\omega_{i}t -k_{i}\bar{R}\cdot
\vec{r}_{i} +\phi_{i}\right)\right] \ ,\label{t156}
\end{eqnarray}
where $k_{i}=\omega_{i}/c$, $\omega_{i}$ is the angular frequency of the
$i$th field and $\phi_{i}$ is its initial phase.

For perfectly correlated fields with equal amplitudes and frequencies,
and fixed the phase difference $\phi_{1}-\phi_{2}$, the average intensity
detected at the point~$\vec{R}$ is given by
\begin{eqnarray}
\langle I(\vec{R},t)\rangle =
2\sigma \langle I_{0}\rangle\left(1+\cos k\bar{R}\cdot \vec{r}_{12}\right)
\ ,\label{t157}
\end{eqnarray}
where $I_{0}=I_{1}=I_{2}$.

Equation~(\ref{t157}) shows that the average intensity depends on the
position $\bar{R}$ of the detector, and small changes in the position
$\vec{R}$ of the detector lead to minima and maxima in the detected
intensity. The usual measure of the minima and maxima of the intensity,
called the interference fringes or interference pattern, is a visibility
defined as
\begin{equation}
{\cal {V}} = \frac{I_{max} -I_{min}}{I_{max} +I_{min}} \ ,\label{t158}
\end{equation}
where $I_{max}$ corresponds to $\cos(k\bar{R}\cdot \vec{r}_{12}) =1$,
whereas $I_{min}$ corresponds to $\cos (k\bar{R}\cdot \vec{r}_{12})
=-1$ of the field intensity~(\ref{t157}). The visibility of the
interference fringes corresponds to the degree of coherence between
two fields. Hence, two classical fields
of equal amplitudes and frequencies, and fixed the phase difference
produce maximum possible interference pattern with the maximum
visibility of 100\%.

For a quantum field, the electric field components can be expressed
in terms of plane waves as
\begin{eqnarray}
\vec{E}^{(+)}\left(\vec{r},t\right) =
\left(\vec{E}^{(-)}\left(\vec{r},t\right)\right)^{\dagger} =
-i\sum_{\vec{k}s}\left(\frac{\hbar\omega_{k}}{2\epsilon_{0}V}\right)
^{\frac{1}{2}}\bar{e}_{\vec{k}s}\hat{a}_{\vec{k}s}e^{i\left(\vec{k}\cdot
\vec{r}-\omega_{k}t\right)} \ ,\label{t159}
\end{eqnarray}
where $V$ is the volume occupied by the field, $\hat{a}_{\vec{k}s}$ is
the annihilation operator for the $\vec{k}$th mode of the field of the
polarization $\bar{e}_{\vec{k}s}$ and $\omega_{k}$ is the angular
frequency of the mode.

It is easily to show, that in the case of interference of quantum
fields, the average intensity detected at the point $\vec{R}$ has the
same form as for the classical fields, Eq.~(\ref{t157}), with
$\langle I_{0}\rangle$ given by
\begin{eqnarray}
        \langle I_{0}\rangle &=& \sum_{\vec{k}s}\frac{\hbar\omega_{k}}
        {2\epsilon_{0}V}\langle\hat{n}_{\vec{k}s}\rangle \
        ,\label{t160}
\end{eqnarray}
where $\langle\hat{n}_{\vec{k}s}\rangle
=\langle\hat{a}_{\vec{k}s}^{\dagger}\hat{a}_{\vec{k}s}\rangle$ is
the average number of photons in the mode $\vec{k}$.

Thus, interference effects involving the first-order coherences cannot
distinguish between the quantum and classical theories of the EM field.

\subsection{Second-order interference}\label{ftsec62}

The second-order correlation function has completely different coherence
properties than the first-order correlation function. An interference
pattern can be observed in the second-order correlation function even if
the fields are produced by two independent sources for which the phase
difference $\phi_{1}-\phi_{2}$ is completely random~\cite{man65,man65a}.
In this case the second-order correlation function, observed at two
points $\vec{R}_{1}$ and $\vec{R}_{2}$, is given by~\cite{rich4}
\begin{eqnarray}
G^{(2)}(\vec{R}_{1},t_{1};\vec{R}_{2},t_{2}) &=&
\langle I_{1}^{2}\left(t_{1}\right)\rangle
+\langle I_{2}^{2}\left(t_{2}\right)\rangle
+2\langle I_{1}\left(t_{1}\right)I_{2}\left(t_{2}\right)\rangle \nonumber \\
&+& 2\langle I_{1}\left(t_{1}\right)I_{2}\left(t_{2}\right)\rangle
\cos\left[ k\vec{r}_{12}\cdot
\left(\bar{R}_{1}-\bar{R}_{2}\right)\right] \ .\label{t161}
\end{eqnarray}
Clearly, the second-order correlation function of two independent
fields exhibits a cosine modulation
with the separation $\vec{R}_{1}-\vec{R}_{2}$ of the two detectors. This is
an interference although it involves a correlation function that is of the
second order in the intensity. Similar to the first-order
correlation function, the sharpness of the fringes depends on the relative
intensities of the fields. For classical fields of equal intensities,
$I_{1}=I_{2}=I_{0}$, the correlation function~(\ref{t161}) reduces to
\begin{equation}
G^{(2)}(\vec{R}_{1},t;\vec{R}_{2},t) =
4\langle I_{0}^{2}\rangle
\left\{1+\frac{1}{2}\cos \left[k\vec{r}_{12}\cdot \left(\bar{R}_{1}
-\bar{R}_{2}\right)\right]\right\} \ .\label{t162}
\end{equation}

In analogy to the visibility in the first-order correlation function, we can
define the visibility of the interference pattern of the intensity
correlations as
\begin{equation}
{\cal{V}}^{(2)} =
\frac{G^{(2)}_{max}-G^{(2)}_{min}}{G^{(2)}_{max}+G^{(2)}_{min}} \ ,
\label{t163}
\end{equation}
and find from Eq.~(\ref{t162}) that in the case of classical fields an
interference pattern can be observed with the maximum possible visibility
of ${\cal{V}}^{(2)}=1/2$. Thus, two independent fields of random and
uncorrelated phases can exhibit an interference pattern in the intensity
correlation with a maximum visibility of $50\%$.

As an example of second-order interference with quantum fields, consider
the simple case of two single-mode fields of equal
frequencies and polarizations. Suppose that there are initially $n$ photons
in the field~$E_{1}$ and $m$ photons in the field~$E_{2}$, and the state
vectors of the fields are the Fock states $|\psi_{1}\rangle =|n\rangle$ and
$|\psi_{2}\rangle =|m\rangle$. The initial state of the two fields is the
direct product of the single-field states, $|\psi\rangle =|n\rangle|m\rangle$.
Inserting Eq.~(\ref{t159}) into Eq.~(\ref{t113}) and taking the expectation
value with respect to the initial state of the fields, we find
\begin{eqnarray}
G^{(2)}\left(\vec{R}_{1},t_{1};\vec{R}_{2},t_{2}\right) &=&
\left(\frac{\hbar\omega}{2\epsilon_{0}V}\right)^{2}\left\{ n\left(n-1\right)
+m\left(m-1\right)\right. \nonumber \\
&&\left. +2nm\left[1+\cos k\vec{r}_{12}\cdot \left(\bar{R}_{1}
-\bar{R}_{2}\right)\right]\right\} \ .\label{t164}
\end{eqnarray}
We note that the first two terms on the right-hand side of
Eq.~(\ref{t164}) vanish when the number of photons in each field is smaller
than 2, i.e.  $n<2$ and $m<2$. In this limit the correlation
function~(\ref{t164}) reduces to
\begin{equation}
G^{(2)}\left(\vec{R}_{1},t_{1};\vec{R}_{2},t_{2}\right) =
2\left(\frac{\hbar\omega}{2\epsilon_{0}V}\right)^{2}
\left[1+\cos k\vec{r}_{12}\cdot \left(\bar{R}_{1}-\bar{R}_{2}
\right)\right] \ .\label{t165}
\end{equation}
Thus, perfect interference pattern with the visibility ${\cal{V}}^{(2)}=1$
can be observed in the second-order correlation function of two quantum fields
each containing only one photon. According to Eq.~(\ref{t162}), the classical
theory predicts only a visibility of ${\cal{V}}^{(2)}=0.5$. For $n,m\gg 1$,
the first two terms on the right-hand side of Eq.~(\ref{t164}) are different
from zero $(m(m-1)\approx n(n-1)\approx n^{2})$, and then the quantum
correlation function (\ref{t164}) reduces to that of the classical field.

The visibility of the interference pattern of the intensity correlations
provides a means of testing for quantum correlations between two light fields.
Mandel~{\it et al.}~\cite{mand1,mand2,mand3} have measured the visibility in
the interference of signal and idler modes simultaneously generated in the
process of degenerate parametric down conversion, and observed a visibility of
about $75\%$, that is a clear violation of the upper bound of $50\%$ allowed by
classical correlations. Richter~\cite{rich90} have extended the analysis of the
visibility into the third-order correlation function, and have also found
significant differences in the visibility of the interference pattern of the
classical and quantum fields.

\subsection{Quantum interference in two-atom systems}\label{ftsec7}

In the Young's interference experiment the slits can be replaced by
two atoms and interference effects can be observed between
coherent or incoherent fields emitted from the atoms. The advantage
of using atoms instead of slits is that a given time each atom cannot
emit more than one photon. Therefore, the atoms can be regarded as
sources of single photon fields.

Using Eq.~(\ref{t116}), we can write the visibility as
\begin{eqnarray}
        V &=& \frac{\left\langle S_{1}^{+}S_{2}^{-}+ S_{2}^{+}S_{1}^{-}
        \right\rangle}{\left\langle S_{1}^{+}S_{1}^{-}+
        S_{2}^{+}S_{2}^{-}\right\rangle} \ ,\label{t166}
\end{eqnarray}
which shows that the interference effects can be studied in terms of
the atomic correlation functions.

There have been several theoretical studies of the fringe visibility in
the fluorescence field emitted by two coupled atoms~\cite{sckd}, and the
Young's interference-type pattern has recently been observed experimentally
in the resonance fluorescence of two trapped ions~\cite{eich}. The
experimental results have been explained theoretically by Wong {\it et al.}
\cite{wong}, and can be understood by treating the ions as independent
radiators which are synchronized by the constant phase of the driving field.
It has been shown that for a weak driving field, the fluorescence field is
predominantly composed of an elastic component and therefore the ions behave
as point sources of coherent light producing an interference pattern. Under
strong excitation the fluorescence field is mostly composed of the incoherent
part and consequently there is no interference pattern. To show this,
we consider a two-atom system driven by a coherent laser field
propagating in the direction perpendicular to the interatomic axis. In
this case, we can use the master equation~(\ref{t42}) and obtain the analytical
formula for the fringe visibility of the steady-state fluorescence field
as~\cite{fr}
\begin{eqnarray}
        V &=& \frac{\left(\Gamma^{2}+4\Delta_{L}^{2}\right)}
        {\left(\Gamma^{2}+4\Delta_{L}^{2}\right)
        +2\Omega^{2}} \ .\label{t167}
\end{eqnarray}
It is seen that in this specific case, the visibility is positive for
all parameter values and is independent of the interatomic
interactions. For a weak driving field, $\Omega \ll \Gamma,
\Delta_{L}$, the fringe visibility $|V|\approx 1$, whereas $|V|\approx 0$
for $\Omega \gg \Gamma, \Delta_{L}$, showing that and there is no
interference pattern when the atoms are driven by a strong field.
For moderate Rabi frequencies, $\Omega \approx \Gamma$, the
visibility may be improved by detuning the laser field from the atomic
resonance. Kochan {\it et al.} \cite{kochan} have shown that the
interference pattern of the strongly driven atoms can also be improved by
placing the atoms inside an optical cavity. The coupling of the atoms to
the cavity mode induces atomic correlations which improves the fringe
visibility.

Here, we derive general criteria for the first- and second-order
interference in the fluorescence field emitted from two two-level
atoms. Using these criteria, we can easily predict conditions for
quantum interference in the two atom system. In this approach, we
apply the collective states of a two-atom system, and write the atomic
correlation functions in terms of the density matrix elements of the
collective system as
\begin{eqnarray}
\left\langle S_{1}^{+}S_{1}^{-}\right\rangle +
\left\langle S_{2}^{+}S_{2}^{-}\right\rangle  &=&
\rho_{ss}+\rho_{aa}+2\rho_{ee} \ ,\nonumber \\
\left\langle S_{1}^{+}S_{2}^{-}\right\rangle &=&
\frac{1}{2}\left(\rho_{ss}-\rho_{aa}+\rho_{as}-\rho_{sa}\right)
\ ,\nonumber \\
\left\langle S_{1}^{+}S_{2}^{+}S_{1}^{-}S_{2}^{-}\right\rangle &=&
\rho_{ee} \ ,\label{t168}
\end{eqnarray}
where $\rho_{ii} (i=a,s,e)$ are the populations of the collective
states and $\rho_{sa}, \rho_{as}$ are coherences.

    From the relations~(\ref{t168}), we find that in terms of the density
matrix elements the first-order correlation function can be written as
\begin{eqnarray}
G^{\left( 1\right) }\left(\vec{R},t\right)
&=& \Gamma u(\vec{R})\left\{ 2\rho_{ee}\left(t\right)
+\rho_{ss}\left(t\right)\left(1
+\cos k\bar{R}\cdot \vec{r}_{12}\right)\right. \nonumber \\
&+&\left. \rho_{aa}\left(t\right)\left(1-\cos
k\bar{R}\cdot \vec{r}_{12}\right)\right. \nonumber \\
&+&\left. i\left(\rho_{sa}\left(t\right)-\rho_{as}\left(t\right)\right)
\sin k\bar{R}\cdot \vec{r}_{12}\right\} \ ,\label{t169}
\end{eqnarray}
and the second-order correlation function takes the form
\begin{eqnarray}
G^{\left( 2\right)}\left( \vec{R}_{1},t;\vec{R}_{2},t\right)
&=& 4\Gamma^{2}u(\vec{R}_{1}) u(\vec{R}_{2})
\nonumber \\
&\times& \rho_{ee}\left(t\right)\left[1+\cos k\left(\bar{R}_{1}
-\bar{R}_{2}\right)\cdot \vec{r}_{12}\right] \ .\label{t170}
\end{eqnarray}
It is evident from Eq.~(\ref{t169}) that first-order correlation
function can exhibit an interference pattern only if $\rho_{ss}\neq
\rho_{aa}$ and/or Im$(\rho_{sa})\neq 0$. This happens when $\langle
e_{1}|\langle g_{2}|\hat{\rho} |e_{2}\rangle |g_{1}\rangle$ and
$\langle g_{1}|\langle e_{2}|\hat{\rho} |g_{2}\rangle |e_{1}\rangle$
are different from zero, i.e. when there are nonzero coherences between
the atoms. Sch\"{o}n and Beige~\cite{sb} have arrived to the same
conclusion using the quantum jump method. On the other hand, the
second-order correlation function is independent of the populations of
the entangled states $\rho_{ss}, \rho_{aa}$ and the coherences, and
exhibit an interference pattern when $\rho_{ee}(t)\neq 0$.

We now examine some specific processes in which one can create unequal
populations of the $|s\rangle$ and $|a\rangle$ states. Dung and
Ujihara~\cite{du} have shown that spontaneous emission from two
identical atoms, with initially only one atom excited, can exhibit an
interference pattern. Their results can be easily interpreted in terms
of the populations $\rho_{ss}(t)$ and $\rho_{aa}(t)$. If initially
only one atom was excited; $\rho_{ee}(0)=0$ and
$\rho_{ss}(0)=\rho_{aa}(0)=\rho_{sa}(0)=\rho_{as}(0)=\frac{1}{2}$.
Using the master
equation~(\ref{t42}) with $\Omega_{1}=\Omega_{2}=0$, we find that the
time evolution of the populations $\rho_{ss}(t)$ and $\rho_{aa}(t)$ is
given by
\begin{eqnarray}
\rho_{ss}\left(t\right) &=& \rho_{ss}\left(0\right) \exp \left[-\left(\Gamma
+\Gamma_{12}\right)t\right] \ ,\nonumber \\
\rho_{aa}\left(t\right) &=& \rho_{aa}\left(0\right) \exp \left[-\left(\Gamma
-\Gamma_{12}\right)t\right] \ .\label{t171}
\end{eqnarray}
Since the populations decay with different rates, the symmetric state
decays with an enhanced rate $\Gamma +\Gamma_{12}$, while the
antisymmetric state decays with a reduced rate $\Gamma -\Gamma_{12}$,
the populations $\rho_{aa}(t)$ is larger than $\rho_{ss}(t)$ for all
$t>0$. Hence, an interference pattern can be observed for $t>0$.
This effect arises from the presence of the interatomic
interactions $(\Gamma_{12}\neq 0)$. Thus, for two independent atoms
the populations decay with the same rate resulting in the
disappearance of the interference pattern.

When the atoms are driven by a coherent laser field, an interference
pattern can be observed even in the absence of the interatomic
interactions. To show this, we consider the steady-state
solutions~(\ref{t150}) for the populations of
the collective atomic states.
It is evident from Eq.~(\ref{t150}) that $\rho_{ss}>\rho_{aa}$ even in
the absence of the interatomic interactions
$(\Gamma_{12}=\Omega_{12}=0)$. Hence, an interference pattern can be
observed even for two independent atoms. In this case the
interference pattern results from the coherent synchronization of the
oscillations of the atoms by the constant coherent phase of the
driving laser field.

We have shown that the first-order coherence is sensitive to the
interatomic interactions and the excitation field. In contrast, the
second-order correlation function can exhibit an interference pattern
independent of the interatomic interactions and the excitation
process~\cite{ftk88,sko,sko1}. According to Eq.(\ref{t170}), to observe
an interference pattern in the second-order correlation function, it is
enough to produce a non-zero population in the state $|e\rangle$. The
interference results from the detection process that a detector does not
distinguish between two simultaneously detected photons. As an
example, consider spontaneous emission from two identical and also
nonidentical atoms with initially both atoms excited.

For two identical atoms, we can easily find from Eqs.~(\ref{t42}) and
(\ref{t170}), and the quantum regression theorem~\cite{lax}, that the
two-time second-order correlation function is given by
\begin{eqnarray}
G^{\left( 2\right)}\left( \vec{R}_{1},t;\vec{R}_{2},t+\tau \right)
&=& \frac{1}{2}\Gamma^{2}u(\vec{R}_{1}) u(\vec{R}_{2})
\exp\left[-\Gamma \left(2t +\tau \right)\right] \nonumber \\
&\times& \left\{\left[1+\cos \left(k\bar{R}_{1}\cdot
\vec{r}_{12}\right) \cos \left(k\bar{R}_{2}\cdot
\vec{r}_{12}\right)\right]\cosh \left(\Gamma_{12}\tau \right)\right.
\nonumber \\
&-&\left. \left[\cos \left(k\bar{R}_{1}\cdot \vec{r}_{12}\right)
+\cos \left(k\bar{R}_{2}\cdot \vec{r}_{12}\right)\right] \sinh
\left(\Gamma_{12}\tau \right)\right. \nonumber \\
&+&\left. \sin \left(k\bar{R}_{1}\cdot
\vec{r}_{12}\right) \sin \left(k\bar{R}_{2}\cdot
\vec{r}_{12}\right) \cos \left(2\Omega_{12}\tau \right)\right\}
\ .\label{t172}
\end{eqnarray}
The above equation shows that the two-time second-order correlation
function exhibits a sinusoidal modulation in space and time.
This modulation can be interpreted both in terms of interference
fringes and quantum beats. The frequency of quantum beats is
$2\Omega_{12}$ and the amplitude of these beats depends on the
direction of observation in respect to the interatomic axis.
The quantum beats vanish for directions $\theta_{1}=90^{o}$ or
$\theta_{2}=90^{o}$, where $\theta_{1}(\theta_{2})$ is the angle
between $\vec{r_{12}}$ and $\bar{R}_{1}(\bar{R}_{2})$, and the
amplitude of the beats has its maximum for two photons detected in
the direction $\theta_{1}=\theta_{2}=0^{o}$. This directional effect
is connected with the fact that the antisymmetric state $\ket a$ does
not radiate in the direction perpendicular to the interatomic axis. We
will discuss this directional effect in more details in
Sec.~\ref{ftsec91}. For independent atoms, $\Gamma_{12}=0,
\Omega_{12}=0$, and then the correlation function~(\ref{t172})
reduces to
\begin{eqnarray}
G^{\left( 2\right)}\left( \vec{R}_{1},t;\vec{R}_{2},t+\tau \right)
&=& \frac{1}{2}\Gamma^{2}u(\vec{R}_{1}) u(\vec{R}_{2})
\exp\left[-\Gamma \left(2t +\tau \right)\right] \nonumber \\
&\times& \left[1+\cos k\left(\bar{R}_{1}
-\bar{R}_{2}\right)\cdot \vec{r}_{12}\right] \ ,\label{t173}
\end{eqnarray}
which shows that the time modulation vanishes. This implies that quantum
beats are absent in spontaneous emission from two independent atoms, but
the spatial modulation is still present.

The situation is different for two nonidentical atoms. In this case,
the two-time second-order correlation function exhibits quantum beats
even if the atoms are independent. For $\Gamma_{12}=0$ and
$\Omega_{12}=0$, the master equation~(\ref{t42}) leads to the
following correlation function
\begin{eqnarray}
G^{\left( 2\right)}\left( \vec{R}_{1},t;\vec{R}_{2},t+\tau \right)
&=& \frac{1}{2}\Gamma^{2}u(\vec{R}_{1}) u(\vec{R}_{2})
\exp\left[-\Gamma \left(2t +\tau \right)\right] \nonumber \\
&\times& \left\{\cosh
\frac{1}{2}\left(\Gamma_{2}-\Gamma_{1}\right)\tau \right. \nonumber \\
&+&\left. \cos \left[ k\left(\bar{R}_{1}
-\bar{R}_{2}\right)\cdot \vec{r}_{12} -2\Delta \tau \right]\right\}
\ .\label{t174}
\end{eqnarray}
Thus, for independent nonidentical atoms, the correlation function
shows a sinusoidal modulation both in space and time. We note that
the modulation term in Eq.~(\ref{t174}) is the same as that obtained
by Mandel~\cite{man64}, who considered the second-order
correlation function for two beams emitted by independent lasers.

\section{Selective excitation of the collective atomic
states}\label{ftsec8}

In the previous section, we have shown that nonclassical effects in
coherently driven two-atom systems reflect the preparation of the system
in a superposition of two collective states. In particular, for the
total photon anticorrelation and maximum squeezing, the two-atom
system is in a superposition of the ground and the entangled
symmetric states. The other states are not populated. We now consider
excitation processes which can lead to a preparation of the two-atom
system in only one of the collective states. In particular, we will focus on
processes which can prepare the two-atom system in the entangled symmetric
state $\ket s$. Our main interest, however, is in the preparation of the
system in the maximally entangled antisymmetric state $\ket a$ which, under
the condition $\Gamma_{12}=\sqrt{\Gamma_{1}\Gamma_{2}}$, is a
decoherence-free state. The central idea is to choose the distance
between the atoms such that the resulting level shift is large enough
to consider the possible transitions between the collective states
separately. This will allow to make a selective excitation of the
symmetric and antisymmetric states and therefore to create controlled
entanglement between the atoms.

\subsection{Preparation of the symmetric state by a pulse
laser}\label{ftsec81}

Beige {\it et al.}~\cite{be} have shown that a system of two identical
two-level atoms may be prepared in the symmetric state $\ket s$ by a short
laser pulse. The conditions for a selective excitation of
the collective atomic states can be analyzed from the interaction
Hamiltonian of the laser field with the two-atom system. We make the
unitary transformation
\begin{equation}
\tilde{H}_{L} = e^{i\hat{H}_{a}t/\hbar}\hat{H}_{L}e^{-i\hat{H}_{a}t/\hbar}
\ ,\label{t175}
\end{equation}
where
\begin{eqnarray}
\hat{H}_{a} &=& \hbar \left\{ \Delta_{L}\left( \ket e\bra e -\ket g
\bra g \right)
+\left(\Delta_{L}+\Omega_{12}\right)\ket s\bra s\right. \nonumber \\
&&+\left.\left(\Delta_{L}-\Omega_{12}\right)\ket a\bra a\right\}
\ ,\label{t176}
\end{eqnarray}
and find that in the case of identical atoms, $\Gamma_{1}=\Gamma_{2}$ and
$\Delta =0$, the transformed interaction Hamiltonian $\tilde{H}_{L}$
is given by
\begin{eqnarray}
\tilde{H}_{L} &=& -\frac{\hbar}{2\sqrt{2}}\left\{
\left(\Omega_{1}+\Omega_{2}\right)\left(S^{+}_{es}
e^{i\left(\Delta_{L}+\Omega_{12}\right)t}
+S^{+}_{sg}e^{i\left(\Delta_{L}-\Omega_{12}\right)t}\right)\right. \nonumber \\
&+&\left. \left(\Omega_{2}-\Omega_{1}\right)\left(S^{+}_{ag}
e^{i\left(\Delta_{L}+\Omega_{12}\right)t}
+S^{+}_{ea}e^{i\left(\Delta_{L}-\Omega_{12}\right)t}\right)
+{\rm H.c.}\right\} \ .\label{t177}
\end{eqnarray}
The Hamiltonian~(\ref{t177}) represents the interaction of the laser field with
the collective two-atom system, and in the transformed form contains terms
oscillating at frequencies $(\Delta_{L}\pm \Omega_{12})$, which correspond to
the two separate groups of transitions between the collective atomic
states at frequencies $\omega_{L}=\omega_{0}+\Omega_{12}$ and
$\omega_{L}=\omega_{0}-\Omega_{12}$.
The $\Delta_{L}+\Omega_{12}$ frequencies are separated from
$\Delta_{L}-\Omega_{12}$ frequencies by $2\Omega_{12}$, and hence the
two groups of the transitions evolve separately when $\Omega_{12}\gg \Gamma$.
Depending on the frequency, the laser can be selectively tuned to one
of the two groups of the transitions.
When $\omega_{L}=\omega_{0}+\Omega_{12}$ $(\Delta_{L}-\Omega_{12}=0)$ the laser
is tuned to exact resonance with the $\ket e -\ket a$ and $\ket g -\ket s$
transitions, and then the terms, appearing in the
Hamiltonian~(\ref{t177}), and corresponding to these transitions have no
explicit time dependence. In contrast,
the $\ket g -\ket a$ and $\ket e -\ket s$ transitions are
off-resonant and the terms corresponding to these transitions have an
explicit time dependence exp$\left(\pm 2i\Omega_{12}t\right)$.
If $\Omega_{12}\gg \Gamma$, the off-resonant terms rapidly oscillate with
the frequency $2\Omega_{12}$, and then we can make a secular
approximation in which we neglect all those rapidly
oscillating terms. The interaction Hamiltonian can then be written in
the simplified form
\begin{equation}
\tilde{H}_{L} = -\frac{\hbar}{2\sqrt{2}}\left[
\left(\Omega_{1}+\Omega_{2}\right)S^{+}_{sg}
+ \left(\Omega_{2}-\Omega_{1}\right)S^{+}_{ea}  + {\rm H.c.}\right]
\ .\label{t178}
\end{equation}
It is seen that the laser field couples to the transitions with significantly
different Rabi frequencies. The coupling strength of the laser to the
$\ket g -\ket s$ transition is proportional to the sum of the Rabi frequencies
$\Omega_{1}+\Omega_{2}$, whereas the coupling strength of the laser to the
$\ket a -\ket e$ transition is proportional to the difference of the Rabi
frequencies $\Omega_{1}-\Omega_{2}$. According to Eq.~(\ref{t46}) the Rabi
frequencies $\Omega_{1}$ and $\Omega_{2}$ of two identical atoms differ only
by the phase factor exp$(i\vec{k}_{L}\cdot \vec{r}_{12})$. Thus, in order to
selectively excite the $\ket g -\ket s$ transition, the driving
laser field should be in phase with both atoms, i.e. $\Omega_{1}=\Omega_{2}$.
This can be achieved by choosing the propagation
vector $\vec{k}_{L}$ of the laser orthogonal to the line joining the atoms.
Under this condition we can make a further simplification and truncate the
state vector of the system into two states $\ket g$ and $\ket s$.
In this two-state approximation we find from the Schr\"{o}dinger
equation the time evolution of the population $P_{s}(t)$ of the state
$\ket s$ as
\begin{equation}
P_{s}\left(t\right) = \sin^{2}\left(\frac{1}{\sqrt{2}}\Omega t\right)
\ ,\label{t179}
\end{equation}
where $\Omega =\Omega_{1}=\Omega_{2}$.

The population oscillates with the Rabi frequency of the $\ket s
-\ket g$ transition and at certain times
$P_{s}(t)=1$ indicating that all the population is in the symmetric state.
This happens at times
\begin{equation}
T_{n} = \left(2n+1\right)\frac{\pi}{\sqrt{2}\Omega} ,\quad n=0,1,\ldots .
\label{t180}
\end{equation}
Hence, the system can be prepared in the state $\ket s$ by simply applying
a laser pulse, for example,  with the duration $T_{0}$, that is a standard
$\pi$ pulse.

The two-state approximation is of course an idealization, and a
possibility that all the transitions can be driven by the laser
imposes significant limits on the Rabi frequency and the duration
of the pulse. Namely, the Rabi frequency cannot be too strong in order to
avoid the
coupling of the laser to the $\ket s -\ket e$ transition, which could lead to
a slight pumping of the population to the state $\ket e$. On the other hand,
the Rabi frequency cannot be too small as for a small $\Omega$ the duration
of the pulse, required for the complete
transfer of the population into the state $\ket s$, becomes longer and
then spontaneous emission can occur during the excitation process. Therefore,
the transfer of the population to the state $\ket s$ cannot be made
arbitrarily fast and, in addition, requires a careful estimation
of the optimal Rabi frequency, which could be difficult to achieve in a real
experimental situation.

\subsection{Preparation of the antisymmetric state}\label{ftsec82}

\subsubsection{Pulse laser}\label{ftsec821}

If we choose the laser frequency such that $\Delta_{L}+\Omega_{12}=0$, the
laser field is then resonant to the $\ket a -\ket g$ and $\ket e -\ket s$
transitions and, after the secular approximation, the
Hamiltonian~(\ref{t177}) reduces to
\begin{equation}
\tilde{H}_{L} = -\frac{\hbar}{2\sqrt{2}}\left[
\left(\Omega_{2}-\Omega_{1}\right)S^{+}_{ag}
+ \left(\Omega_{1}+\Omega_{2}\right)S^{+}_{es}  + {\rm H.c.}\right]
\ .\label{t181}
\end{equation}
Clearly, for $\Omega_{1}=-\Omega_{2}$ the laser couples only to the
$\ket a -\ket g$ transition. Thus, in order to selectively excite the
$\ket g -\ket a$ transition, the atoms should experience opposite phases
of the laser field. This can be achieved by choosing the propagation
vector $\vec{k}_{L}$ of the laser along the interatomic axis, and the
atomic separations such that
\begin{equation}
\vec{k}_{L}\cdot \vec{r}_{12} = \left(2n+1\right)\pi ,\quad n=0,1,2,\ldots
\ ,\label{t182}
\end{equation}
which corresponds to a situation that the atoms are separated by a distance
$r_{12}=(2n+1)\lambda/2$.

The smallest distance at which the atoms could experience opposite phases
corresponds to $r_{12}=\lambda/2$. However, at this particular separation
the dipole-dipole interaction parameter $\Omega_{12}$ is small, see
Fig.~\ref{ftfig1}, and then all of the transitions between the collective
states occur at approximately the same frequency. In this case the secular
approximation is not valid and we cannot separate the transitions at
$\Delta_{L}+\Omega_{12}$ from the transitions at $\Delta_{L}-\Omega_{12}$.

One possible solution to the problem of the selective excitation
with opposite phases is to use a standing laser field instead of the
running-wave field.
If the laser amplitudes differ by the sign, i.e. $\vec{E}_{L_{1}}=
-\vec{E}_{L_{2}}=\vec{E}_{0}$, and $\vec{k}_{L_{1}}\cdot \vec{r}_{1} =
-\vec{k}_{L_{2}}\cdot \vec{r}_{2}$, the Rabi frequencies experienced by the
atoms are
\begin{eqnarray}
\Omega_{1} &=& \frac{2i}{\hbar} \vec{\mu}_{1} \cdot \vec{E}_{0}
\sin \left(\frac{1}{2}\vec{k}_{L}\cdot \vec{r}_{12}\right) , \nonumber \\
\Omega_{2} &=& -\frac{2i}{\hbar} \vec{\mu}_{2}\cdot \vec{E}_{0}
\sin \left(\frac{1}{2}\vec{k}_{L}\cdot \vec{r}_{12}\right) \
,\label{t183}
\end{eqnarray}
where $\vec{k}_{L}=\vec{k}_{L_{1}}=\vec{k}_{L_{2}}$ and
we have chosen the reference frame such that $\vec{r}_{1}=\frac{1}{2}
\vec{r}_{12}$ and $\vec{r}_{2}=-\frac{1}{2}\vec{r}_{12}$. It follows
from Eq.~(\ref{t183}) that the Rabi frequencies oscillate with opposite
phases independent of the separation between the atoms. However, the
magnitude of the Rabi frequencies decreases with decreasing $r_{12}$.

\subsubsection{Indirect driving through the symmetric
state}\label{ftsec822}

We now turn to the situation of non-identical atoms and consider
different possible processes of the population transfer to the
antisymmetric state which could be present even if the antisymmetric
state does not decay to the ground level. This can happen when
$\Gamma_{12}=\sqrt{\Gamma_{1}\Gamma_{2}}$, i.e. when the separation
between the atoms is negligible small.
Under this condition the antisymmetric state is also decoupled from
the driving field. According to Eq.~(\ref{t95}), the antisymmetric state can
still be coupled, through the coherent interaction $\Delta_{c}$, to the
symmetric state $\ket s$. However, this coupling appears only for nonidentical
atoms.

    From the master equation~(\ref{t42}), we find that under the condition
$\Gamma_{12}=\sqrt{\Gamma_{1}\Gamma_{2}}$ the equation of motion for the
population of the antisymmetric state $\ket a$ is given by~\cite{afs}
\begin{eqnarray}
\dot{\rho}_{aa} &=& \frac{\left(\Gamma_{1}-\Gamma_{2}\right)^{2}}
{\Gamma_{1}+\Gamma_{2}}\rho_{ee}
+i\Delta _{c}\left( \rho _{sa}-\rho _{as}\right)
\nonumber \\
&&-\frac{1}{2}i\Omega \frac{\left(\Gamma_{1}-\Gamma_{2}\right)}
{\sqrt{\Gamma_{1}^{2}+\Gamma_{2}^{2}}}\left(
\rho _{ea}-\rho_{ae}\right) \ .  \label{t184}
\end{eqnarray}
The equation (\ref{t184}) shows that the non-decaying antisymmetric state
$\ket a$ can be populated by spontaneous emission from the
upper state $\ket e$ and also by the coherent interaction
with the state $\ket s$. The first condition is satisfied
only when $\Gamma_{1}\neq \Gamma_{2}$, while the other condition is satisfied
only when $\Delta _{c}\neq 0$. Thus, the transfer of population to the state
$\ket a$ from the upper state $\ket e$ and the symmetric
state $\ket s$ does not appear when the atoms are identical,
but is possible for nonidentical atoms.
\begin{figure}[t]
\begin{center}
\includegraphics[width=10cm]{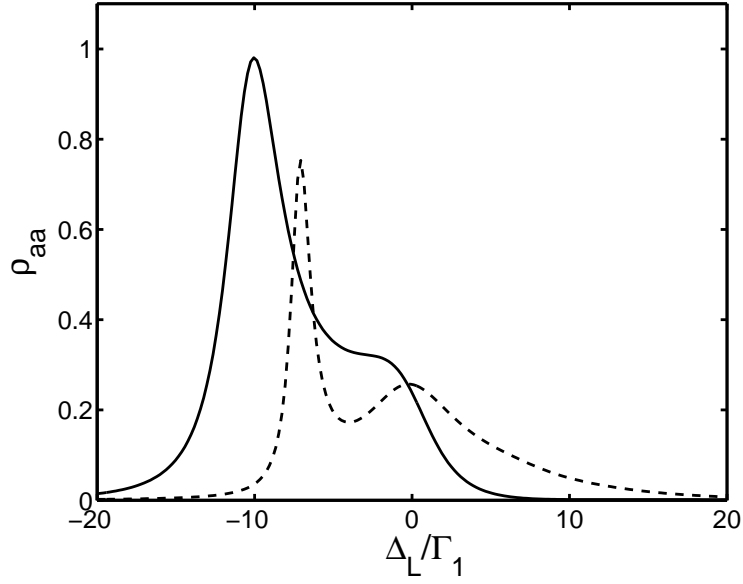}
\end{center}
\caption{The steady-state population of the maximally entangled
antisymmetric state
$\ket a$ for $\Omega =10\Gamma_{1}, \Omega_{12}=10\Gamma_{1}$ and
$\Gamma_{2}=\Gamma_{1}, \Delta =\Gamma_{1}$ (solid line),
$\Gamma_{2}= 2\Gamma_{1}, \Delta =0$ (dashed line).}
\label{ftfig13}
\end{figure}

\begin{figure}[ht]
\begin{center}
\includegraphics[width=10cm]{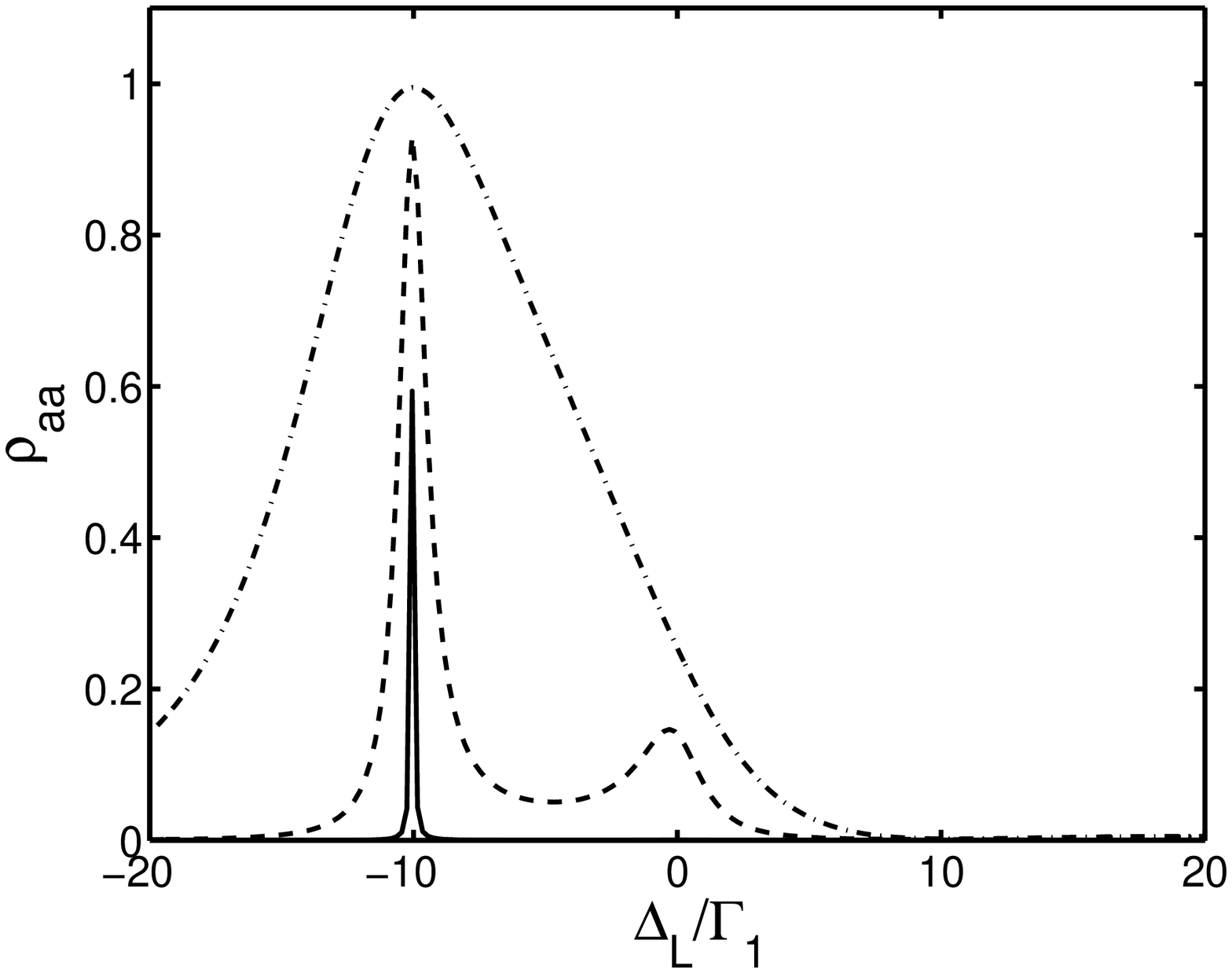}
\end{center}
\caption{The steady-state population of the antisymmetric state $\ket a$
for $\Gamma_{2}=\Gamma_{1}, \Omega_{12}=10\Gamma_{1}, \Delta
=\Gamma_{1}$ and different $\Omega$: $\Omega =\Gamma_{1}$ (solid
line), $\Omega =5\Gamma_{1}$ (dashed line), $\Omega =20\Gamma_{1}$
(dashed-dotted line). }
\label{ftfig15}
\end{figure}

We illustrate this effect in Fig.~\ref{ftfig13}, where we plot the
steady-state population of the maximally entangled state
$\ket a$ as a function of $\Delta _{L}$ for two different types
of nonidentical atoms. In the first case the atoms have the same damping
rates $(\Gamma_{1}=\Gamma_{2})$ but different transition frequencies
$(\Delta \neq 0)$, while in the
second case the atoms have the same frequencies $(\Delta =0)$ but different
damping rates $(\Gamma_{1}\neq \Gamma_{2})$. It is seen from
Fig.~\ref{ftfig13} that in both
cases the antisymmetric state can be populated even if is not directly driven
from the ground state. The population is transferred to $\ket a$
through the coherent interaction $\Delta_{c}$ which leaves the other excited
states completely unpopulated.
\begin{figure}[t]
\begin{center}
\includegraphics[width=10cm]{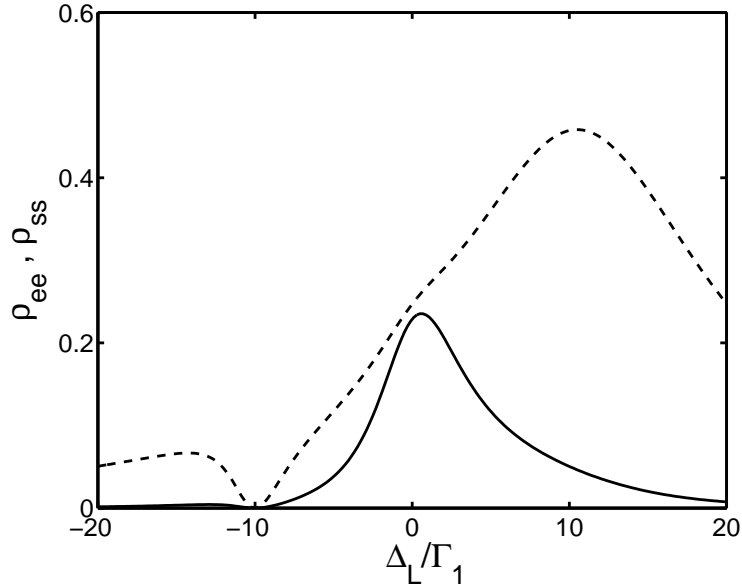}
\end{center}
\caption{The steady-state populations of the upper state $\ket e$ (solid line)
and the symmetric state $\ket s$ (dashed line) for
$\Gamma_{2}=\Gamma_{1}, \Omega =10\Gamma_{1}, \Omega_{12}=10\Gamma_{1}$
and $\Delta =\Gamma_{1}$.}
\label{ftfig14}
\end{figure}
This is shown in Fig.~\ref{ftfig14}, where we plot the steady-state populations
$\rho _{ss}$ and $\rho_{ee}$ of the states $\ket s$ and $\ket e$. It is
apparent from Fig.~\ref{ftfig14} that at $\Delta_{L}= -\Omega_{12}$ the states
$\ket s$ and $\ket e$ are not populated. However, the population is not
entirely trapped in the antisymmetric state $\ket a$, but rather in a linear
superposition of the antisymmetric and ground states. This is
illustrated in Fig.~\ref{ftfig15}, where we plot the steady-state population
$\rho_{aa}$ for the same parameters as in Fig.~\ref{ftfig14}, but different
$\Omega $. Clearly, for a small $\Omega$ the steady-state population
$\rho_{aa}\approx \frac{1}{2}$, and the amount of the population increases with
increasing $\Omega$. The population $\rho_{aa}$ attains the maximum value
$\rho_{aa}\approx 1$ for a very strong driving field.

This result shows that we can relatively easily prepare two nonidentical atoms
in the maximally entangled antisymmetric state. The
closeness of the prepared state to the ideal one is measured by the fidelity
$F$. Here $F$ is equal to the obtained maximum population in the state $%
\left| a\right\rangle $. For $\Omega \gg \Gamma$ the fidelity of the prepared
state is maximal, equal to~1. As we have already mentioned, the system has the
advantage that the maximally entangled state $\left| a\right\rangle $ does not
decay, i.e. is a decoherence-free state.

\subsubsection{Atom-cavity-field interaction}\label{ftsec823}

There have been several proposals to generate the antisymmetric
state $\ket a$ in a system of two identical atoms interacting with a
single-mode cavity field. In this case, collective effects arise
from the interaction between the atoms induced by a strong coupling
of the atoms to the cavity mode. An excited atom emits a photon
into the cavity mode that is almost immediately absorbed by the
second atom.
Plenio {\it et al.}~\cite{phbk}
have considered a system of two atoms trapped inside an optical
cavity and separated by a distance much larger than the optical
wavelength. This allows for the selective excitation of only one of
the atoms. In this scheme the generation of the antisymmetric state
relies on the concept of conditional dynamics due to continuous
observation of the cavity field. If only one atom is excited and no
photon is detected outside the cavity, the atoms are prepared in a dark
state~\cite{yzm}, which is equivalent to the antisymmetric state
$\ket a$.

In earlier studies, Phoenix and Barnett~\cite{pb}, Kudrayvtsev and
Knight~\cite{kk} and
Cirac and Zoller~\cite{cz} have analyzed two-atom Jaynes-Cummings
models for a violation of Bell's inequality, and have shown that the
atoms moving across a single-mode cavity can be prepared in the
antisymmetric state via the interaction with the cavity field. In this
scheme, the preparation of the antisymmetric state takes place in two
steps. In the first step, one atom initially prepared in its excited
state $\ket {e_{1}}$ is sent through a single-mode cavity being in the
vacuum state $\ket 0_{c}$. During the interaction with the cavity mode,
the atomic population undergoes the vacuum Rabi oscillations, and the
interaction time was varied by selecting different atomic velocities.
If the velocity of the atom is such that the interaction time of the
atom with the cavity mode is equal to quarter of the vacuum Rabi
oscillations, the state of the combined system, the atom plus the
cavity mode, is a superposition state
\begin{equation}
\ket {a_{1}c} =\frac{1}{\sqrt{2}}\left(\ket {e_{1}}\ket 0_{c} -\ket
{g_{1}}\ket 1_{c}\right) \ .\label{t185}
\end{equation}
Hence, the state of the total system, two atoms plus the cavity mode,
after the first atom has crossed the cavity is
\begin{equation}
\ket {\Psi_{1}} =\frac{1}{\sqrt{2}}\left(\ket {e_{1}}\ket 0_{c}
-\ket {g_{1}}\ket 1_{c}\right)\ket {g_{2}} \ .\label{t186}
\end{equation}
If we now send the second atom, being in its ground state, with the
selected velocity such that during the interaction with the cavity
mode the atom undergoes half of the vacuum Rabi oscillation, the final
state of the system becomes
\begin{eqnarray}
\ket {\Psi_{12}c} &=& \frac{1}{\sqrt{2}}\left(\ket {e_{1}}\ket
0_{c}\ket {g_{2}} -\ket {g_{1}}\ket 0_{c}\ket {e_{2}}\right)
\nonumber \\
&=& \frac{1}{\sqrt{2}}\left(\ket {e_{1}}\ket {g_{2}}
-\ket {g_{1}}\ket {e_{2}}\right)\ket 0_{c} = \ket a \ket 0_{c}
\ .\label{t187}
\end{eqnarray}
Thus, the final state of the system is a product state of the atomic
antisymmetric state $\ket a$ and the vacuum state of the cavity mode.
In this scheme the cavity mode is left in the vacuum state which
prevents the antisymmetric state from any noise of the cavity. The
scheme to entangle two atoms in a cavity, proposed by Cirac and
Zoller~\cite{cz}, has recently been realized experimentally by Hagly
{\it et al.}~\cite{hag}.

Gerry~\cite{ge} has proposed a similar method based on a dispersive
interaction of the atoms with a cavity mode prepared in a coherent
state $\ket \alpha$. The atoms enter the cavity in superposition states
\begin{eqnarray}
\ket {a_{1}} &=& \frac{1}{\sqrt{2}}\left(\ket {e_{1}} +i\ket
{g_{1}}\right) \ , \nonumber \\
\ket {a_{2}} &=& \frac{1}{\sqrt{2}}\left(\ket {e_{2}} -i\ket
{g_{2}}\right) \ .\label{t188}
\end{eqnarray}
After passage of the second atom, the final state of the system is
\begin{eqnarray}
\ket {\Psi_{12}c} &=& \frac{1}{2}\left\{ \left(\ket {g_{1}}\ket {g_{2}}
+\ket {e_{1}}\ket {e_{2}}\right)\ket {-\alpha}\right. \nonumber \\
&&+\left. i\left(\ket {e_{1}}\ket {g_{2}} -\ket {g_{1}}\ket
{e_{2}}\right)\ket {\alpha}\right\} \ .\label{t189}
\end{eqnarray}
Thus, if the cavity field is measured and found in the state
$\ket {\alpha}$,
the atoms are in the antisymmetric state. If the cavity field is
found in the state $\ket {-\alpha}$, the atoms are in the entangled
state
\begin{equation}
\ket {\Psi_{12}(-\alpha)} = \frac{1}{2}\left(\ket {g_{1}}\ket {g_{2}}
+\ket {e_{1}}\ket {e_{2}}\right) \ .\label{t190}
\end{equation}
The state~(\ref{t190}) is called as a two photon entangled state. In
section~\ref{ftsec10}, we will discuss another method of preparing
the system in the two-photon entangled state based on the interaction
of two atoms with a squeezed vacuum field.

\subsection{Entanglement of two distant atoms}\label{ftsec83}

In the previous subsection, we have discussed different excitation
processes which can prepare two atoms in the antisymmetric state.
The analysis involved single mode cavities, but ignored spontaneous
emission from the atoms and the cavity damping. Here, we will extend
this analysis to include spontaneous emission from the atoms and the
cavity damping~\cite{wan01}. We will show that two atoms separated by
an arbitrary distance $r_{12}$ and interacting with a strongly damped
cavity mode can behave as the Dicke model even if there is no assumed
interaction between the atoms.

Consider two identical atoms separated by a large distance such that
$\Gamma_{12}\approx 0$ and $\Omega_{12}\approx 0$. The interatomic
axis is oriented perpendicular to the direction of the cavity mode
(cavity axis) which is driven by an external coherent laser field of the
Rabi frequency $\Omega$. The atoms are coupled to the cavity mode with
coupling constant $g$, and damped at the rate $\Gamma$ by spontaneous
emission to modes other than the privileged cavity mode. For
simplicity, we assume that the frequencies of the laser field
$\omega_{L}$ and the cavity mode $\omega_{c}$ are both equal to the
atomic transition frequency $\omega_{0}$. The master equation for the
density operator $\hat{\rho}_{ac}$ of the system of two atoms plus
cavity field has the form
\begin{eqnarray}
        \frac{\partial \hat{\rho}_{ac} }{\partial t} &=&
        -\frac{1}{2}i\Omega \left[\hat{a}+\hat{a}^{\dagger},
        \hat{\rho}_{ac}\right]
        -\frac{1}{2}ig\left[S^{-}\hat{a}^{\dagger}+\hat{a}S^{+},
        \hat{\rho}_{ac}\right] \nonumber \\
        && -\frac{1}{2}\Gamma
        \hat{L}_{a}\hat{\rho}_{ac}
        -\frac{1}{2}\Gamma_{c}\hat{L}_{c}\hat{\rho}_{ac} \ ,\label{t191}
\end{eqnarray}
where
\begin{eqnarray}
        \hat{L}_{a}\hat{\rho}_{ac} &=& \sum_{i=1}^{2}\left( \hat{\rho}_{ac}
S_{i}^{+}S_{i}^{-}+S_{i}^{+}S_{i}^{-}\hat{\rho}_{ac}
-2S_{i}^{-}\hat{\rho}_{ac} S_{i}^{+}\right) \ ,\nonumber \\
        \hat{L}_{c}\hat{\rho}_{ac} &=&
        \hat{a}^{\dagger}\hat{a}\hat{\rho}_{ac}
        +\hat{\rho}_{ac}\hat{a}^{\dagger}\hat{a}
        -2\hat{a}\hat{\rho}_{ac}\hat{a}^{\dagger} \ ,\label{t192}
\end{eqnarray}
are operators representing the damping of the atoms by spontaneous
emission and of the field by cavity decay, respectively; $\hat{a}$
and $\hat{a}^{\dagger}$ are the cavity-mode annihilation and creation
operators, $S^{\pm}= S^{\pm}_{1}+S^{\pm}_{2}$ are collective atomic
operators, and $\Gamma_{c}$ is the cavity damping rate.

To explore the dynamics of the atoms, we assume the "bad-cavity''
limit of $\Gamma_{c}\gg g \gg \Gamma$. This enables us to
adiabatically eliminate the cavity mode and obtain a master equation
for the reduced density operator of the atoms. We make the unitary
transformation
\begin{eqnarray}
        \hat{\rho}_{T} &=& \hat{D}(-\eta )\hat{\rho}_{ac}\hat{D}(\eta )
        \ ,\label{t193}
\end{eqnarray}
where
\begin{eqnarray}
        \hat{D}(\eta ) = e^{\eta \left(\hat{a}+\hat{a}^{\dagger}\right)}
        \label{t194}
\end{eqnarray}
is the displacement operator, and $\eta =\Omega/\Gamma_{c}$.

The master equation for the transformed operator reduces to
\begin{eqnarray}
        \frac{\partial \hat{\rho}_{T} }{\partial t} &=&
        \frac{1}{2}ig\eta \left[S^{+}+S^{-},
        \hat{\rho}_{T}\right]
        -\frac{1}{2}ig\left[S^{-}\hat{a}^{\dagger}+\hat{a}S^{+},
        \hat{\rho}_{T}\right] \nonumber \\
        && -\frac{1}{2}\Gamma
        \hat{L}_{a}\hat{\rho}_{T}
        -\frac{1}{2}\Gamma_{c}\hat{L}_{c}\hat{\rho}_{T} \ .\label{t195}
\end{eqnarray}
We now introduce the photon number representation for the density
operator $\hat{\rho}_{T}$ with respect to the cavity mode
\begin{eqnarray}
\hat{\rho}_{T} = \sum_{m,n=0}^{\infty} \rho_{mn}\ket m\bra n
\ ,\label{t196}
\end{eqnarray}
where $\rho_{mn}$ are the density matrix elements in the basis of the
photon number states of the cavity mode. Since the cavity mode is
strongly damped, we can neglect populations of the highly excited
cavity modes and limit the expansion to $m,n=1$. Under this
approximation, the master equation~(\ref{t195}) leads to the
following set of coupled equations of motion for the density matrix
elements
\begin{eqnarray}
        \dot{\rho}_{00} &=& \hat{L}\rho_{00}
        -\frac{1}{2}ig\left(S^{+}\rho_{10}-\rho_{01}S^{-}\right)
        +\Gamma_{c}\rho_{11} \ ,\nonumber \\
        \dot{\rho}_{10} &=& \hat{L}\rho_{10}
        -\frac{1}{2}ig\left(S^{-}\rho_{00}-\rho_{11}S^{-}\right)
        -\frac{1}{2}\Gamma_{c}\rho_{10} \ ,\nonumber \\
        \dot{\rho}_{11} &=& \hat{L}\rho_{11}
        -\frac{1}{2}ig\left(S^{-}\rho_{01}-\rho_{10}S^{+}\right)
        -\Gamma_{c}\rho_{11} \ ,\label{t197}
\end{eqnarray}
where $\hat{L}\rho_{ij}=\frac{1}{2}ig\eta \left[S^{+}+S^{-},
        \rho_{ij}\right] -\frac{1}{2}\Gamma \hat{L}_{a}\rho_{ij}$.

We note that the field-matrix elements $\rho_{mn}$ are still operators
with respect to the atoms. Moreover
\begin{eqnarray}
        \rho_{00}+\rho_{11} ={\rm Tr}_{F}\left(\hat{\rho}_{T}\right)
        = \hat{\rho} \label{t198}
\end{eqnarray}
is the reduced density operator of the atoms.

For the case of a strong cavity damping the most populated state of
the cavity field is the ground state $\ket 0$, and then we can assume
that the coherence $\rho_{10}$ changes slowly in time, so that we can
take $\dot{\rho}_{10}=0$. Hence, we find that
\begin{eqnarray}
        \rho_{10}\approx -\frac{ig}{\Gamma_{c}}\left(S^{-}\rho_{00}
        -\rho_{11}S^{-}\right) \ .\label{t199}
\end{eqnarray}
Substituting this solution to $\dot{\rho}_{00}$ and
$\dot{\rho}_{11}$, we get
\begin{eqnarray}
        \dot{\rho}_{00} &=& \hat{L}\rho_{00}
        +\Gamma_{c}\rho_{11}
        -\frac{g^{2}}{2\Gamma_{c}}\left(S^{+}S^{-}\rho_{00}
        +\rho_{00}S^{+}S^{-} -2S^{+}\rho_{11}S^{-}\right)
        \ ,\nonumber \\
        \dot{\rho}_{11} &=& \hat{L}\rho_{11}
        -\Gamma_{c}\rho_{11}
        +\frac{g^{2}}{2\Gamma_{c}}\left(2S^{-}\rho_{00}S^{+}
        -S^{-}S^{+}\rho_{11} -\rho_{11}S^{-}S^{+}\right)
        \ .\label{t200}
\end{eqnarray}
Adding these two equations together and neglecting population of the
state $\ket 1$, we obtain the master equation for the reduced density
operator of the atoms as
\begin{eqnarray}
        \frac{\partial \hat{\rho} }{\partial t} &=&
        \frac{1}{2}ig\eta \left[S^{+}+S^{-},
        \hat{\rho}\right]
        -\frac{1}{2}\Gamma
        \hat{L}_{a}\hat{\rho} \nonumber \\
        &&-\frac{g^{2}}{2\Gamma_{c}}\left(S^{+}S^{-}\hat{\rho}
        +\hat{\rho}S^{+}S^{-} -2S^{-}\hat{\rho}S^{+}\right)
        \ .\label{t201}
\end{eqnarray}
The first term in Eq.~(\ref{t201}) describes the interaction of the
atoms with the driving field of an effective Rabi frequency
$g\eta$. The second term represents the usual damping of
the atoms by spontaneous emission, whereas the last term describes the
damping of the collective system with an effective damping rate
$g^{2}/\Gamma_{c}$. If we choose the parameters such that the collective
damping is much larger than the spontaneous rates of the single atoms,
the second term in Eq.~(\ref{t201}) can be ignored, and then the
master equation~(\ref{t201}) describes the time evolution of the
collective two-atom system. Thus, two independent atoms located
inside a strongly damped one-mode cavity behave as a single collective
small sample model (Dicke model) with the damping rate
$g^{2}/\Gamma_{c}$. This model, however, requires that the atoms are
strongly coupled to the cavity mode $(g\gg \Gamma)$ and are
located inside the cavity such that the interatomic axis is
perpendicular to the direction of the cavity mode and the driving
field.

\subsection{Preparation of a superposition of the antisymmetric and the
ground states}\label{ftsec84}

In the section~\ref{ftsec822}, we have shown that two nonidentical
two-level atoms can be
prepared in an arbitrary superposition of the maximally entangled
antisymmetric state $\ket a$ and the ground state $\ket g$
\begin{equation}
\ket \Phi = \gamma \ket a +\sqrt{1-|\gamma|^{2}}\ket g \ .\label{t202}
\end{equation}
However, the preparation of the superposition state requires that the
atoms have different transition frequencies. Recently, Beige {\it et
al.}~\cite{bmk} have proposed a scheme in which the superposition
state $\ket \Phi$ can be prepared in a system of two identical atoms
placed at fixed positions inside an optical cavity.

\begin{figure}[t]
\begin{center}
\includegraphics[width=10cm]{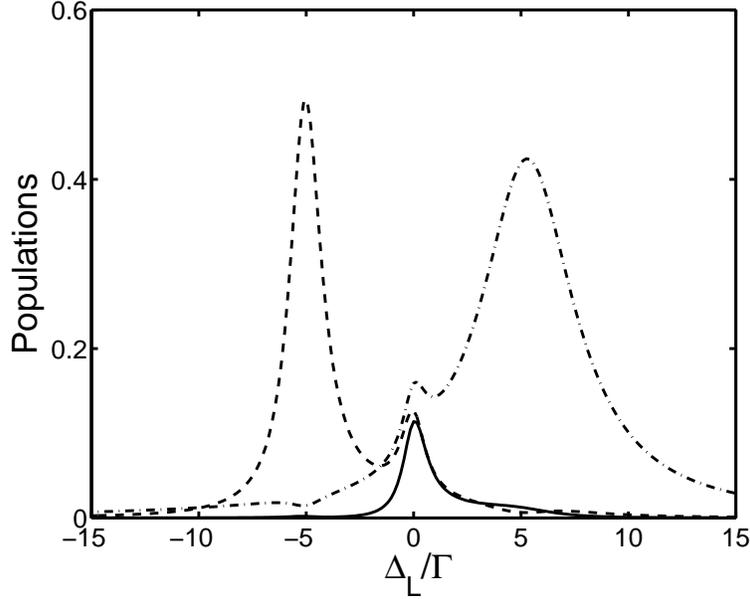}
\end{center}
\caption{The steady-state populations of the collective atomic states
of two identical atoms as a
function of $\Delta_{L}$ for the driving field propagating in the
direction of the interatomic axis, $\Omega =2.5\Gamma$,
$r_{12}/\lambda =0.08$ and $\bar{\mu} \perp
\bar{r}_{12}$: $\rho_{ee}$ (solid line), $\rho_{aa}$ (dashed line),
$\rho_{ss}$ (dashed-dotted line).}
\label{ftfig16}
\end{figure}

Here, we discuss an alternative scheme where the superposition state
$\ket \Phi$ can be generated in two identical atoms driven in free space
by a coherent laser field. This can happen when the atoms are in
nonequivalent positions in the driving field, i.e. the atoms
experience different intensities and phases of the driving field.
For a comparison, we first consider a specific geometry for the driving
field, namely that
the field is propagated perpendicularly to the atomic axis
$(\vec{k}_{L}\cdot \vec{r}_{12}=0)$. We find from Eq.~(\ref{t150}), that
in this case the collective states are populated with the
population distribution $\rho_{ee}=\rho_{aa}<\rho_{ss}$.
The population distribution
changes dramatically when the driving field propagates in directions
different from perpendicular to the interatomic axis~\cite{hsf82,fs,rfd}.
In this situation the populations strongly depend on the interatomic
separation and the detuning $\Delta_{L}$. This can produce the interesting
modification that the collective states can be selectively populated.
We show this by solving numerically the system of~15 equations for
the density matrix elements. The populations are plotted against the
detuning $\Delta_{L}$ in Fig.~\ref{ftfig16} for the laser field
propagating in the direction of the interatomic axis. We see from
Fig.~\ref{ftfig16} that the collective states $\ket s$ and $\ket e$ are
populated at $\Delta_{L}=0$ and $\Delta_{L} = \Omega_{12}$.
The antisymmetric state is significantly populated at
$\Delta_{L} = -\Omega_{12}$, and at this detuning the populations of the
states $\ket s$ and $\ket e$ are close to zero. Since
$\rho_{aa}<1$, the population is distributed between the antisymmetric
and the ground states, and therefore at $\Delta_{L} =-\Omega_{12}$ the
system is in a superposition of the maximally entangled state $\ket a$ and
the ground state $\ket g$.

Turchette {\it et al.}~\cite{tur} have recently realized experimentally
a superposition state of the ground state and a non-maximally
entangled antisymmetric state in two trapped ions. In the experiment
two trapped barium ions were sideband cooled to their motional ground
states. Transitions between the states of the ions were induced by
Raman pulses using co-propagating lasers. The ions were at positions
that experience different Rabi frequencies $\Omega_{1}$ and
$\Omega_{2}$ of the laser fields. By preparing the initial motional
ground state with one ion excited $\ket {e_{1}}\ket {g_{2}}\ket 0$,
and applying the laser fields for a time $t$, the following entangled
state $\ket {\Psi \left(t\right)}$ was created
\begin{eqnarray}
\ket {\Psi \left(t\right)} &=& -\frac{i\Omega_{2}}{\Omega}\sin
\left(\Omega t\right)\ket g \ket 1
+ \left\{\left[\frac{\Omega_{2}^{2}}{\Omega^{2}}\left(\cos \Omega t
-1\right) +1\right]\ket {e_{1}}\ket {g_{2}}\right. \nonumber \\
&& +\left. \left[\frac{\Omega_{1}\Omega_{2}}{\Omega^{2}}\left(\cos \Omega t
-1\right)\right]\ket {g_{1}}\ket {e_{2}}\right\}\ket 0 \ ,\label{t203}
\end{eqnarray}
where $\Omega^{2}=\Omega_{1}^{2}+\Omega_{2}^{2}$.

For $\Omega t =\pi$ the entangled state~(\ref{t203}) reduces to a
non-maximally entangled antisymmetric state
\begin{equation}
\ket {\Psi_{a}} = \left[\frac{\Omega_{1}^{2}-\Omega_{2}^{2}}{\Omega^{2}}
\ket {e_{1}}\ket {g_{2}} -\frac{2\Omega_{1}\Omega_{2}}{\Omega^{2}}
\ket {g_{1}}\ket {e_{2}}\right]\ket 0 \ .\label{t204}
\end{equation}

Franke {\it et al.}~\cite{fhb} have proposed to use the non-maximally
entangled state~(\ref{t204}) to demonstrate the intrinsic difference
between quantum and classical information transfers. The difference
arises from the different ways in which the probabilities occur and is
particularly clear in terms of entangled states.

\section{Detection of the entangled states}\label{ftsec9}

In this section we describe two possible methods for detection of
entangled states of two interacting atoms. One is the observation of
angular intensity distribution of the fluorescence field emitted by
the system of two interacting atoms. The other is based on quantum
interference in which one observes interference pattern of the
emitted field. Beige {\it et al.}~\cite{bbtk} have proposed a scheme,
based on the quantum Zeno effect, to observe a decoherence-free state
in a system of two three-level atoms located inside an optical cavity.
The two schemes discussed here involve two two-level atoms in free
space.

\subsection{Angular fluorescence distribution}\label{ftsec91}

It is well known that the fluorescence field emitted from a two-atom
system exhibits strong directional properties~\cite{ftk87,leh,rfd,du}.
This property can be used to detect an internal state of two interacting
atoms. To show this, we consider the fluorescence intensity, defined
in Eq.~(\ref{t69}), that in terms of the density matrix elements of
the collective atomic system can be written as
\begin{eqnarray}
I\left(\vec{R},t\right) &=& u(\vec{R}) \left\{
\left(\rho_{ee}+\rho_{ss}\right)\left[1 +\cos\left(kr_{12}\cos\theta
\right)\right]\right. \nonumber \\
&&+\left. \left(\rho_{ee}+\rho_{aa}\right)\left[1 -\cos\left(kr_{12}
\cos\theta \right)\right]\right. \nonumber \\
&&+\left. i\left(\rho_{sa}-\rho_{as}\right)\sin\left(kr_{12}\cos\theta
\right)\right\} \ ,\label{t205}
\end{eqnarray}
where $\theta$ is the angle between the observation direction $\vec{R}$
and the vector $\vec{r}_{12}$.

The first term in Eq.~(\ref{t205}) arises from the fluorescence emitted
on the $\ket e \rightarrow \ket s \rightarrow \ket g$ transitions,
which involve the symmetric state.
The second term arises from the $\ket e \rightarrow \ket a \rightarrow
\ket g$ transitions through the antisymmetric state. These two terms
describe two different channels of transitions for which the angular
distribution is proportional to $\left[1 \pm \cos\left(kr_{12}\cos\theta
\right)\right]$. The last term in Eq.~(\ref{t205}) originates from
interference between these two radiation channels. It is evident from
Eq.~(\ref{t205}) that the angular distribution of the fluorescence field
depends on the population of the entangled states $\ket s$ and $\ket
a$. Moreover, independent of the interatomic separation~$r_{12}$, the
antisymmetric state does not radiate in the direction perpendicular to
the atomic axis, as for $\theta =\pi/2$ the factor
$\left[1 -\cos\left(kr_{12}\cos\theta \right)\right]$ vanishes. In
contrast, the symmetric state radiates in all directions.
Hence, the spatial distribution of the fluorescence field is
not spherical unless $\rho_{ss}=\rho_{aa}$ and then the angular
distribution is spherically symmetric independent of the interatomic
separation. Therefore, an asymmetry in the angular distribution of the
fluorescence field  would be a compelling evidence that
the entangled states $\ket s$ and $\ket a$ are not equally populated.
If the fluorescence is detected in the direction perpendicular to the
interatomic axis the observed intensity (if any) would correspond to the
fluorescence field emitted from the symmetrical state $\ket s$ and/or
the upper state $\ket e$. On the other hand, if there is no
fluorescence detected in the direction perpendicular to the atomic
axis, the population is entirely in a superposition of the
antisymmetric state $\ket a$ and the ground state $\ket g$.

Guo and Yang~\cite{gy1,gy2} have analyzed spontaneous decay from two atoms
initially prepared in an entangled state. They have shown that the
time evolution of the population inversion, which is proportional to
the total radiation intensity~(\ref{t71}), depends on the degree of
entanglement of the initial state of the system. In Sections~\ref{sec411}
and~\ref{sec412}, we have shown that in the case of two non-identical
atoms the time evolution of the total radiation intensity
$I\left(t\right)$ can exhibit quantum beats
which result from the presence of correlations between the symmetric
and antisymmetric states. In fact, quantum beats are present only if
initially the system is in a non-maximally entangled state, and no
quantum beats are predicted for maximally entangled as well as
unentangled states.

\subsection{Interference pattern with a dark center}\label{ftsec92}

An alternative way to detect entangled states of a two-atom system
is to observe an interference pattern of the fluorescence field
emitted in the direction $\vec{R}$, not necessary perpendicular to
the interatomic axis.

This scheme is particularly useful for detection of the symmetric or
the antisymmetric state. To show this, we consider the visibility in
terms of the density matrix elements of the collective atomic system
as
\begin{equation}
{\cal {V}} =
\frac{\rho_{ss}-\rho_{aa}}{\rho_{ss}+\rho_{aa}+2\rho_{ee}} \
.\label{t206}
\end{equation}
This simple formula shows that the sign of ${\cal {V}}$ depends on the
population difference between the symmetric and antisymmetric states.
For $\rho_{ss}>\rho_{aa}$ the visibility ${\cal {V}}$ is positive, and
then the interference pattern exhibits a maximum (bright center),
whereas for $\rho_{ss}<\rho_{aa}$ the visibility ${\cal {V}}$ is
negative and then there is a minimum (dark center). The optimum
positive (negative) value is ${\cal {V}} = 1$ $({\cal {V}} = -1)$, and
there is no interference pattern when ${\cal{V}}=0$. The later happens
when $\rho_{ss}=\rho_{aa}$.

Similar to the fluorescence intensity distribution, the visibility can
provide an information about the entangled states of a two-atom
system. When the system is prepared in the antisymmetric state or in a
superposition of the antisymmetric and the ground states,
$\rho_{ss}=\rho_{ee}=0$, and then the visibility has the optimum
negative value ${\cal {V}} = -1$. On the other hand, when the system
is prepared in the symmetric state or in a linear superposition of the
symmetric and ground states, the visibility has the maximum positive
value ${\cal {V}} = 1$.
\begin{figure}[t]
\begin{center}
\includegraphics[width=10cm]{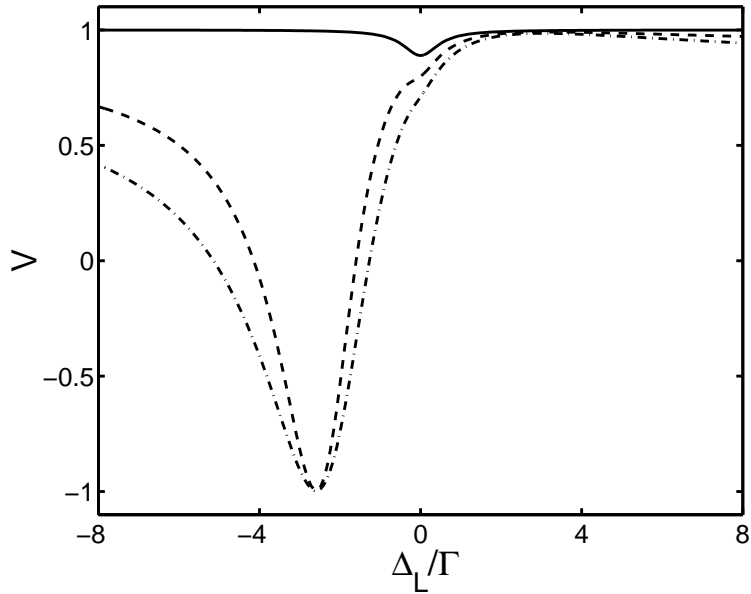}
\end{center}
\caption{The visibility $V$ as a function of $\Delta_{L}$ for
$r_{12}=0.1\lambda, \Omega =0.5\Gamma$ and various angles
$\theta_{L}$; $\theta_{L}=\pi/2$ (solid line), $\theta_{L}=\pi/4$
(dashed line), $\theta_{L}=0$ (dashed-dotted line).}
\label{ftfig17}
\end{figure}

\begin{figure}[t]
\begin{center}
\includegraphics[width=10cm]{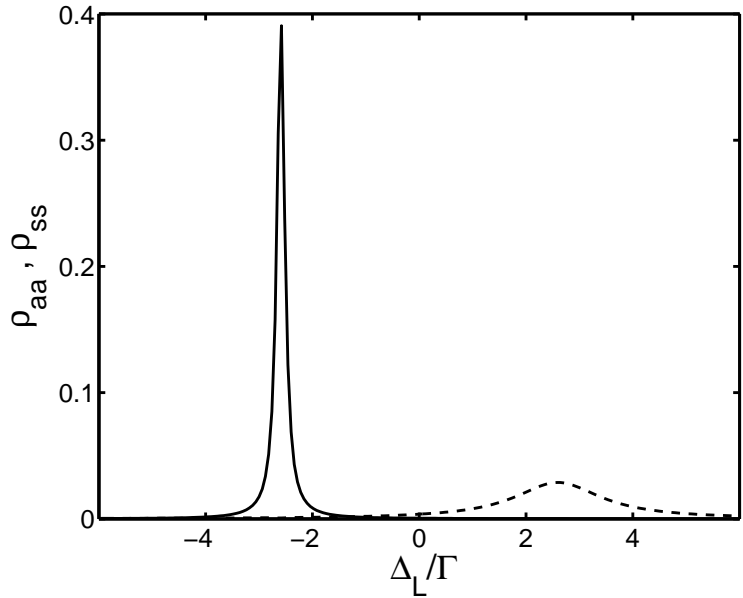}
\end{center}
\caption{Populations of the symmetric and antisymmetric states for
the same parameters as in Fig.~\ref{ftfig17}, with $\theta_{L}=0$. }
\label{ftfig18}
\end{figure}

The earliest theoretical studies of the fringe visibility involved
a coherent driving field which produces an interference pattern with a
bright center.
Recently, Meyer and Yeoman \cite{meyer} have shown that in contrast to the
coherent excitation, the incoherent field produces an interference pattern
with a dark center. Dung and Ujihara~\cite{du} have shown that the
fringe contrast factor can be negative for spontaneous
emission from two undriven atoms, with initially one atom excited.
Interference pattern with a dark center can also be obtained with a
coherent driving field~\cite{fr}. This happens when the atoms experience
different phases and/or intensities of the driving field. To show this,
we solve numerically the master equation~(\ref{t42}) for the
steady-state density matrix elements of the driven system of two atoms.
The visibility $V$ is plotted in Fig.~\ref{ftfig17} as a function of
the detuning $\Delta_{L}$ for $r_{12}=0.1\lambda,
\Omega =0.25\Gamma$ and various angles $\theta_{L}$ between the
interatomic axis and the direction of propagation of the laser field.
The visibility $V$ is positive for most values of $\Delta_{L}$, except
$\Delta_{L}\approx -\Omega_{12}$. At this detuning the parameter $V$
is negative and reaches the optimum negative value $V=-1$ indicating
that the system produces interference pattern with a dark center. In
Fig.~\ref{ftfig18}, we plot the populations of the symmetric and
antisymmetric states for the same parameters as in Fig~\ref{ftfig17}.
It is evident from Fig.~\ref{ftfig18}
that at $\Delta_{L}=-\Omega_{12}$ the antisymmetric state is
significantly populated, whereas the population of the symmetric
state is close to zero. This population difference leads to negative
values of $V$, as predicted by Eq.~(\ref{t206}) and seen in
Fig.~\ref{ftfig17}. Experimental observation of the interference pattern
with a dark center would be an interesting demonstration of the
controled excitation of a two-atom system to the entangled antisymmetric
state.

\section{Two-photon entangled states}\label{ftsec10}

In our discussions to date on entanglement creation in two-atom
systems, we have focused on  different methods of creating entangled
states of the form
\begin{eqnarray}
\ket \Psi = c_{1}\ket {e_{1}}\ket {g_{2}} \pm c_{2}\ket {g_{1}}
\ket {e_{2}} \ .\label{t207}
\end{eqnarray}
As we have shown in Sec.~\ref{ftsec31}, entangled states of the above
form are generated by the dipole-dipole interaction
between the atoms and the preparation of these states is sensitive to
the difference $\Delta$ between the atomic transition frequencies
and to the atomic decay rates. These states are better known as the
symmetric and antisymmetric collective atomic states.

Apart from the symmetric and antisymmetric states,
there are two other collective states of the two-atom system: the
ground state $\ket g =\ket {g_{1}}\ket {g_{2}}$ and the upper state
$\ket e =\ket {e_{1}}\ket {e_{2}}$, which are also product states of
the individual atomic states. These states are
not affected by the dipole-dipole interaction $\Omega_{12}$.

In this section, we discuss a method of creating entanglement between
these two states of the general form
\begin{equation}
        \ket {\Upsilon} = c_{g}\ket g \pm c_{e}\ket e \ ,\label{t208}
\end{equation}
where $c_{g}$ and $c_{e}$ are transformation parameters such that
$|c_{g}|^{2}+|c_{e}|^{2}=1$. The entangled states of the form
(\ref{t208}) are known in literature as pairwise atomic
states~\cite{pas1,pas2,pas3,pas4} or multi-atom squeezed
states~\cite{mas}.  According to Eq.~(\ref{t63}), the collective
ground and excited states are separated in energy by $2\hbar \omega_{0}$,
and therefore we can call the states $\ket \Upsilon$ as two-photon
entangled (TPE) states.

The two-photon entangled states cannot be generated by a simple coherent
excitation. A coherent field applied to the two-atom system couples
to one-photon transitions. The problem is that coherent excitation
populates not only the upper state $\ket e$ but also the
intermediate states $\ket s$ and $\ket a$, see Eq.~(\ref{t150}).
The two-photon entangled states (\ref{t208}) are superpositions of
the collective ground and excited states with no contribution from the
intermediate collective states $\ket s$ and $\ket a$.

The two-photon behavior of the entangled states (\ref{t208}) suggests
that the simplest technique for generating the TPE
states would be by applying a two-photon excitation process. An
obvious candidate is a squeezed vacuum field which is characterized
by strong two-photon correlations which would enable the transition
$\ket g \rightarrow \ket e$ to occur effectively in a single step
without populating the intermediate states. We will illustrate this
effect by analyzing the populations of the collective atomic states of
a two-atom system interacting with a squeezed vacuum field.

\subsection{Populations of the entangled states in a squeezed vacuum}
\label{ftsec101}

The general master equation~(\ref{t36}) allows us to calculate the
populations of the collective atomic states and coherences, which
gives information about the stationary state of a two-atom system.
We first consider a system of two identical atoms, separated by an
arbitrary distance $r_{12}$ and interacting with a squeezed vacuum field.
For simplicity, we assume that the carrier frequency $\omega_{s}$ of the
squeezed vacuum field is resonant to the atomic transition frequency
$\omega_{0}$, and the squeezed field is perfectly matched to the
atoms, $D(\omega_{s})=1$ and $\theta_{s}=\pi$.

    From the master equation~(\ref{t36}), we find the following equations
of motion for the populations of the collective states and the
two-photon coherences of the collective system of two identical atoms
\begin{eqnarray}
\dot{\rho}_{ee} &=&
-2\Gamma\left(\tilde{N}+1\right)\rho_{ee}
+\tilde{N}\left[\left(\Gamma
+\Gamma_{12}\right)\rho_{ss} +\left(\Gamma
-\Gamma_{12}\right)\rho_{aa}\right]
+\Gamma_{12}|\tilde{M}|\rho_{u} \ ,\nonumber \\
\dot{\rho}_{ss} &=&
\left(\Gamma +\Gamma_{12}\right)\left\{\tilde{N}
-\left(3\tilde{N}+1\right)\rho_{ss} -\tilde{N}\rho_{aa}
+\rho_{ee} -|\tilde{M}|\rho_{u}\right\} \ ,\nonumber \\
\dot{\rho}_{aa} &=&
\left(\Gamma -\Gamma_{12}\right)\left\{\tilde{N}
-\left(3\tilde{N}+1\right)\rho_{aa} -\tilde{N}\rho_{ss}
+\rho_{ee}
+|\tilde{M}|\rho_{u}\right\} \ ,\nonumber \\
\dot{\rho}_{u} &=& 2\Gamma_{12}|\tilde{M}|
-\left(2\tilde{N}+1\right)\Gamma \rho_{u} \nonumber \\
&& -2|\tilde{M}|\left[\left(\Gamma
+2\Gamma_{12}\right)\rho_{ss} -\left(\Gamma
-2\Gamma_{12}\right)\rho_{aa}\right] \ ,\label{t209}
\end{eqnarray}
where $\tilde{N}=\tilde{N}\left(\omega_{0}\right)$, $\tilde{M}
=\tilde{M}\left(\omega_{0}\right)$ and
$\rho_{u}=\rho_{eg}\exp (-i\phi_{s}) +\rho_{ge}\exp (i\phi_{s})$.

It is seen from Eq.~(\ref{t209}) that the evolution of the populations
depends on the two-photon coherencies $\rho_{eg}$ and $\rho_{ge}$,
which can transfer the population from the ground state $\ket g$
directly to the upper state $\ket e$ leaving the states $\ket s$
and $\ket a$ unpopulated. The evolution of the populations depends
on $\Gamma_{12}$, but is completely independent of the dipole-dipole
interaction $\Omega_{12}$.

Similar to the interaction with the ordinary vacuum, discussed in
Sec.~\ref{ftsec31}, the steady-state solution of Eqs.~(\ref{t209})
depends on whether $\Gamma_{12}=\Gamma$ or $\Gamma_{12}\neq \Gamma$.
Assuming that $\Gamma_{12}=\Gamma$ and setting the left-hand side of
equations~(\ref{t209}) equal to zero, we obtain the steady-state
solutions for the populations and the two-photon coherence in the
Dicke model. A straightforward algebraic manipulation of
Eqs.~(\ref{t209}) leads to the following steady-state solutions
\begin{eqnarray}
\rho_{ee} &=& \frac{\tilde{N}^{2}\left(2\tilde{N}+1\right)
-\left(2\tilde{N}-1\right)|\tilde{M}|^{2}}{\left(2\tilde{N}+1\right)
\left(3\tilde{N}^{2} +3\tilde{N} +1
-3|\tilde{M}|^{2}\right)} \ ,\nonumber \\
\rho_{ss} &=& \frac{\tilde{N}\left(\tilde{N}+1\right)
-|\tilde{M}|^{2}}{3\tilde{N}^{2} +3\tilde{N} +1
-3|\tilde{M}|^{2}} \ ,\nonumber \\
\rho_{u} &=& \frac{2|\tilde{M}|}{\left(2\tilde{N}+1\right)
\left(3\tilde{N}^{2} +3\tilde{N} +1
-3|\tilde{M}|^{2}\right)} \ .\label{t210}
\end{eqnarray}

The steady-state populations depend strongly on the squeezing correlations
$\tilde{M}$. For a classical squeezed
field with the maximal correlations $\tilde{M}=\tilde{N}$ the steady-state
populations reduces to
\begin{eqnarray}
\rho_{ss} &=& \frac{\tilde{N}}{3 \tilde{N} +1} \ ,\nonumber \\
\rho_{ee} &=& \frac{2\tilde{N}^{2}}{\left( 2\tilde{N}+1\right)
\left(3 \tilde{N} +1\right)} \ .\label{t211}
\end{eqnarray}
It is easily to check that in this case the populations obey a Boltzmann
distribution with $\rho_{gg}>\rho_{ss}>\rho_{ee}$. Moreover, in the
limit of low intensities $(\tilde{N}\ll 1)$ of the field, the
population $\rho_{ee}$ is proportional to $\tilde{N}^{2}$, showing
that in classical fields the population exhibits a quadratic  dependence
on the intensity.

The population distribution is qualitatively different for a quantum
squeezed field with the maximal correlations $|\tilde{M}|^{2}
=\tilde{N}(\tilde{N}+1)$. In this case, the stationary populations of
the excited collective states are
\begin{eqnarray}
\rho_{ss} &=& 0 \ , \nonumber \\
\rho_{ee} &=& \frac{ \tilde{N}}{\left(2 \tilde{N}+1\right)}
\ .\label{t212}
\end{eqnarray}
Clearly, the symmetric state is not populated. In this case the
populations no longer obey the Boltzmann distribution. The
population is distributed only between the ground state $\ket g$ and
the upper state~$\ket e$. Moreover, it can be seen from Eq.~(\ref{t212})
that for a weak quantum squeezed field the population $\rho_{ee}$ depends
linearly on the intensity. This distinctive feature reflects the
direct modifications of the two-photon absorption that the
nonclassical photon correlations enable the transition
$\ket g \rightarrow \ket e$ to occurs in a "single step" proportional
to $\tilde{N}$. In other words, the nonclassical two-photon correlations
entangle the ground state $\ket g$ and the upper state $\ket e$ with
no contribution from the symmetric state $\ket s$.

The question we are interested in concerns the final state of the
system and its purity. To answer this question, we apply
Eq.~(\ref{t210}) and find that in the steady-state, the density
matrix of the system is given by
\begin{eqnarray}
\hat{\rho} &=& \left(
\begin{array}{ccc}
\rho_{gg} & 0 & \rho_{ge} \\
0 & \rho_{ss} & 0 \\
\rho_{eg} & 0 & \rho_{ee}
\end{array}
\right) \ , \label{t213}
\end{eqnarray}
where $\rho_{ij}$ are the non-zero steady-state density matrix
elements.

It is evident from Eq.~(\ref{t213}) that in the squeezed vacuum the
density matrix of the system is not diagonal due to the presence of
the two-photon coherencies $\rho_{ge}$ and $\rho_{eg}$. This
indicates that
the collective states $\ket g$, $\ket s$ and $\ket e$ are no longer
eigenstates of the system. The density matrix can be rediagonalized
by including $\rho_{eg}$ and $\rho_{ge}$ to give the new (entangled)
states
\begin{eqnarray}
\ket {\Upsilon_{1}} &=& \left[\left(P_{1}-\rho_{ee}\right)\ket g +
\rho_{eg}\ket e
\right]/\left[\left(P_{1}-\rho_{ee}\right)^{2}
+\left|\rho_{eg}\right|^{2}\right]^{\frac{1}{2}} \ ,\nonumber \\
\ket {\Upsilon_{2}} &=& \left[\rho_{ge}\ket g +
\left(P_{2}-\rho_{gg}\right)\ket e
\right]/\left[\left(P_{2}-\rho_{gg}\right)^{2}
+\left|\rho_{eg}\right|^{2}\right]^{\frac{1}{2}} \ ,\nonumber \\
\ket {\Upsilon_{3}} &=& \ket s \ ,\label{t214}
\end{eqnarray}
where the diagonal probabilities are
\begin{eqnarray}
P_{1} &=& \frac{1}{2}\left(\rho_{gg}+\rho_{ee}\right)
+\frac{1}{2}\left[\left(\rho_{gg}-\rho_{ee}\right)^{2}
+4\left|\rho_{eg}\right|^{2}\right]^{\frac{1}{2}} \ ,\nonumber \\
P_{2} &=& \frac{1}{2}\left(\rho_{gg}+\rho_{ee}\right)
-\frac{1}{2}\left[\left(\rho_{gg}-\rho_{ee}\right)^{2}
+4\left|\rho_{eg}\right|^{2}\right]^{\frac{1}{2}} \ ,\nonumber \\
P_{3} &=& \rho_{ss} \ .\label{t215}
\end{eqnarray}
In view of Eqs.~(\ref{t212}) and~(\ref{t214}), it is easy to see
that the squeezed vacuum causes the system to decay into entangled
states which are linear superpositions of the
collective ground state $\ket g$ and the upper state $\ket e$.
The intermediate symmetric state remains unchanged under the squeezed
vacuum excitation.
In general, the states~(\ref{t214}) are mixed states. However, for
perfect correlations $|\tilde{M}|^{2}=\tilde{N}(\tilde{N}+1)$ the
populations $P_{2}$ and $P_{3}$ are zero
leaving the population only in the state $\ket {\Upsilon_{1}}$. Hence,
the state $\ket {\Upsilon_{1}}$
is a pure state of the system of two atoms driven by a squeezed vacuum
field. From Eqs.~(\ref{t214}), we find that the pure entangled state
$\ket {\Upsilon_{1}}$ is given by~\cite{pk90}
\begin{equation}
\ket {\Upsilon_{1}} = \frac{1}{\sqrt{2\tilde{N}+1}}
\left[\sqrt{\tilde{N}+1}\ket g
+\sqrt{\tilde{N}}\ket e\right] \ .\label{t216}
\end{equation}
The pure state~(\ref{t216}) is non-maximally entangled state, it
reduces to a maximally entangled state for $\tilde{N}\gg 1$. The entangled
state is analogous to the pairwise atomic
state~\cite{pas1,pas2,pas3,pas4} or the multi-atom squeezed
state~\cite{mas}, (see also Ref.~\cite{park}),
predicted in the small sample model of two coupled atoms.

The pure entangled state $\ket {\Upsilon_{1}}$ is characteristic not
only of the two-atom Dicke model, but in general of the Dicke model of
an even number of atoms~\cite{ap90}. The $N$-atom Dicke system
interacting with a squeezed vacuum can decay to a state which the
density operator is given by
\begin{eqnarray}
        \hat{\rho} &=& C_{n}\left(\mu S^{-}+\nu S^{+}\right)^{-1}
        \left(\mu S^{+}+\nu S^{-}\right)^{-1} \ ,\label{t217}
\end{eqnarray}
if $N$ is odd, or
\begin{eqnarray}
        \hat{\rho} =\ket \Upsilon \bra \Upsilon \ ,\label{t218}
\end{eqnarray}
if $N$ is even, where $C_{n}$ is the normalization constant,
$S^{\pm}$ are the collective atomic operators, $\mu^{2} =\nu^{2}+1
=\tilde{N}+1$, and $\ket \Upsilon$ is defined by
\begin{eqnarray}
        \left(\mu S^{-}+\nu S^{+}\right)\ket \Upsilon =0 \ .\label{t219}
\end{eqnarray}
Thus, for an even number of atoms the stationary state of the system
is the pure pairwise atomic state.

\subsection{Effect of the antisymmetric state on the purity of the
system}\label{ftsec102}

The pure entangled state $\ket {\Upsilon_{1}}$ can be obtained for
perfect matching of the squeezed modes to the atoms and interatomic
separations much smaller than the optical
wavelength. To achieve perfect matching, it is necessary
to squeeze of all the modes to which the atoms are coupled. That is,
the squeezed modes must occupy the whole $4\pi$ solid angle of the
space surrounding the atoms. This is not possible to achieve with the
present experiments in free space, and in order to avoid the difficulty
cavity environments have been suggested~\cite{kimb1,kimb2}. Inside
a cavity the atoms interact strongly only with the privileged cavity
modes. By the squeezing of these cavity modes, which occupy only a
small solid angle about the cavity axis, it would be possible to achieve
perfect matching of the squeezed field to the atoms.

There is, however, the practical problem to fulfil the second requirement
that interatomic separations should be much smaller than the
resonant wavelength. This assumption may prove difficult in
experimental realization as with the present techniques two atoms can
be trapped within distances of the order of a resonant
wavelength~\cite{eich,deb,ber,tos}. As we have shown in
Sec.~\ref{ftsec31}, the dynamics of such a system involve the
antisymmetric state and are significantly different from the
dynamics of the Dicke model.

For two atoms separated by an arbitrary distance $r_{12}$, the
collective damping $\Gamma_{12}\neq \Gamma$, and then the steady-state
solutions of Eqs.~(\ref{t209}) are
\begin{eqnarray}
\rho_{ee} &=&\frac{\tilde{N}^{2}}{\left(2\tilde{N}+1\right)^{2}}
+\frac{a^{2}|\tilde{M}|^{2}\left(
4\tilde{N}+1\right) }{G} \ ,  \nonumber \\
\rho_{ss} &=&\frac{\tilde{N}\left(\tilde{N}+1\right) }
{\left(2\tilde{N}+1\right)^{2}}-\frac{a|\tilde{M}|^{2}
\left[2\left(2\tilde{N}+1\right)^{2}-a\right] }{G} \ ,  \nonumber \\
\rho_{aa} &=& \frac{\tilde{N}\left(\tilde{N}+1\right) }
{\left(2\tilde{N}+1\right)^{2}}
+\frac{a|\tilde{M}|^{2}
\left[2\left(2\tilde{N}+1\right)^{2}+a\right] }{G},  \nonumber \\
\rho_{u} &=& \frac{2a\left(2\tilde{N}+1\right)^{3}|\tilde{M}|}{G}
\ ,\label{t220}
\end{eqnarray}
where $a=\Gamma_{12}/\Gamma$, and
\begin{equation}
G = \left(2\tilde{N}+1\right)^{2}\left\{ \left(2\tilde{N}+1\right)^{4}
+4|\tilde{M}|^{2}\left[ a^{2}-\left(2\tilde{N}+1\right)^{2}\right]
\right\} \ .\label{t221}
\end{equation}
\begin{figure}[t]
\begin{center}
\includegraphics[width=13cm]{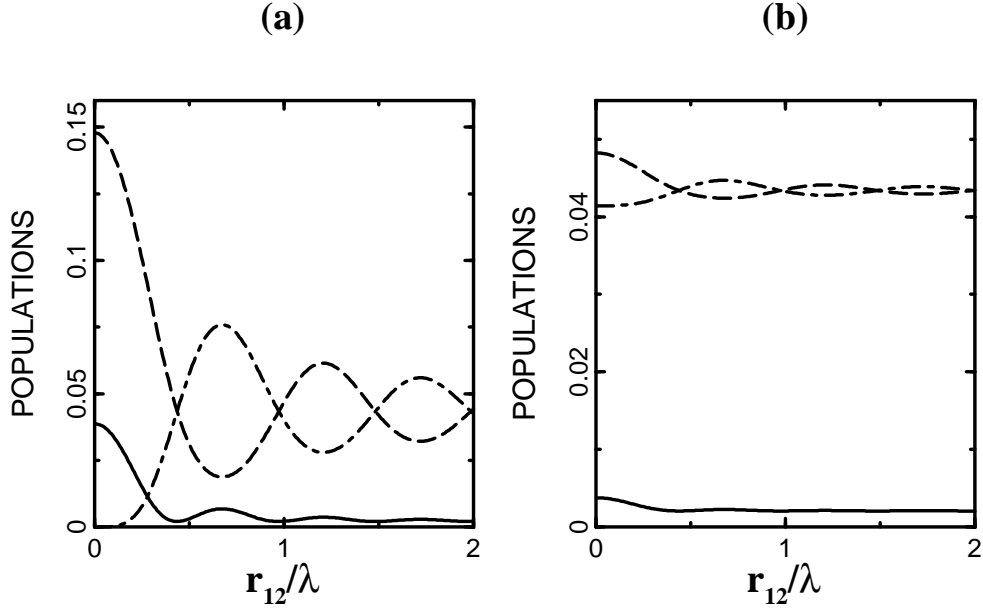}
\end{center}
\caption{The steady-state populations of the collective atomic states
as a function of $r_{12}$ for (a) quantum squeezed field with
$|\tilde{M}|^{2}=\tilde{N}(\tilde{N}+1)$, (b) classical
squeezed field with  $|\tilde{M}|=\tilde{N}$, and $\tilde{N}=0.05$,
$\bar{\mu} \perp \bar{r}_{12}$: $\rho_{ee}$ (solid line),
$\rho_{aa}$ (dashed line), $\rho_{ss}$ (dashed-dotted line).}
\label{ftfig19}
\end{figure}

This result shows that the antisymmetric state is populated in the
steady-state even for small interatomic separations $(\Gamma_{12}\approx
\Gamma)$. For large interatomic separations $\Gamma_{12}\approx 0$, and
then the symmetric
and antisymmetric states are equally populated. When the interatomic
separation decreases, the population of the state $\ket a$ increases,
whereas the population of the state $\ket s$ decreases and
$\rho_{ss}=0$ for very small interatomic separations. These features
are illustrated in Fig.~\ref{ftfig19}(a), where we plot the steady-state
populations as a function of the interatomic separation for the
maximally correlated quantum squeezed field. We see that the collective
states are unequally populated and in the case of small $r_{12}$, the
state $\ket a$ is the most populated state of the system, whereas the
state $\ket s$ is not populated. In Fig.~\ref{ftfig19}(b), we show the
populations for the equivalent maximally correlated classical
squeezed field, and in this case all states are populated independent
of $r_{12}$.

\begin{figure}[t]
\begin{center}
\includegraphics[width=10cm]{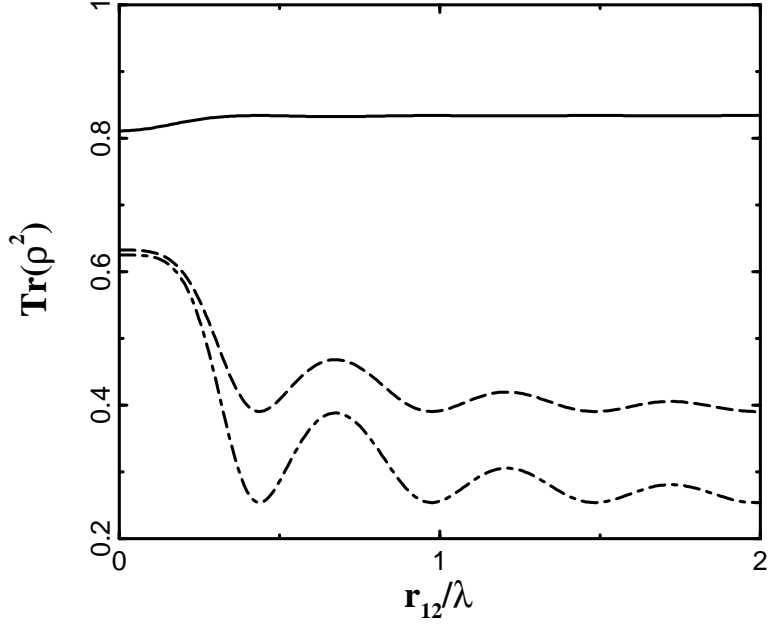}
\end{center}
\caption{Tr$(\hat{\rho}^{2})$ as a function of the interatomic separation
for $|\tilde{M}|^{2}=\tilde{N}(\tilde{N}+1)$,
$\bar{\mu} \perp \bar{r}_{12}$ and different $\tilde{N}$: $\tilde{N}=0.05$
(solid line), $\tilde{N}=0.5$ (dashed line), $\tilde{N}=5$
(dashed-dotted line). }
\label{ftfig20}
\end{figure}

This fact can lead to a destruction of the purity of the stationary
state of the system. To show this, we calculate the quantity
\begin{equation}
{\rm Tr}\left(\hat{\rho}^{2}\right) =\rho_{gg}^{2}+\rho_{ss}^{2}
+\rho_{aa}^{2}+\rho_{ee}^{2}+\left|\rho_{u}\right|^{2} \ ,
\label{t222}
\end{equation}
which determines the purity of the system. Tr$(\hat{\rho}^{2})=1$
corresponds to a pure state of the system, while Tr$(\hat{\rho}^{2})<1$
corresponds to a mixed state. Tr$(\hat{\rho}^{2})=1/4$ describes a
completely mixed state of the system. In Fig.~\ref{ftfig20}, we display
Tr$(\hat{\rho}^{2})$ as a function of the interatomic separation $r_{12}$
for perfect correlations $|\tilde{M}|^{2}=\tilde{N}(\tilde{N}+1)$
and various $\tilde{N}$. Clearly, the system is in a mixed state
independent of the interatomic separation. Moreover, the purity decreases
as $\tilde{N}$ increases.

For small interatomic separation, the mixed state of the system is
composed of two states: the TPE state $\ket {\Upsilon_{1}}$ and the
antisymmetric state $\ket a$. We illustrate this by diagonalizing
the steady-state density matrix of the system
\begin{eqnarray}
\hat{\rho} &=& \left(
\begin{array}{cccc}
\rho_{gg} & 0 & 0 & \rho_{ge} \\
0 & \rho_{aa} & 0 & 0 \\
0 & 0 & \rho_{ss} & 0 \\
\rho_{eg} & 0 & 0 & \rho_{ee}
\end{array}
\right) \ , \label{t223}
\end{eqnarray}
from which we find the new (entangled) states
\begin{eqnarray}
\ket {\Upsilon_{1}} &=& \left[\left(P_{1}-\rho_{ee}\right)\ket g +
\rho_{eg}\ket e
\right]/\left[\left(P_{1}-\rho_{ee}\right)^{2}
+\left|\rho_{eg}\right|^{2}\right]^{\frac{1}{2}} \ ,\nonumber \\
\ket {\Upsilon_{2}} &=& \left[\rho_{ge}\ket g +
\left(P_{2}-\rho_{gg}\right)\ket e
\right]/\left[\left(P_{2}-\rho_{gg}\right)^{2}
+\left|\rho_{eg}\right|^{2}\right]^{\frac{1}{2}} \ ,\nonumber \\
\ket {\Upsilon_{3}} &=& \ket s \ ,\nonumber \\
\ket {\Upsilon_{4}} &=& \ket a \ ,\label{t224}
\end{eqnarray}
where the diagonal probabilities (populations of the entangled states) are
\begin{eqnarray}
P_{1} &=& \frac{1}{2}\left(\rho_{gg}+\rho_{ee}\right)
+\frac{1}{2}\left[\left(\rho_{gg}-\rho_{ee}\right)^{2}
+4\left|\rho_{eg}\right|^{2}\right]^{\frac{1}{2}} \ ,\nonumber \\
P_{2} &=& \frac{1}{2}\left(\rho_{gg}+\rho_{ee}\right)
-\frac{1}{2}\left[\left(\rho_{gg}-\rho_{ee}\right)^{2}
+4\left|\rho_{eg}\right|^{2}\right]^{\frac{1}{2}} \ ,\nonumber \\
P_{3} &=& \rho_{ss} \ ,\nonumber \\
P_{4} &=& \rho_{aa} \ .\label{t225}
\end{eqnarray}
\begin{figure}[t]
\begin{center}
\includegraphics[width=13cm]{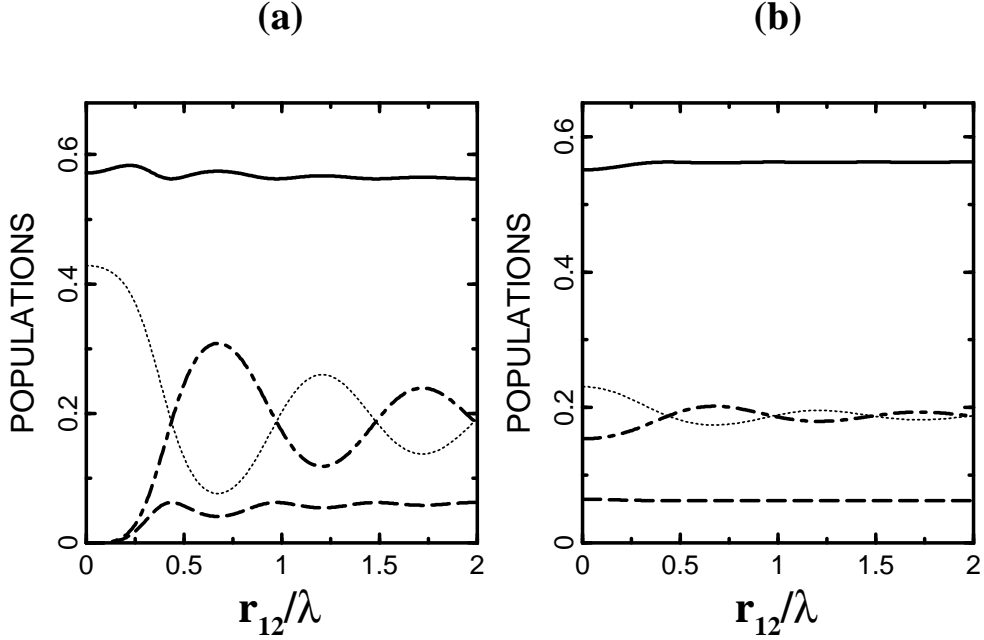}
\end{center}
\caption{Populations of the entangled states~(\ref{t224})
as a function of the interatomic separation for (a) quantum squeezed
field with $|\tilde{M}|^{2}=\tilde{N}(\tilde{N}+1)$, (b)
classical squeezed field with $|\tilde{M}|=\tilde{N}$, and $\bar{\mu}
\perp \bar{r}_{12}$, $\tilde{N}=0.5$. In both frames $P_{1}$
(solid line), $P_{2}$ (dashed line), $P_{3}$ (dashed-dotted line),
$P_{4}$ (dotted line). }
\label{ftfig21}
\end{figure}

Note, that the states $\ket {\Upsilon_{1}}, \ket {\Upsilon_{2}}$ and
$\ket {\Upsilon_{3}}$ are the same as for the small sample model,
discussed in the preceding section. This means that the presence of
the antisymmetric state does not affect the two-photon entangled
states, but it can affect the population distribution between the
states and the purity of the system. In Fig.~\ref{ftfig21}, we plot
the populations $P_{i}$ of the states $\ket {\Upsilon_{i}}$ as a
function of the interatomic separation. The figure demonstrates that
in the case of a quantum squeezed field the atoms are driven into a
mixed state composed of {\it only} two entangled states
$\ket {\Upsilon_{1}}$ and $\ket a$, and there is a vanishing
probability that the system is in the states $\ket {\Upsilon_{2}}$
and $\ket s$. In contrast, for a classical squeezed field, shown in
Fig.~\ref{ftfig21}(b), the atoms are driven to a mixed state composed
of all the entangled states.

Following the discussion presented in Sec.~\ref{ftsec31}, we can argue
that the system can decay to the pure TPE state $\ket {\Upsilon_{1}}$
with the interatomic separation included. This can happen when
the observation time is shorter than $\Gamma ^{-1}$.
The antisymmetric state $\left|a\right\rangle $ decays on a time scale
$\sim \left( \Gamma -\Gamma_{12}\right) ^{-1}$, and for
$\Gamma _{12}\approx \Gamma $ the decay rate of the antisymmetric state
is much longer than $\Gamma ^{-1}$.  By contrast, the
state $\left| s\right\rangle $ decays on a time scale $\sim \left( \Gamma
+\Gamma _{12}\right) ^{-1},$ which for $\Gamma _{12}\approx \Gamma $ is
shorter than $\Gamma ^{-1}$. Clearly, for observation times shorter
than  $\Gamma^{-1},$ the antisymmetric state does not participate in the
interaction and the system reaches the steady-state only between the
triplet states. Thus, for perfect matching of the squeezed
modes to the atoms the symmetric state is not populated and then the
system is in the pure TPE state $\ket {\Upsilon_{1}}$.

\subsection{Two-photon entangled states for two non-identical
atoms}\label{ftsec103}

We now extend the analysis of the population distribution in a
squeezed vacuum to the case of two nonidentical atoms.
For two nonidentical atoms with $\Delta \neq 0$ and
$\Gamma_{1}=\Gamma_{2}=\Gamma$, the master equation~(\ref{t36}) leads
to the following equations of motion for the density matrix elements
\begin{eqnarray}
\dot{\rho}_{ee} &=&
-2\Gamma\left(\tilde{N}+1\right)\rho_{ee}
+\tilde{N}\left[\Gamma \left(\rho_{ss}
+\rho_{aa}\right) +
\Gamma_{12}\left(\rho_{ss}-\rho_{aa}\right)e^{i\Delta t}\right] \nonumber \\
&&+\Gamma_{12}|\tilde{M}|\left(\rho_{eg}
e^{-i\left[2\left(\omega_{s}-\omega_{0}\right)t +\phi_{s} \right]}
+\rho_{ge}
e^{i\left[2\left(\omega_{s}-\omega_{0}\right)t +\phi_{s} \right]}
\right) \ ,\nonumber \\
\dot{\rho}_{ss} &=&
\left(\Gamma +\Gamma_{12}e^{i\Delta t}\right)\left[\tilde{N}
-\left(3\tilde{N}+1\right)\rho_{ss} -\tilde{N}\rho_{aa}
+\rho_{ee}\right] \nonumber \\
&& -\Gamma |\tilde{M}|\left(\rho_{eg}
e^{-i\left[2\left(\omega_{s}-\omega_{1}\right)t +\phi_{s} \right]}
+\rho_{ge}
e^{i\left[2\left(\omega_{s}-\omega_{1}\right)t +\phi_{s} \right]}
\right) \nonumber \\
&& -\Gamma_{12} |\tilde{M}|\left(\rho_{eg}
e^{-i\left[2\left(\omega_{s}-\omega_{0}\right)t +\phi_{s} \right]}
+\rho_{ge}
e^{i\left[2\left(\omega_{s}-\omega_{0}\right)t +\phi_{s} \right]}
\right) \ ,\nonumber \\
\dot{\rho}_{aa} &=&
\left(\Gamma -\Gamma_{12}e^{i\Delta t}\right)\left[\tilde{N}
-\left(3\tilde{N}+1\right)\rho_{aa} -\tilde{N}\rho_{ss}
+\rho_{ee}\right] \nonumber \\
&& +\Gamma |\tilde{M}|\left(\rho_{eg}
e^{-i\left[2\left(\omega_{s}-\omega_{1}\right)t +\phi_{s} \right]}
+\rho_{ge}
e^{i\left[2\left(\omega_{s}-\omega_{1}\right)t +\phi_{s} \right]}
\right) \nonumber \\
&& -\Gamma_{12} |\tilde{M}|\left(\rho_{eg}
e^{-i\left[2\left(\omega_{s}-\omega_{0}\right)t +\phi_{s} \right]}
+\rho_{ge}
e^{i\left[2\left(\omega_{s}-\omega_{0}\right)t +\phi_{s} \right]}
\right) \ ,\nonumber \\
\dot{\rho}_{eg} &=&
\left(\dot{\rho}_{ge}\right)^{\ast} =
\Gamma_{12}|\tilde{M}|
e^{i\left[2\left(\omega_{s}-\omega_{1}\right)t +\phi_{s} \right]}
-\left(2\tilde{N}+1\right)\Gamma \rho_{eg} \nonumber \\
&& -\Gamma |\tilde{M}|
e^{i\left[2\left(\omega_{s}-\omega_{1}\right)t +\phi_{s} \right]}
\left(\rho_{ss} -\rho_{aa}\right) \nonumber \\
&& -2\Gamma_{12}|\tilde{M}|
e^{i\left[2\left(\omega_{s}-\omega_{0}\right)t +\phi_{s} \right]}
\left(\rho_{ss}+\rho_{aa}\right) \ ,\label{t226}
\end{eqnarray}
where $\omega_{0}=\frac{1}{2}\left(\omega_{1}+\omega_{2}\right)$.

Equations~(\ref{t226}) contain time-dependent terms which oscillate
at frequencies exp$(\pm i\Delta t)$ and exp$[\pm
2i(\omega_{s}-\omega_{0})t +\phi_{s}]$. If we tune the squeezed
vacuum field to the middle of the frequency difference between the
atomic frequencies, i.e. $\omega_{s} =(\omega_{1}+\omega_{2})/2$, the
terms proportional to exp$[\pm 2i(\omega_{s}-\omega_{0})t +\phi_{s}]$
become stationary in time. None of the other time dependent components
is resonant with the frequency of the squeezed vacuum field.
Consequently, for $\Delta \gg \Gamma$, the
time-dependent components oscillate rapidly in time
and average to zero over long times. Therefore, we can make a
secular approximation in which we ignore the rapidly oscillating
terms, and find the following steady-state solutions~\cite{fw97}
\begin{eqnarray}
\rho_{ee}
&=&\frac{1}{4}\left\{\frac{\left(2\tilde{N}-1\right)}{2\tilde{N}+1}+
\frac{1}{\left[\left(2\tilde{N}+1\right)^{2}
-4a^{2}|\tilde{M}|^{2}\right] }\right\}
\ ,\nonumber \\
\rho_{ss} &=& \rho_{aa} =\frac{1}{4}\left\{ 1-\frac{1}{\left[
\left(2\tilde{N}+1\right)^{2}-4a^{2}|\tilde{M}|^{2}\right] }\right\}
\ ,\nonumber \\
\rho_{u} &=&
\frac{2a|\tilde{M}|}{\left(2\tilde{N}+1\right)
\left[\left(2\tilde{N}+1\right)^{2}
-4a^{2}|\tilde{M}|^{2}\right]}
\ .\label{t227}
\end{eqnarray}
Equations (\ref{t227}) are quite different from Eqs.~(\ref{t220})
and show that in the case of non-identical atoms
the symmetric and antisymmetric states are equally
populated independent of the interatomic separation. These are,
however, some similarities to the steady-state solutions of the Dicke
model that for small interatomic separations
$\rho_{ss}=\rho_{aa} \approx 0$, and then only the collective ground
and the upper states are populated.

To conclude this section, we point out that by employing two spatially
separated non-identical atoms of significantly different
transition frequencies $(\Delta \gg \Gamma)$, it is possible to achieve
the pure TPE state with the antisymmetric state fully participating in
the interaction.

\section{Mapping of entangled states of light on
atoms}\label{ftsec104}

The generation of the pure TPE state is an example of mapping of a
state of quantum correlated light onto an atomic system. The
two-photon correlations contained in the squeezed vacuum field can be
completely transferred to the atomic system and can be measured, for
example, by detecting fluctuations of the fluorescence field emitted by
the atomic system. Squeezing in the fluorescence field is proportional to
the squeezing in the atomic dipole operators (spin squeezing) which, on
the other hand, can be found from the steady-state solutions for the
density matrix elements.

\subsection{Mapping of photon correlations}\label{ftsec1041}

Equation~(\ref{t227}) shows that the collective damping parameter
$\Gamma_{12}$ plays the role of a degree of the correlation
transfer from the squeezed vacuum to the atomic system. For large
interatomic separations, $\Gamma_{12} \approx 0$, and there is no
transfer of the correlations to the system. In contrast, for very
small separations, $\Gamma_{12}\approx \Gamma$, and then the
correlations are completely transferred to the atomic system.

However, the complete transfer of the correlations does not necessary
mean that the two-photon correlations are stored in the pure TPE state.
This happens only for two nonidentical atoms in the Dicke model, where
the steady-state is the pure
TPE state. For identical atoms separated by a finite distance $r_{12}$
only a part of the correlations can be stored in the antisymmetric state.
This can be shown, for example, by calculating of the interference
pattern of the fluorescence field emitted by the system. Using the
steady-state solutions~(\ref{t220}), we find that the visibility in
the interference pattern is given by~\cite{ft}
\begin{equation}
{\cal{V}} =
-\frac{2a|\tilde{M}|^{2}}{\tilde{N}\left(2\tilde{N}+1\right)^{3}
+2|\tilde{M}|^{2}\left[a^{2}+\left(2\tilde{N}+1\right)
-\left(2\tilde{N}+1\right)^{2}\right]} \ .\label{t228}
\end{equation}

\begin{figure}[t]
\begin{center}
\includegraphics[width=13cm]{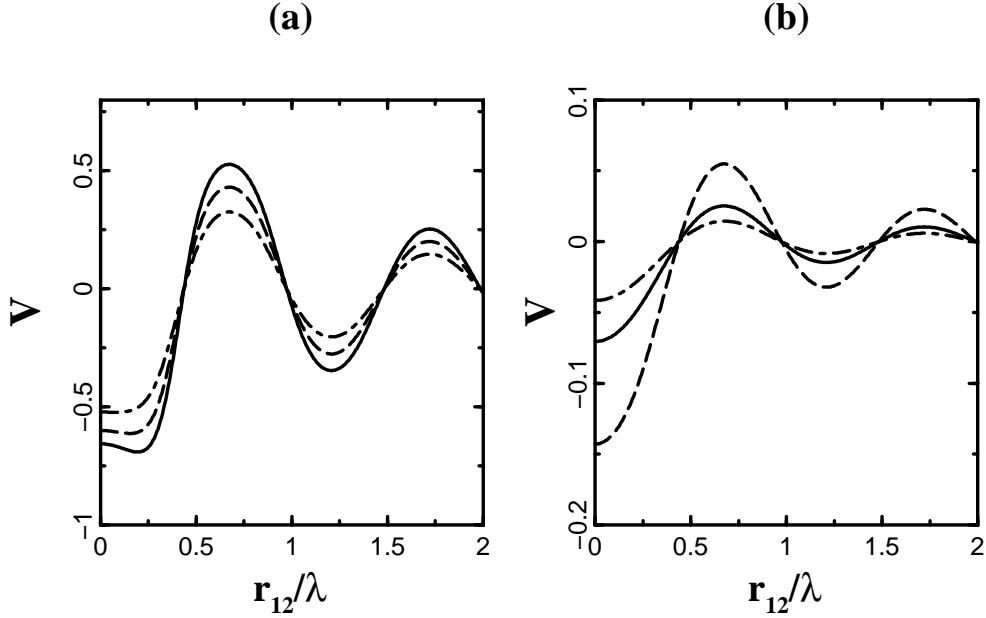}
\end{center}
\caption{The visibility ${\cal {V}}$ as a function of the interatomic
separation $r_{12}$ for (a) a quantum squeezed field with
$|\tilde{M}|^{2}=\tilde{N}(\tilde{N}+1)$, (b) a classical
squeezed field with $|\tilde{M}|=\tilde{N}$,
$\bar{\mu} \perp \bar{r}_{12}$ and different $\tilde{N}$:
$\tilde{N}=0.05$ (solid line), $\tilde{N}=0.5$ (dashed line),
$\tilde{N}=5$ (dashed-dotted line). }
\label{ftfig22}
\end{figure}
The visibility is negative indicating that the squeezing
correlations are mostly stored in the antisymmetric state.
In Fig.~\ref{ftfig22}, we plot the visibility ${\cal{V}}$ as a function
of the interatomic separation for a quantum squeezed field with
$|\tilde{M}|^{2}=\tilde{N}(\tilde{N}+1)$, Fig.~\ref{ftfig22}(a), and a
classical squeezed field with $|\tilde{M}|=\tilde{N}$,
Fig.~\ref{ftfig22}(b). An interference pattern with a dark center is
observed for small interatomic separations $(r_{12}<\lambda/2 )$ and
with the quantum squeezed field the visibility attains
the maximal negative value of
${\cal{V}}\approx -0.7$ for $r_{12}< 0.3\lambda$. According to
Eq.~(\ref{t222}), at these interatomic separations the antisymmetric
state is the most populated state of the system. The value
${\cal{V}}= -0.7$ compared to the possible negative
value ${\cal{V}}=-1$ indicates that $70\%$  of the squeezing
correlations are stored in the antisymmetric state.
In Fig.~\ref{ftfig22}(b), we show the visibility for a classical
squeezed field with $|\tilde{M}|=\tilde{N}$. The visibility is much
smaller than that in the quantum squeezed field and vanishes when
$\tilde{N}\rightarrow \infty$. In contrast, for the quantum squeezed
field $V$ approaches $-1/2$ when $\tilde{N}\rightarrow \infty$. Thus,
the visibility can provide the information about the degree of
nonclassical correlations stored in the entangled state $\ket a$.

\subsection{Mapping of the field fluctuations}\label{ftsec1042}

The fluctuations of the electric field are determined by the normally
ordered variance of the field operators as~\cite{park,fd1,fd2}
\begin{equation}
\langle :\left(\Delta E_{\theta}\right)^{2}:\rangle =
\sum_{\vec{k}s}E_{k}\left( 2\langle \hat{a}_{\vec{k}s}^{\dagger}
\hat{a}_{\vec{k}s}\rangle +\langle \hat{a}_{\vec{k}s}\hat{a}_{\vec{k}s}
\rangle e^{2i\theta}
+ \langle \hat{a}_{\vec{k}s}^{\dagger}\hat{a}_{\vec{k}s}^{\dagger} \rangle
e^{-2i\theta}\right) \ .\label{t229}
\end{equation}
Using the correlation functions~(\ref{t17}) of the three-dimensional
squeezed vacuum field and choosing $\theta
=\pi/2$, the variance of the incident squeezed vacuum field can be
written as
\begin{equation}
\left\langle :\left(\Delta E^{in}_{\pi/2}\right)^{2}:\right\rangle =
2E_{0}\left(\tilde{N} -|\tilde{M}|\right) \ ,\label{t230}
\end{equation}
where $E_{0}$ is a constant. Since
$|\tilde{M}|=\sqrt{\tilde{N}(\tilde{N}+1)}> \tilde{N}$, the
variance~(\ref{t230}) is negative indicating that the incident field
is in a squeezed state.

On the other hand, the normally ordered variance of the emitted
fluorescence field can be expressed in terms of the density matrix
elements of the two-atom system as
\begin{equation}
\left\langle :\left(\Delta E^{out}_{\theta}\right)^{2}:\right\rangle =
E_{0}\left(2\rho_{ss}+2\rho_{ee} +|\rho_{u}|\cos 2\theta \right)
\ .\label{t231}
\end{equation}
Using the steady-state solutions~(\ref{t210}) and choosing $\theta
=\pi/2$, we find
\begin{equation}
\left\langle :\left(\Delta E^{out}_{\pi/2}\right)^{2}:\right\rangle =
2E_{0}\frac{\left(\tilde{N} -|\tilde{M}|\right)}{2\tilde{N}+1}
\ .\label{t232}
\end{equation}
Thus, at low intensities of the squeezed vacuum field $(\tilde{N}\ll 1)$
the fluctuations in the incident field are perfectly
mapped onto the atomic system. For large intensities $(\tilde{N}>1)$, the
thermal fluctuations of the atomic dipoles dominate over the squeezed
fluctuations resulting in a reduction of squeezing in the fluorescence
field.

The idea of mapping the field fluctuations on the collective system of
two atoms have been extended to multi-level atoms. For example,
Kozhekin {\it et al.}~\cite{koz} proposed a method of mapping of
quantum states onto an atomic system based on the stimulated Raman
absorption of propagating quantum light by a cloud of three-level atoms.
Hald {\it et al.}~\cite{hal1,hal2} have experimentally observed
the squeezed spin states of trapped three-level atoms in the $\Lambda$
configuration, irradiated by a squeezed field. The observed squeezed
spin states have been generated via entanglement exchange with the
squeezed field that was completely absorbed by the atoms. The exchange
process was, however, accomplished by spontaneous emission and only
a limited amount of spin squeezing was achieved.
Fleischhauer {\it et al.}~\cite{flei} have considered
a similar system of three-level atoms and have found that quantum
states of single-photon fields can be mapped onto entangled states
of the field and the collective states of the atoms. This effect
arises from a substantial reduction of the group velocity of the field
propagating through the atomic system, which results in a temporary
storage of a quantum state of the field in atomic spins.
These models are two examples of the continuing fruitful investigation
of entanglement and reversible storage of information in collective
atomic systems.

\section{Conclusions}\label{ftsec11}

In this paper, we have reviewed the recent work on entanglement and
nonclassical effects in two-atom systems. We have discussed different
schemes for generation of nonclassical states of light and preparation
of two interacting atoms in specific entangled state. In particular,
we have presented different methods of preparing two atoms in the
antisymmetric state which is an example of a decoherence-free entangled
state. The ability to prepare two-atom system in the decoherence-free
state represents the ultimate quantum control of a physical system and
opens the door for a number of applications ranging from quantum
information, quantum computing to high-resolution spectroscopy.
However, the practical implementation of entanglement in information
processing and quantum computation requires coherent manipulation of a
large number of atoms, which is not an easy task. Although the two-atom
systems, discussed in this review, are admittedly elementary models, they
offers some advantages over the multiatom problem. Because of their
simplicity, we have obtained detailed and almost exact solutions
that can be easily interpreted physically, and thus provide insight into
the behavior of more complicated multiatom systems. Moreover, many results
discussed in this review is analogous to phenomena that one would expect
in multiatom systems. For example, the nonexponential decay of the
total radiation intensity from two nonidentical atoms is an elementary
example of superradiant pulse formation, and a manifestation of the
presence of coherences between the collective entangled states.
A number of theoretical studies have been performed recently on
entanglement and irreversible dynamics of a large number of
atoms~\cite{bbtk,koz,hal1,hal2,flei,sm01,b1,b2,nat1,nat2,nat3}.
These studies, however, have been limited to the Dicke model that
ignores antisymmetric states of a multi-atom system. Nevertheless, the
calculations have shown that population (information) can be stored in
the collective atomic states or in the so called
dark-state polaritons~\cite{fl00,fg02}, which are quasiparticles
associated with electromagnetically induced transparency in
multi-level atomic systems.

\section*{Acknowledgments}
This work was supported by the Australian Research Council.

\end{document}